\documentclass[11pt,a4paper]{article}
\usepackage[T1]{fontenc}
\usepackage[utf8]{inputenc}
\usepackage{lmodern}
\usepackage[margin=29mm]{geometry}
\usepackage{amsmath,amssymb,amsthm,mathtools,mathrsfs}
\usepackage{microtype}
\usepackage{lastpage}
\usepackage[all,defaultlines=3]{nowidow}
\usepackage{enumitem}
\usepackage{booktabs}
\usepackage{placeins}
\usepackage{float}
\usepackage{needspace}
\usepackage{xcolor}
\usepackage{tikz}
\usetikzlibrary{arrows.meta,positioning,calc}
\tikzset{manuscript illustration/.style={x=1cm,y=1cm,font=\small,line width=.65pt,
  box/.style={draw=black!75,rounded corners=2pt,fill=blue!10,inner xsep=5pt,inner ysep=3pt,align=center},
  key/.style={draw=black!75,rounded corners=2pt,fill=orange!16,inner xsep=5pt,inner ysep=3pt,align=center},
  dot/.style={draw=black!75,fill=white,circle,inner sep=2pt},
  edge/.style={draw=black!75,semithick},
  arr/.style={edge,-{Stealth[length=2mm]}},
  note/.style={align=center,inner sep=2pt,font=\footnotesize}}}
\usepackage[colorlinks=true,linkcolor=blue!50!black,citecolor=blue!50!black,urlcolor=blue!50!black]{hyperref}
\usepackage{aliascnt}
\usepackage[nameinlink,noabbrev,capitalise]{cleveref}
\crefname{appendix}{Appendix}{Appendices}
\Crefname{appendix}{Appendix}{Appendices}
\newtheorem{theorem}{Theorem}[section]
\newaliascnt{lemma}{theorem}
\newtheorem{lemma}[lemma]{Lemma}
\aliascntresetthe{lemma}
\newaliascnt{proposition}{theorem}
\newtheorem{proposition}[proposition]{Proposition}
\aliascntresetthe{proposition}
\newaliascnt{corollary}{theorem}
\newtheorem{corollary}[corollary]{Corollary}
\aliascntresetthe{corollary}
\newaliascnt{conjecture}{theorem}

\aliascntresetthe{conjecture}
\theoremstyle{definition}
\newaliascnt{definition}{theorem}

\aliascntresetthe{definition}
\newaliascnt{example}{theorem}

\aliascntresetthe{example}
\newaliascnt{question}{theorem}

\aliascntresetthe{question}
\AddToHook{env/theorem/before}{\Needspace{6\baselineskip}}
\AddToHook{env/lemma/before}{\Needspace{6\baselineskip}}
\AddToHook{env/proposition/before}{\Needspace{6\baselineskip}}
\AddToHook{env/corollary/before}{\Needspace{6\baselineskip}}
\theoremstyle{remark}
\newaliascnt{remark}{theorem}

\aliascntresetthe{remark}
\usepackage{thm-restate}
\renewcommand{\phi}{\varphi}

\newcommand{\col}{\operatorname{col}}

\title{Definable decompositions for graphs of bounded cliquewidth}
\author{Sebastian Siebertz\\[4pt]
\small University of Bremen\\[3pt]
\texttt{\href{mailto:siebertz@uni-bremen.de}{siebertz@uni-bremen.de}}}
\hypersetup{pdftitle={Definable decompositions for graphs of bounded cliquewidth},pdfauthor={Sebastian Siebertz}}
\date{}
\begin{document}
\maketitle
\begin{abstract}
\noindent The cliquewidth of a graph is the minimum number of colors needed by a clique expression that constructs it.
For every graph class of bounded cliquewidth, we construct a transduction in monadic second-order logic (MSO) that maps each graph in the class to at least one clique expression of uniformly bounded width for that graph.
\end{abstract}

\tableofcontents

\paragraph*{AI Declaration.}
OpenAI's Astra found the initial sketch of the proof in a conversation with me.
The first full draft of the proof had 41 pages.
In further conversations with Astra, I worked through the arguments and developed the exposition into the present paper of~\pageref*{LastPage} pages.
I give Astra full credit for the technical work.
\bigskip

\section{Introduction}

\subsection{Motivation and our results}

Structural decompositions are an important tool for understanding the logical and algorithmic properties of graphs.
In many applications, one first describes a graph by simpler pieces and then combines the information from these pieces along a tree.
For example, tree decompositions of bounded width, which originated in the work of Halin~\cite{halin} and were later developed by Robertson and Seymour~\cite{treewidthorigin}, allow us to process a graph via dynamic programming while keeping only a bounded amount of information at each bag.
Clique expressions construct a graph from labeled vertices by taking disjoint unions, adding edges between label classes, and changing labels.
Their expression trees allow a similar dynamic programming approach to graphs that may be dense, with the number of labels controlling the amount of information needed at each node.
The minimum number of labels needed to construct a graph is its cliquewidth.
The parameter was implicit in the graph operations of Courcelle, Engelfriet, and Rozenberg~\cite{cliquewidthorigin} and was defined explicitly by Courcelle and Olariu~\cite{courcelleolariu}.
Although treewidth and cliquewidth are defined in different ways, Courcelle and Olariu showed that every class of bounded treewidth has bounded cliquewidth~\cite{courcelleolariu}.
Cliquewidth is often viewed as a dense analogue of treewidth~\cite{linear,stable,pilipczuklens,siebertzvigny,generalizedcoloring}.

These tree-like descriptions are particularly useful in connection with monadic second-order logic (MSO) and counting monadic second-order logic~(CMSO), which extends MSO by predicates testing set cardinalities modulo fixed integers.
On finite ranked trees, where the node label fixes the number of children and the child positions are distinguished, CMSO, MSO, and tree automata have the same expressive power~\cite{courcelleengelfriet}.
When a bounded-width decomposition is given as such a tree, a CMSO property of the graph can be translated into a property of the tree and checked by a tree automaton~\cite[Theorem~6.3]{courcelleengelfriet}.
Courcelle used this approach to prove his famous model-checking theorem for graphs of bounded treewidth~\cite{courcelle1990}, and Courcelle, Makowsky, and Rotics proved the corresponding algorithmic result for MSO properties of graphs given by clique expressions of bounded width~\cite[Theorem~4]{cmr2000}.

This also raises a converse question.
Suppose a tree automaton recognizes a graph property on bounded-width decompositions and its answer depends only on the graph represented.
Can we then define the property by a CMSO sentence on the graph itself?
Courcelle conjectured that recognizability and CMSO-definability coincide on graphs of bounded treewidth~\cite[Conjecture~1 and its reformulation in~4.12]{courcellegap}.
One way to prove this conjecture was to use a CMSO transduction to construct a tree decomposition of bounded width from the graph itself.
Such a transduction describes its output by formulas on finitely many colored copies of the input.
One could then translate the CMSO acceptance condition on the decomposition back to the graph.
Courcelle also asked more generally under what conditions logical formulas could construct a hierarchical description of a graph \mbox{\cite[p.~87, question~(3)]{courcelleorder}}.
Boja\'nczyk and Pilipczuk proved Courcelle's conjecture by constructing such a decomposition with an MSO transduction for every fixed treewidth bound~\cite{treewidthlogic}.
In subsequent work, they gave an MSO transduction producing a decomposition of optimum width~\cite[Corollary~2.2]{optimizing}.
Boja\'nczyk, Grohe, and Pilipczuk considered graphs of bounded linear cliquewidth, which admit bounded-width expressions that add vertices one at a time.
For every fixed bound on linear cliquewidth, they constructed a cliquewidth decomposition of bounded width by an MSO transduction~\cite{linear}.
In both cases, the authors constructed the decomposition without counting and deduced CMSO-definability from recognizability by translating the CMSO acceptance condition on the decomposition back to the graph.

Rankwidth measures the complexity of cuts by the ranks of their adjacency matrices over the two-element field~$\mathbb{F}_2$.
Oum and Seymour showed that a graph class has bounded rankwidth if and only if it has bounded cliquewidth~\cite{rankwidth}.
Recently, Kalampakas showed that on graphs of rankwidth at most two, the restrictions of recognizable graph properties are exactly the properties definable there in CMSO~\cite[Theorem~1.1]{ranktwo}.
A further special case arises when a class of bounded cliquewidth excludes a fixed half-graph as a semi-induced subgraph.
This means that for some~$t$ no graph in the class has distinct vertices~$a_1,\ldots,a_t,b_1,\ldots,b_t$ with $a_i$ adjacent to $b_j$ exactly when~$i\leq j$, regardless of the edges within the two sides.
Ne\v{s}et\v{r}il et al.\ showed that such a class has structurally bounded treewidth, which means that every graph in the class is an output of a fixed first-order~(FO) transduction on some graph of bounded treewidth~\cite{stable}.
Their proof also gave a polynomial-time algorithm that constructs such a bounded-treewidth representation from the input graph~\cite{stable}.
Dreier, Gajarsk\'y, and Pilipczuk considered the more general setting of classes of structurally bounded expansion, where the class of bounded treewidth in this definition is replaced by a class of bounded expansion.
They gave an efficient procedure that constructs a colored graph from which a fixed FO interpretation recovers the input.
These colored graphs belong to a class of bounded expansion that is itself FO-transducible from the input class~\cite{reverse}.
In the bounded-cliquewidth setting, the representations have bounded cliquewidth and their bounded expansion then gives bounded treewidth~\cite{reverse,gw}.
For background on structural sparsity and sparsification we refer to~\cite{structuralexpansion,structuralnowheredense,pilipczuklens}.

Boja\'nczyk, Grohe, and Pilipczuk left open whether their decomposition theorem extended from bounded linear cliquewidth to bounded cliquewidth~\cite{linear}.
We prove that it does.
We use their version of clique expressions, in which a join operation in the cliquewidth expression may have any positive number of unordered inputs~\cite{linear}.

\begin{theorem}\label{thm:main}
For every positive integer $k$, there are a positive integer $\ell$ and an MSO transduction $\mathsf D_k$ that maps every graph of cliquewidth at most $k$ to at least one clique expression of width at most $\ell$.
Every output of $\mathsf D_k$ on a graph $G$ is such an expression, and the map sending each leaf to the input vertex of which it is a copy is an isomorphism from the graph defined by the expression to $G$.
\end{theorem}

\input{figures/main-dense-expression}

In \cref{fig:main-dense-expression}, the two children of the root construct the two sides of a complete bipartite graph.
The transduction may color the vertices with a fixed number of colors and make a fixed number of copies of each vertex.
The number of colors and the formulas may depend on $k$.
Its formulas do not use counting quantifiers, and the transduction takes the graph alone as its input.
\medskip

We work with rankwidth because it lets us describe highly connected regions of a graph by the sides they choose at cuts of small rank.
Such a consistent choice of sides is called a \emph{tangle}.
We follow the approach of Kalampakas for rankwidth two~\cite{ranktwo}.
The main case of his proof concerned split-prime graphs of rankwidth two.
In these graphs, every cut with at least two vertices on each side has rank at least two.
Each such graph has a unique order-two tangle, which Kalampakas used to construct a tree and obtain a linear ordering of the vertices in each local piece.
For arbitrary rankwidth, we obtain a linear ordering of blocks that may themselves need to be decomposed by induction.
We also have to separate several tangles, and for this we use a second induction.
We next give a detailed proof sketch. 

\subsection{Sketch of the proof}
\label{sec:proof-sketch}

Given a graph~$G$ of rankwidth at most~$k$, our main goal is to construct by MSO transduction a sparse graph~$S$ from which a second MSO transduction can recover~$G$.
In fact, we can give~$S$ finitely many unary colors and keep a designated copy of each vertex of~$G$.
A fixed MSO interpretation (not transduction because the colors are given now), the \emph{decoder}, selects these copies and can define the original adjacency on them.
When we later join several local encodings, these copies tell us where to attach them.

We prove a bound on the degeneracy of $S$, which is a uniform measure of its sparsity. 
That means, for some~$d$ depending only on~$k$, we show that every nonempty subgraph of~$S$ has a vertex of degree at most~$d$.
Since MSO transductions preserve bounded cliquewidth, the graphs~$S$ also have bounded cliquewidth~\cite{courcelleengelfriet}.
Together these two bounds give bounded treewidth~\cite{gw}.
From a representation of bounded treewidth with this decoder we can then obtain a clique expression for~$G$ (\cref{prop:boundedtwencoder,lem:treecliquecompiler}).
This last step is standard and we present the proof in the appendix for completeness. 

\paragraph{Induction on rankwidth.}
A rank decomposition of~$G$ is a tree of maximum degree three whose leaves are the vertices of~$G$.
Deleting a tree edge partitions the leaves into two sets and thus defines a cut of~$G$.
The \emph{cut-rank} of a set~$X$, written~$\rho(X)$, is the rank over the two-element field of the adjacency matrix between~$X$ and its complement.
We call this rank the \emph{order} of the cut.
The \emph{rankwidth} of~$G$ is the least~$k$ for which some rank decomposition has all its cuts of order at most~$k$.
Our construction uses two nested inductions, shown together in \cref{fig:proof-outline}.
The outer induction is on the rankwidth bound~$k$.
Oum characterized the graphs of rankwidth at most one as the distance-hereditary graphs, in which every connected induced subgraph preserves distances between its vertices~\cite{rwone}.
Each connected component has a \emph{split tree} whose leaves are the original vertices and whose internal nodes represent cliques or stars~\cite{gioanpaul}.
Campbell, Guillon, Kant\'e, Kim, and K\"ohler constructed a split decomposition by a CMSO transduction~\cite{cmsosplit}.
For our base case we adapt the MSO split-decomposition transduction of Campbell and K\"ohler~\cite[Corollary~7(3)]{laminar} to obtain the required encoding for graphs of rankwidth at most one (\cref{lem:rankonebase}).
For the induction step, fix~$k\geq 2$ and assume that a fixed MSO transduction constructs a clique expression of uniformly bounded width for every graph of rankwidth less than~$k$.
The inner induction keeps~$k$ fixed and separates tangles until we can use the outer induction again.

Tangles, introduced by Robertson and Seymour~\cite{tanglesorigin}, describe highly connected regions by the sides of cuts that contain them.
An \emph{order-$k$ tangle} chooses one side of every cut of order less than~$k$ so that any three chosen sides have a common vertex and no chosen side is a singleton.
We call cuts of order less than~$k$ \emph{low-order cuts}.
The side chosen by the tangle is \emph{big}, and the opposite side is \emph{small}.
Equivalently, the condition on triples of big sides says that no three small sides cover the vertex set.
By tangle duality, a graph has rankwidth less than~$k$ if and only if it has no order-$k$ tangle~\cite{ggrwduality, ggwtangles}.
So if~$G$ has no order-$k$ tangle, the outer induction applies directly to~$G$.

Now suppose first that~$G$ has exactly one tangle~$P$ of order~$k$.
A set~$X$ is called \emph{$P$-weak} if it is contained in a side~$A$ that~$P$ regards as small.
Suppose that the subgraph induced by~$X$ had an order-$k$ tangle.
We could orient each cut of~$G$ of order less than~$k$ by asking which side this tangle chooses when we restrict the cut to~$X$.
This gives a tangle of~$G$ in which~$A$ is big, since~$A\cap X=X$ is the whole vertex set of the induced graph (\cref{lem:inducedtanglelift}).
This would give a second tangle of~$G$.
Thus every nonempty~$P$-weak set induces a graph of rankwidth less than~$k$, and the outer induction gives a decomposition of each such graph.
To finish this case, we must still combine the decompositions of the weak pieces into one for~$G$.
This is what \cref{thm:weakpieceencoding} gives us.
Given a fixed MSO transduction that decomposes each nonempty weak piece on its own, the theorem gives a transduction that constructs a labeled tree for the whole graph.
The decoder recovers~$G$ on its designated leaves, and \cref{lem:treecliquecompiler} turns this tree into a clique expression.

\paragraph{Combining the weak pieces around one tangle (proof of \cref{thm:weakpieceencoding}).}
To carry out the construction around the unique order-$k$ tangle~$P$, we need to describe its big sides in MSO.
Grohe and Schweitzer showed that there exists a set~$W$ of bounded size that meets the intersection of every three big sides~\cite{triplecovers}.
In particular, a big side and a small side cannot have the same intersection with~$W$, since~$W$ meets the intersection of the big side with the complement of the small side.
For this test, it already suffices that~$W$ meet the intersection of every two big sides.
Adapting an argument of Seibt~\cite{witnesses}, we show that we can choose such a pair witness with at most~$2^k-1$ vertices inside any prescribed big side.
To describe the tangle, it therefore suffices to specify the vertices of~$W$ and list all subsets of~$W$ that occur as intersections of big sides with~$W$ (\cref{lem:localpairwitness,cor:tangleparameters}).

If at most~$8$ small sides~$A_1,\ldots,A_t$ cover the graph, we consider the blocks~$B_i=A_i\setminus\bigcup_{j<i}A_j$, omitting empty blocks.
Each~$B_i$ is weak because it is contained in~$A_i$, but its own cut-rank may be at least~$k$.
The blocks up to~$B_i$ have union~$A_1\cup\cdots\cup A_i$.
By subadditivity of cut-rank, $\rho(X\cup Y)\leq\rho(X)+\rho(Y)$.
Since each~$A_j$ has cut-rank at most~$k-1$, the prefix has cut-rank at most~$i(k-1)\leq 8(k-1)$.
We show in \cref{lem:linearclosure} that under this bound on the cut-rank of the prefixes, an MSO transduction can combine the block decompositions into a clique expression of bounded width for~$G$.
Its expression tree gives the required encoding, with one leaf for each original vertex, which completes this case.

Otherwise~$P$ is called \emph{robust}~\cite{flowers}.
The tangle~$P$ orients every cut of order strictly less than~$k$, so each such cut has a small side and hence a weak side.
At order exactly~$k$, neither side need be contained in a~$P$-small set.
We call a cut of this order with two such sides a \emph{strong separation}.
Starting from one (either one) side~$X$, we add nonempty~$P$-weak sets from the opposite side whenever the enlarged side still has cut-rank at most~$k$.
If we can reach the whole vertex set in this way from at least one side, the separation is \emph{sequential}.
If neither side can reach the whole vertex set, the separation is \emph{non-sequential}.
All maximal sequences of such additions from~$X$ give the same set, the \emph{full closure} of the starting side~$X$, which we write as~$c(X)$~\cite[Lemma~3.6]{flowers}.

Two non-sequential separations are equivalent if they have the same unordered pair of full closures.
Two classes are \emph{compatible} if they have representatives~$(A,B)$ and~$(C,D)$ whose sides can be named so that~$A\subseteq C$.
We call a class \emph{strong} if it is compatible with every other class.
Clark and Whittle proved that every robust tangle~$P$ admits a tree with bag nodes and \emph{flower nodes} that displays a representative of every non-sequential separation class~\cite{flowers}.
The bags partition~$V(G)$.
Deleting a tree edge gives two components, and the union of the bags in each component gives one side of the displayed cut.
At a flower node, the incident branches can be grouped in different ways, so a single node can display several crossing cuts.

We construct an inclusion tree from the strong classes.
To orient such a class, we choose which of its two full closures comes first.
A side represents this orientation if it belongs to a cut in the class and has the chosen full closure.
Among all sides representing a given orientation, one is contained in every other such side.
We call it the \emph{core} of that orientation.
For a representative with first side~$X$ and opposite side~$Y$, the core is~$V(G)\setminus c(Y)$ \mbox{\cite[Lemmas~3.7 and~3.8]{flowers}}.
As in the rankwidth-two argument of Kalampakas~\cite{ranktwo}, we choose an orientation of one strong class as an anchor, if such a class exists, and orient the other strong classes relative to it.
These cores form a \emph{laminar} family, meaning that any two are disjoint or one contains the other (\cref{lem:generalcorelaminar}).
We add the whole vertex set and the singletons, and use the transduction of Campbell and K\"ohler~\cite{laminar} to construct the inclusion tree.
The children of a set are its maximal proper subsets in this family, so the singleton leaves represent the original vertices.
If there is no strong class, the inclusion tree has a single internal node.
For each internal node~$u$, write~$S_u$ for the set of original vertices at leaves below~$u$.
Each edge from~$u$ to another internal node represents a separation class.
At~$u$ we take the orientation pointing across this edge to its other endpoint and call its core the \emph{outward core} at~$u$ for that edge.
Thus the same edge gives opposite orientations at its two endpoints.
For an edge to an internal child~$v$, this is the orientation whose core~$S_v$ defines the child node in the inclusion tree.
If~$u$ has a parent, the outward core towards the parent is the core of the reverse orientation of the separation class at~$S_u$.
For this orientation, the opposite side is~$S_u$, so the parent-side core is~$V(G)\setminus c(S_u)$.
It is strictly smaller than~$V(G)\setminus S_u$ whenever the full closure~$c(S_u)$ strictly contains~$S_u$.
At a fixed node the outward cores are disjoint.
We contract each to a single element and measure a cut of the new ground set by the cut-rank of the original vertices on its two sides.
The new elements still admit a decomposition tree of maximum degree three in which every cut has order at most~$k$ (\cref{cor:simultaneouscores}).
To prove this, we adapt a construction of Hall, Oxley, Semple, and Whittle~\cite{hosw}.

We now construct a graph at each internal node of the inclusion tree.
Our aim is to decompose these graphs separately and then join their decompositions along the tree.
The tree of Clark and Whittle tells us how to order the blocks at a node~$u$.
We can choose this tree so that each strong class is displayed by an edge.
Cut these edges and replace each remaining component by one node.
When several edges display the same class, they form a path that we replace by one edge.
The resulting tree is the inclusion tree with its singleton leaves omitted (\cref{thm:generalclassnormal}).
For the component at~$u$, we keep the vertices of~$G$ outside the outward cores at~$u$, together with the edges between them.
We replace each outward core~$F$ by a set of auxiliary vertices called \emph{ports}.
The ports describe how vertices in~$F$ are adjacent to the vertices we keep and to vertices in the other outward cores.
We call the resulting graph the \emph{port graph} of the component.
Following the port construction of Kalampakas for rank two~\cite{ranktwo}, we choose a basis of the rows across the cut around~$F$ and introduce one port for each nonzero binary combination of its~$k$ basis rows.
This combination gives the adjacency of the port to each original vertex we keep.
The component contains at most one flower node.
Using its structure and the width bound after contraction, we obtain an ordered partition of the port graph into blocks (\cref{lem:localweakblocks}).
Each block is either a~$P$-weak set of original vertices or the whole port set for one outward core.
In this graph the union of the first~$i$ blocks has cut-rank at most~$2k$ for every~$i$.

The local graph we actually encode at~$u$ needs only the original vertices at its leaf children and the ports for its internal children.
At a nonroot node, we therefore delete the ports for the parent direction and any original vertices outside~$S_u$.
Restricting the blocks to this graph preserves the prefix bound.
For each internal child we add a marker adjacent to all its ports and put it in the same block, so it contributes no edge across a completed prefix.
Since~$P$ is the unique order-$k$ tangle in the case at hand, each weak block induces a graph of rankwidth less than~$k$.
We apply the outer induction to decompose these blocks.
The port blocks have bounded size and can be decomposed directly.
Boja\'nczyk, Grohe, and Pilipczuk used Simon's factorization forest theorem~\cite{simon} to combine single vertices along a linear decomposition~\cite{linear}.
In \cref{lem:linearclosure} we adapt their argument to blocks that already have decompositions.
One MSO transduction can therefore decompose all these local graphs in parallel, without being given the block orders.
We join the local trees by linking the marker for each child to the root of its tree.
The resulting labeled tree encodes the whole graph~$G$ on its designated leaves (\cref{thm:weakpieceencoding}).
By \cref{lem:treecliquecompiler}, we obtain a clique expression of bounded width for~$G$, completing the construction when~$P$ is the unique order-$k$ tangle.

\paragraph{Induction at fixed rankwidth.}
Suppose now that~$G$ has several order-$k$ tangles.
If we fix one such tangle~$P$, a tangle in a~$P$-weak piece may lift to a different tangle of~$G$, so the piece may still have rankwidth~$k$.
A cut \emph{distinguishes} two tangles if they choose opposite sides.
We fix one tangle as an anchor and consider, for each other tangle~$Q$, the cuts of minimum order distinguishing it from the anchor.
Among the sides of minimum distinguishing cuts that are big in~$Q$ and small in the anchor, there is a unique smallest one under inclusion.
This is called the \emph{closest set} of~$Q$, following the leftmost minimum separation of Grohe and Schweitzer~\cite[Lemma~2.8]{computingtangles}.
By their uncrossing argument, these sets are nested or disjoint (\cref{lem:closestunique}), so we can again arrange them in an inclusion tree with~$V(G)$ as its root and the individual vertices as leaves.
For each internal node, write~$A$ for the set of vertices below it.
To construct the local graph~$H_A$, we keep the vertices of~$A$ outside its internal child sets, with all edges of~$G$ between them.
The sets at its internal children, together with~$V(G)\setminus A$ when~$A$ is not the root, are the \emph{directions} at~$A$.
We replace each direction by a graph of bounded size.
The vertices of these replacements represent neighborhoods across the corresponding cuts.
Here a replacement must preserve the tangles that make all the replaced sides small, as well as the adjacencies across the cut.
For this purpose we partition each direction into a bounded number of \emph{titanic} pieces.
In every partition of such a piece into at most three parts, some part has order at least that of the piece.
Contracting these pieces preserves the width bound by a result of Hlin\v{e}n\'y and Oum \mbox{\cite[Lemma~3.4]{hlinenyoum}}, and the tangle lifting theorem of Geelen, Gerards, and Whittle describes the tangles after contraction~\cite[Lemmas~4.3 and~4.4]{ggwtangles}.
Hlin\v{e}n\'y and Oum used uniform matroids to realize replacements with prescribed connectivity~\cite[Section~4]{hlinenyoum}.
For cut-rank we replace each titanic piece by a clique with one port for each vector in the row space of the adjacency matrix between the piece and its complement.
The set of ports is again titanic (\cref{lem:affineporttitanic}).
Each cut comes from a parent-child edge of the inclusion tree, and we choose a basis of its rows on the child side.
For fixed~$k$, a finite table then describes the edges among the ports of each direction and their adjacencies across the cut in these coordinates.
We test which tables come from a suitable partition and use the first one that does.
This lets us define the replacements throughout the tree without having to choose all the partitions at once (\cref{lem:titanicpartition,lem:finitegadgettype}).
The original vertices we keep and the ports for all the directions together form~$H_A$.
By \cref{prop:virtualtitanicports},~$H_A$ has rankwidth at most~$k$, but it may still have several order-$k$ tangles.

A tangle of~$H_A$ may have a big side of order less than~$k$ contained entirely in one of the replacements.
To recognize such tangles, we mark the vertex set of each replacement and call it a \emph{support}.
A support \emph{captures} a tangle if it contains a low-order side that is big in that tangle.
We show below that once at most one tangle remains uncaptured, we can finish the construction by applying the outer induction to suitable induced subgraphs.

We therefore formulate the inner induction for a graph~$H$ of rankwidth at most~$k$ together with its supports.
We represent each support by an extra vertex, its \emph{index}, joined to the vertices in that support.
These membership edges alone form the support incidence graph.
A fixed MSO transduction can construct~$H$ and all the support incidences together from the original input, and the size of each support is bounded.
Initially~$H=G$ and the support family is empty.
After a separating step,~$H$ is one of the local graphs, with the supports from its new replacements and those carried over from the preceding step.
Let~$r$ be a lower bound on the minimum order distinguishing any two uncaptured tangles, so that they agree on every cut of order less than~$r$.
Initially we take~$r=0$.
By induction on~$k-r$, we construct a sparse encoding of~$H$ from which we can recover both~$H$ and its support incidences (\cref{lem:recursiveencoding}).
If~$r=k$, any two uncaptured tangles agree on every cut they orient, so at most one remains.
More generally, the following argument applies as soon as at most one tangle remains uncaptured.

\paragraph{The terminal case.}
Suppose first that every order-$k$ tangle of~$H$ is captured.
The incidence graph has bounded cliquewidth by construction.
It is also sparse, since every edge has one end at a support index and each index has bounded degree.
It therefore has bounded treewidth~\cite{gw}.
On the vertices of~$H$, form the \emph{conflict graph} by joining two vertices whenever they belong to a common support.
In a tree decomposition of the incidence graph, replace each support index in a bag by its bounded set of neighbors.
This gives a bounded-width tree decomposition of the conflict graph, as in the construction of Samer and Szeider \mbox{\cite[Lemma~2(3)]{incidenceprimal}}.
We can therefore properly color the conflict graph with a bounded number of colors, giving different colors to all vertices in any one support (\cref{lem:supportincidence}).
We claim that the subgraph induced by any one or two colors has rankwidth less than~$k$.
An order-$k$ tangle of this subgraph would lift to a tangle of~$H$ by orienting each low-order cut according to its restriction to the subgraph (\cref{lem:inducedtanglelift}).
The lifted tangle has a big side contained in a support.
Since the colors are distinct on the support, this big side restricts to at most two vertices.
It would still be big in the tangle of the induced subgraph.
Since~$k\geq2$, each singleton has order less than~$k$ and is small.
If a side of size at most two were big, its complement and the singletons in that side would cover the graph with at most three small sides (\cref{lem:rainbowregions}).
The first induction now gives decompositions for all the one- and two-color subgraphs.
Every pair of vertices occurs together in one of them.
Following the finite-cover construction of Gajarsk\'y et al.\ in their sparsification theorem~\cite[Section~5.3]{structuralexpansion}, we link each vertex to its copies in these encodings to obtain a sparse representation of~$H$ (\cref{thm:finite-cover-encoding}).
As at the start of the sketch, this representation has bounded treewidth, and its decoder lets us turn it into a clique expression for~$H$.

Suppose next that one tangle~$P$ remains uncaptured.
We can then apply the same coloring argument inside each~$P$-weak set.
Indeed, a tangle in a one- or two-color restriction of this set would lift to a tangle different from~$P$, since any~$P$-small side containing the weak set is big in the lifted tangle.
Every such lift is captured.
The outer induction gives decompositions for the subgraphs on one or two colors inside each weak piece.
The same finite-cover construction then gives a decomposition of the weak piece itself (\cref{lem:weakrainbowblocks,lem:rainbowblockdcw}).
Combining these decompositions around~$P$ as before gives a cliquewidth decomposition of~$H$ (\cref{thm:terminalsupport}).
In either terminal case we use the resulting expression tree as the sparse encoding and join a copy of each support index to the designated leaves of its members.
These indices have bounded degree, and deleting them gives back the expression tree.

\paragraph{Separating the uncaptured tangles.}
Suppose now that at least two order-$k$ tangles of~$H$ remain uncaptured.
Choose one of them as an anchor and form the closest sets of the other uncaptured tangles.
As before, these sets are nested or disjoint (\cref{lem:closestunique}).
Together with the whole vertex set and the singletons, they form an inclusion tree~$D$.
We repeat the construction of the local graph~$H_A$ at each internal node~$A$, this time starting from~$H$.
We say that~$A$ \emph{owns} a vertex of~$H$ if it is a leaf child of~$A$ in~$D$.
These are exactly the vertices we keep in~$H_A$ when we replace the directions at~$A$ by their ports.
For the induction we need to know which tangles survive in~$H_A$ and at what order they can be distinguished.
The replacements preserve the tangles that regard all replaced regions as small, together with the minimum order at which any two of them can be distinguished (\cref{prop:virtualtitanicports} and \cref{cor:gadgetdistinguishingorder}).

We keep the vertex set of each replacement graph as a new support.
We carry an old support to a node if the node owns a member of the support or if the support meets at least three directions there.
At each such node we keep the local members and include the replacements for the regions the support meets (\cref{lem:supportlocations,lem:supportcapture,lem:supportcopies}).
Indeed, if the support lies in just one or two directions that a tangle regards as small, it cannot contain a big side of that tangle.
Otherwise those directions and the complement of the big side would cover the graph with at most three small sides.
With these supports, every uncaptured tangle of~$H_A$ corresponds to an uncaptured tangle of~$H$ that regards every replaced region at~$A$ as small.
Among the uncaptured tangles of~$H$, only the anchor makes every direction at the root small.
At a nonroot node the corresponding tangles all have closest set~$A$.

This gives the increase in~$r$ needed for the second induction.
The cut around~$A$ distinguishes each of these tangles from the anchor, so its order is at least~$r$.
Why must two tangles with closest set~$A$ agree on every cut of order at most the order of~$A$?
If they differed on such a cut, one would choose the side that the anchor regards as small.
The cut would then have minimum order for that tangle and the anchor, so its side that is small in the anchor would contain~$A$.
Both tangles regard~$A$ as big, so neither can choose the opposite side, which is disjoint from~$A$.
Their minimum distinguishing order is preserved in~$H_A$ and is therefore at least~$r+1$ (\cref{cor:ownerthreshold}).
We apply the terminal construction at the root and the second induction with this new lower bound at the other internal nodes.
We can again apply the transduction in parallel to all nonroot local graphs by the parallel application lemma of Boja\'nczyk, Grohe, and Pilipczuk.
Along each branch of the recursion there are at most~$k-r$ separating stages, although the inclusion tree at any one stage may have arbitrarily large depth.

\paragraph{Assembling the sparse representation.}
For each internal node~$v$ of~$D$, we now have a sparse encoding~$S_v$ of its local graph~$H_v$.
Its decoder recovers~$H_v$ on distinguished copies of the vertices owned by~$v$ and of the ports for its incident cuts.
Taking the disjoint union of these encodings allows us to recover every local graph.
To recover the input graph~$H$, we also need to determine adjacency between original vertices owned by different nodes.

In \cref{sec:induction} we keep the internal nodes of~$D$ with the edges between them and attach each local encoding~$S_v$ to its node~$v$.
More precisely, we use~$v$ as a hub and join each vertex of~$S_v$ to~$v$ by a marked path of length two.
These paths identify the domain of the local decoder even when~$S_v$ is disconnected.
At each nonroot node we choose a basis for the rows of the cut between its vertices and the rest of~$H$.
When a basis vertex at a parent lies below a child, we keep that vertex in the same position in the child's basis.
We use one label at each node to say which positions are occupied.
For each of the~$k$ positions, one unary predicate marks the child leading to the chosen basis vertex.
This extends the two-predicate construction of Kalampakas~\cite[Lemma~5.1]{ranktwo} (\cref{lem:coherentgeneralbases}).
For each parent-child edge we add a vertex adjacent to the two hubs and to the ports for this edge at either end.
Its labels tell us which replacements occur at the two ends and give a transition matrix that expresses the child rows, restricted to vertices outside the parent set, in the parent basis (\cref{eq:generaltransition}).
We also keep the incidences of the current supports with the distinguished copies of their members.
Together, the tree, the local encodings, and these attachments form the sparse representation of~$H$.
Indeed, the tree edges and the membership paths form two forests, and there are only boundedly many attachments for each tree edge and each support index.
Their union with the disjoint local encodings therefore has bounded degeneracy.

To recover adjacency between two original vertices, consider the least common ancestor~$v$ of their owners.
Suppose first that the vertices lie in different child sets of~$v$.
For each vertex we start with its row coordinates at its owner and apply the transition matrices up to the child of~$v$ that contains it.
The resulting coordinates describe its neighborhood outside that child set.
The port rows span this space, so we can express each vertex's row as a sum of port rows.
Adding the corresponding entries of the adjacency matrix between the two port sets in~$H_v$, over the two-element field, then gives the adjacency of the original pair.
If one vertex is owned by~$v$, we use that vertex directly in~$H_v$ and carry only the coordinates of the other vertex.
If both are owned by~$v$, their adjacency is already given by~$H_v$.
For fixed~$k$ there are only finitely many coordinate vectors, so an MSO formula can check the changes of coordinates along the two paths.

This completes the induction on~$k-r$.
We now apply \cref{lem:recursiveencoding} to the original graph~$G$ with an empty support family, so every order-$k$ tangle is initially uncaptured.
We may take~$r=0$ because every cut distinguishing two such tangles has nonnegative order.
The lemma therefore gives the sparse representation~$S$ of~$G$ by a fixed MSO transduction.
Using the tree-decomposition transduction of Boja\'nczyk and Pilipczuk~\cite{treewidthlogic}, we encode~$S$ by a labeled forest and join its components under a new root (\cref{prop:boundedtwencoder}).
We compose the interpretation on this tree with the decoder for~$S$.
The tree-to-expression construction of Courcelle and Engelfriet~\cite[Theorem~7.34(1)]{courcelleengelfriet}, in the form of \cref{lem:treecliquecompiler} for unordered trees, gives a clique expression with the original vertices as its leaves.
We have thus constructed clique expressions of bounded width for graphs of rankwidth at most~$k$, using the outer induction only on graphs of smaller rankwidth.

\input{figures/proof-outline}
\FloatBarrier

\section{Preliminaries: graphs, transductions, and encodings}
\label{sec:preliminaries}

All graphs in this paper are finite, nonempty, undirected, and simple.
We write $V(G)$ for the vertex set of a graph $G$, $E(G)$ for its edge set, and $G[X]$ for its subgraph induced by a set $X\subseteq V(G)$.

Starting with labeled vertices, we may take disjoint unions, join all vertices with one label to all vertices with another label, and change labels.
The least number of labels needed to construct $G$ is its \emph{cliquewidth}, denoted by $\operatorname{cw}(G)$.

For the decompositions we construct here, we use the version of cliquewidth expressions of Boja\'nczyk, Grohe, and Pilipczuk with unbounded joins~\cite{linear}.
Fix a positive integer $k$ and write $[k]=\{1,\ldots,k\}$.
A \emph{$k$-colored graph} is a graph $G$ together with a map $c:V(G)\to[k]$.
We use the following operations on these graphs.
\begin{itemize}
\item \emph{Constant.}
For each $i\in[k]$, the operation $\mathsf{Constant}_i$ has no inputs and gives a graph with one vertex of color $i$.
\item \emph{Recolor.}
For each map $f:[k]\to[k]$, the unary operation $\mathsf{Recolor}_f$ replaces the coloring~$c$ by $f\circ c$ and leaves the graph unchanged.
\item \emph{Join.}
Let $S\subseteq\{X\subseteq[k]:1\leq |X|\leq2\}$.
The operation $\mathsf{Join}_S$ takes any finite nonempty family of $k$-colored graphs, first takes their disjoint union, and then adds an edge between vertices $u,v$ from different input graphs if and only if $\{c(u),c(v)\}\in S$.
It keeps all vertex colors and all edges within each input graph.
A singleton $\{i\}\in S$ permits edges between vertices of the same color $i$ in different inputs, and $S=\varnothing$ gives disjoint union.
\cref{fig:cliquewidth-join} illustrates this operation.
\end{itemize}

\input{figures/cliquewidth-operations}

A \emph{cliquewidth decomposition of width $k$} is a rooted tree whose nodes are labeled by these operations.
Each Constant node is a leaf, each Recolor node has one child, and each Join node has a finite nonempty set of children.
We evaluate the tree from the leaves to the root, applying each operation to the colored graphs obtained at its children.
The decomposition represents $G$ if the graph obtained at the root is isomorphic to $G$ after forgetting the colors.
Since Join does not depend on the order of its inputs, we do not impose an order on the children of an expression node.
The minimum width in this version differs from the usual cliquewidth by at most a factor of two, so the two definitions give the same classes of bounded cliquewidth~\cite{linear}.

\medskip

In the \emph{adjacency representation} of a graph, the elements are its vertices and the only relation is adjacency.
In the \emph{incidence representation}, both vertices and edges are elements, and the incidence relation says which vertices are the endpoints of each edge.
We use \emph{monadic second-order logic} (MSO) on the adjacency representation unless we specify otherwise.
It permits quantification over vertices and sets of vertices.
The transductions we construct use neither edge-set quantification nor modulo-counting predicates.
A formula $\phi(x_1,\ldots,x_k)$ \emph{defines} a relation $R$ in a structure $\mathfrak A$ when
\[
(a_1,\ldots,a_k)\in R
\quad\Longleftrightarrow\quad
\mathfrak A\models\phi(a_1,\ldots,a_k).
\]
\medskip

We use the following presentation of MSO transductions~\cite{linear,optimizing}.
An \emph{MSO transduction} is a fixed finite composition of the following operations.
\begin{itemize}
\item \emph{Coloring.}
Add a unary predicate with an arbitrary interpretation, allowing a different output for each choice.
\item \emph{Filtering.}
An MSO sentence determines whether the structure obtained so far is kept or rejected.
\item \emph{Copying.}
For a fixed positive integer $c$, replace each element $v$ by the elements $(v,1),\ldots,(v,c)$.
Unary predicates distinguish the layers, and a binary predicate identifies pairs with the same original element.
Each original relation holds on a tuple of copies if and only if it holds on their original elements, regardless of their layers.
\item \emph{Interpretation.}
The universe stays unchanged, and each output relation $R$ is specified by an MSO formula $\varphi_R(x_1,\ldots,x_r)$, where $r$ is its arity.
A tuple belongs to $R$ if and only if it satisfies this formula in the current structure.
\item \emph{Universe restriction.}
An MSO formula $\delta(x)$ selects the elements to keep, and all relations are restricted to these elements.
\end{itemize}

These operations may be composed in any fixed order and may be used more than once.
After coloring and copying the input, the formulas used in subsequent operations may refer to the layer predicates, the relation between copies, and the chosen colors.
In the end, a single formula $\eta(x,y)$ can use this information to specify the output adjacency relation.
In intermediate structures we may use several relations, for instance separate relations for tree edges and for membership in a support.
Their names and arities form the \emph{signature} of the structure.
Every MSO transduction has a normal form in which we first choose the colors and apply a filter, then copy, interpret the output relations, and restrict the universe \mbox{\cite[Theorem~6.2]{optimizing}}.
At the end we may rename relations and discard auxiliary ones.

We will use MSO transductions to encode an input graph $G$ by an output structure $H$ from which we can recover $G$.
The \emph{origin} of an element of $H$ is the vertex of $G$ from which it was copied.
We call the encoding \emph{source-preserving} when $H$ contains one distinguished copy~$\widehat v$ of each vertex $v\in V(G)$ and a fixed MSO interpretation $\mathsf I$, applied to $H$, recovers~$G$ on these copies.
We call~$\mathsf I$ the \emph{decoder}, or an \emph{MSO inverse}, of the encoding.
Here we allow the interpretation to restrict its domain to the distinguished copies after defining adjacency in~$H$.
The resulting graph $\mathsf I(H)$ has an edge between $\widehat u$ and $\widehat v$ if and only if $uv\in E(G)$.
Thus, the origin map $\widehat v\mapsto v$ is an isomorphism from $\mathsf I(H)$ to the input graph $G$.
The same interpretation $\mathsf I$ must work for every accepted output of the transduction, and every intended input graph must have at least one accepted output.
If the input vertices carry a fixed finite collection of unary labels, we also keep these labels on their distinguished copies and recover them when applying $\mathsf I$.

\input{figures/source-preserving-example}

We use the backwards translation theorem for MSO transductions~\cite[Theorem~7.10]{courcelleengelfriet}.
Let $\mathsf T$ be an MSO transduction in normal form, and let $\psi$ be an MSO sentence over its output signature.
There is an MSO formula $\psi^{\mathsf T}(\bar P)$ over the input signature such that, for every input $\mathfrak A$ and every choice of the unary parameters $\bar P$ that passes the filter, the resulting output $\mathfrak B=\mathsf T(\mathfrak A,\bar P)$ satisfies
\[
\mathfrak B\models\psi
\quad\Longleftrightarrow\quad
\mathfrak A\models\psi^{\mathsf T}(\bar P).
\]
To express that some output satisfies~$\psi$, we ask whether there is a choice of~$\bar P$ that passes the filter and satisfies~$\psi^{\mathsf T}(\bar P)$ on the input.

The \emph{Gaifman graph} of a relational structure has the same vertex set, and two distinct vertices are adjacent exactly when they occur together in a tuple of one of its relations.
For a signature with only unary and binary relations, it is the undirected graph obtained by combining all binary relations and ignoring loops.
The unary relations do not add edges.
In a final tree encoding, the tree adjacency is the only binary relation, and all other labels are unary.

Courcelle and Engelfriet proved the following preservation property, including the case of graphs with finitely many labels~\cite[Corollary~7.38(2)]{courcelleengelfriet}.
\begin{lemma}
\label{lem:transduction-preserves-cliquewidth}
For every MSO transduction $\mathsf T$ from graphs to graphs and every integer $k$, there is an integer $k'$ such that, if $\operatorname{cw}(G)\leq k$ and $H\in\mathsf T(G)$, then $\operatorname{cw}(H)\leq k'$.
The same statement holds when the input graphs have a fixed finite collection of unary labels.
\end{lemma}

When we enrich an intermediate structure by several relations, we always keep the stronger condition that all of them are jointly obtained by a fixed MSO transduction from a bounded-cliquewidth graph class.
Thus, every later graph projection has bounded cliquewidth, by composition and \cref{lem:transduction-preserves-cliquewidth}.

Following Boja\'nczyk, Grohe, and Pilipczuk~\cite{linear}, a \emph{decomposer} of width $k$ is an MSO transduction such that, for every graph $G$, each output on $G$ is a cliquewidth expression of width at most $k$ and the map sending each leaf to its origin is an isomorphism from the graph defined by the expression to $G$.
We say that a decomposer \emph{accepts} $G$ if it outputs at least one cliquewidth decomposition of $G$.
We measure the \emph{size} of a transduction as in~\cite{linear}, by summing the sizes of its operations.
For each operation, we count the sizes of its input and output signatures, including the arities, and add the largest quantifier rank of its formulas (zero if it uses no formulas).
The quantifier rank of a formula is the maximum nesting depth of its quantifiers.
The \emph{definable cliquewidth} of a graph $G$, denoted by $\operatorname{dcw}(G)$, is the minimum size of a decomposer that accepts $G$.
All bounds on definable cliquewidth also apply when we keep a fixed finite collection of unary labels, by refining the construction colors according to these labels.

In this terminology, the transduction in \cref{thm:main} is a decomposer.
For each cliquewidth bound $k$, its size gives a uniform bound on definable cliquewidth.

\begin{restatable}{theorem}{generaldecomposition}
\label{thm:generaldecomposition}
For every positive integer $k$, there is an MSO decomposer $\mathsf D_k$ that accepts every graph of cliquewidth at most $k$.
Consequently, there is a function $f:\mathbb N\to\mathbb N$ such that
\[
 \operatorname{dcw}(G)\leq f(\operatorname{cw}(G))
\]
for every graph $G$.
\end{restatable}

\subsection{Sparse encodings}
\label{sec:sparse-encodings}

A graph is \emph{$d$-degenerate} if every nonempty induced subgraph has a vertex of degree at most~$d$.
We call a graph class \emph{uniformly sparse} if its graphs have bounded degeneracy.

A \emph{tree decomposition} assigns a set of vertices, called a \emph{bag}, to each node of a tree.
The bags containing each vertex must form a nonempty connected subtree, and the endpoints of every edge must occur together in a bag.
The width is the largest bag size minus one, and the \emph{treewidth} of a graph is the minimum width of a tree decomposition.

We use the following characterization throughout the proof.
A graph class has bounded treewidth if and only if it has bounded cliquewidth and is uniformly sparse.
Indeed, graphs of treewidth at most~$w$ are $w$-degenerate and have cliquewidth bounded in terms of~$w$~\cite{courcelleolariu}.
Conversely, a $d$-degenerate graph excludes $K_{d+1,d+1}$ as a subgraph, and Gurski and Wanke proved that a class of bounded cliquewidth excluding a fixed complete bipartite graph as a subgraph has bounded treewidth~\cite{gw}.
We will use this characterization whenever a fixed MSO transduction of a class of bounded cliquewidth gives uniformly sparse outputs, since the outputs also have bounded cliquewidth by \cref{lem:transduction-preserves-cliquewidth}.

When we add several relations to an encoding, it suffices to bound their degeneracy separately and define them all by one transduction.
A union of $m$ graphs of degeneracy at most $d$ has degeneracy at most $2md$.
Indeed, every induced subgraph on $n$ vertices has at most $dn$ edges from each of the $m$ graphs, so its average degree is at most $2md$ and it has a vertex of degree at most $2md$.
Here graphs on different vertex sets can first be extended to their common vertex set by adding isolated vertices.
Joint definability gives us the cliquewidth bound for the union.
The union of two independently chosen trees on the same vertex set can have arbitrarily large treewidth.

For disjoint domains, we use the parallel application lemma of Boja\'nczyk, Grohe, and Pilipczuk~\cite[Lemma~5.6]{linear}.
If the input is partitioned into definable disjoint domains, then one fixed transduction can be run on all domains at once, by relativizing its formulas.
A unary parameter on the union chooses its restrictions independently in all domains.
To keep track of the domains in a sparse output, we may mark one representative in each domain and use an extra copy of it as a hub.
We then join every output vertex to the hub of its domain by a membership edge.
These edges form a star forest.
Thus, the dense equivalence relation of belonging to the same domain can be discarded after the construction.

For source-preserving encodings, correctness of every accepted output can be enforced by an MSO sentence, using the filter construction of Boja\'nczyk, Grohe, and Pilipczuk \mbox{\cite[Lemma~5.3]{linear}}.
We temporarily keep the input graph $G$, the proposed output $H$, and the origin relation given by copying.
An MSO sentence can then check that the origin map identifies the graph recovered from $H$ with $G$, including any unary labels.
To check a fixed degeneracy bound~$d$ in MSO, we ask whether every nonempty vertex set in~$H$ contains a vertex with at most~$d$ neighbors in that set.

\subsection{From sparse encodings to cliquewidth decompositions}
\label{sec:encoding-to-decomposition}

We first encode a graph of bounded treewidth by a labeled forest and then use its MSO decoder to obtain a cliquewidth decomposition.
These steps are standard, and we state their source-preserving versions here because the recursive construction will need to attach further information to individual vertices.
We give the proofs in \cref{app:boundedtwencoder,app:treecliquecompiler}.

\begin{restatable}{proposition}{boundedtwencoder}
\label{prop:boundedtwencoder}
For every $w$ and every fixed finite set of unary labels, there is an MSO transduction from graphs of treewidth at most $w$ with these labels to finitely labeled forests.
Every accepted output keeps one designated leaf for each input vertex, with its exact origin, and a fixed MSO interpretation on these leaves recovers the input graph and its labels.
There is at least one accepted output for every input of treewidth at most $w$.
\end{restatable}
\begin{restatable}{lemma}{treecliquecompiler}
\label{lem:treecliquecompiler}
Fix a finite unary signature on trees and an MSO graph interpretation $\delta$ whose vertex domain is a distinguished set of tree vertices.
There is an MSO transduction that maps every such tree~$T$ with nonempty interpreted domain to a cliquewidth decomposition of~$\delta(T)$ of bounded width, with one leaf of origin $x$ for each interpreted vertex $x$.
\end{restatable}

\begin{corollary}
If a graph class has a source-preserving MSO encoding by finitely labeled trees, then it has bounded definable cliquewidth.
The same holds for labeled forests.
\end{corollary}

\section{Combining definable decompositions}
\label{sec:closure}

We will need to combine decompositions in two situations.
In the first, finitely many induced subgraphs cover every pair of vertices.
In the second, the vertices are partitioned into a sequence of blocks, and after each block the cut between the vertices already added and the remaining vertices has bounded rank.
The number of blocks may be arbitrarily large.

\subsection{Combining finitely many local encodings}

For the first case, we adapt the finite-cover argument of Gajarsk\'y et al.\ from the proof of their sparsification theorem~\cite[Section~5.3]{structuralexpansion}.
A family of vertex sets is a \emph{$2$-cover} if every pair of vertices lies together in some member of the family.
In their setting, the subgraphs of the cover had bounded shrubdepth, so they could encode them by labeled trees of bounded height and combine the trees along their common leaves.
Our cover comes from a coloring with a fixed number of colors.
We take each color class and each union of two color classes, and assume that we have a source-preserving MSO encoding of each of the resulting nonempty induced subgraphs into a sparse graph.
We keep the local encodings on separate copies and use a star to link each original vertex to the vertices that represent it in those encodings.
We can then recover whether two original vertices are adjacent by using the inverse on a piece that contains both of them.
\pagebreak

\begin{restatable}{theorem}{finitecoverencoding}\label{thm:finite-cover-encoding}
Fix $m$.
For every nonempty $I\subseteq[m]$ of size at most two, let $\mathscr D_I$ be a class with a source-preserving MSO encoding into uniformly sparse colored graphs and an MSO inverse.
Let $\mathscr C$ be a class in which every graph has a coloring $c:V(G)\to[m]$ such that $G[c^{-1}(I)]\in\mathscr D_I$ whenever $c^{-1}(I)$ is nonempty.
Then $\mathscr C$ also has a source-preserving MSO encoding into uniformly sparse colored graphs, with an MSO inverse.
If $\mathscr C$ has bounded cliquewidth, its encoded graphs have bounded treewidth.
\end{restatable}

The proof is given in \cref{app:finitecover}.

\subsection{Linear products of definably decomposable blocks}

Boja\'nczyk, Grohe, and Pilipczuk considered constructions that add one vertex at a time~\cite{linear}.
Here we add a whole block at each step and assume that one fixed MSO decomposer works for all the blocks.
If each prefix of the block order has cut-rank at most~$r$, its vertices have at most~$2^r$ distinct neighborhoods among the remaining vertices.
This bound lets us adapt their argument to combine the block decompositions.
It suffices that such a block order exists, since the transduction takes only the graph as input.

In the statement, $A_G[X,Y]$ denotes the adjacency matrix between~$X$ and~$Y$, with rows indexed by~$X$ and columns by~$Y$.
We give the proof in \cref{app:linearclosure}.

\input{figures/prefix-hypothesis}

\begin{restatable}{lemma}{linearclosure}
\label{lem:linearclosure}
For all nonnegative integers $r,m$, there is an integer $F(r,m)$ with the following property.
Suppose the vertices of $G$ have a partition into nonempty sets $W_1,\ldots,W_t$ and, writing $P_i=W_1\cup\cdots\cup W_i$ with $P_0=\varnothing$, we have
\[
 \operatorname{rank}_{\mathbb F_2} A_G[P_i,V(G)\setminus P_i]\leq r
 \quad(0\leq i\leq t),\qquad
 \operatorname{dcw}(G[W_i])\leq m\quad(1\leq i\leq t).
\]
Then $\operatorname{dcw}(G)\leq F(r,m)$.
In particular, one fixed MSO decomposer works for all graphs admitting such a partition, without being given the partition or its order.
\end{restatable}

\section{Induction on rankwidth and the description of tangles}
\label{sec:tangles}

\subsection{Branch decompositions and rankwidth}
\label{sec:branch-rankwidth}

Most of the time we measure a cut by its cut-rank.
Later we will replace a set of vertices by a single element and consider only cuts that keep the set on one side.
For this reason we work with the more general notion of a connectivity function~\cite{computingtangles}.

A \emph{connectivity function} on a finite set $V$ is a nonnegative integer function $\lambda:2^V\to\mathbb N$ with $\lambda(\varnothing)=0$ that is \emph{symmetric} and \emph{submodular}, meaning that
\[
 \lambda(X)=\lambda(V\setminus X),\qquad
 \lambda(X\cap Y)+\lambda(X\cup Y)
 \leq\lambda(X)+\lambda(Y).
\]
Symmetry and submodularity also give \emph{posimodularity},
\[
 \lambda(X\setminus Y)+\lambda(Y\setminus X)
 \leq\lambda(X)+\lambda(Y).
\]
The pair~$(V,\lambda)$ is called a \emph{connectivity system}.
\cref{fig:connectivity-inequalities} shows the sets in the two inequalities.
\input{figures/connectivity-inequalities}

A \emph{cut} of $V$ is a division into two complementary sets $X$ and $V\setminus X$, which we call its \emph{sides}.
Its \emph{order} is $\lambda(X)=\lambda(V\setminus X)$.
We also call~$\lambda(X)$ the order of the set~$X$ and use \emph{separation} for the cut given by its two sides.

A \emph{branch decomposition} of $(V,\lambda)$ is a tree of maximum degree three whose leaves are in bijection with $V$.
Deleting a tree edge gives two components, and the leaves in these components determine a cut of~$V$.
We say that the edge \emph{displays} this cut, or that it displays either of its sides.
We always suppress vertices of degree two, since the two edges at such a vertex give the same cut.
The \emph{width} of the decomposition is the largest order of a cut given by one of its edges, and the \emph{branchwidth} of $(V,\lambda)$ is the minimum width of a branch decomposition.
For a one-element ground set, we take the one-vertex tree and give it width zero.

For a graph $G$, we take $V=V(G)$ and use the \emph{cut-rank} function $\rho_G$ of Oum and Seymour~\cite{rankwidth}.
For $X\subseteq V(G)$, its value is $\rho_G(X)=\operatorname{rank}_{\mathbb F_2} A_G[X,V(G)\setminus X]$, where the rows are indexed by $X$, the columns by $V(G)\setminus X$, and an entry is one if and only if its row vertex and its column vertex are adjacent.
Cut-rank is symmetric and submodular, so it is a connectivity function~\cite{rankwidth}.
A branch decomposition for this function is a \emph{rank decomposition} of $G$, and its minimum width is the \emph{rankwidth} of $G$, denoted by $\operatorname{rw}(G)$.
We illustrate a rank decomposition in \cref{fig:rankwidth}.
\input{figures/rankwidth}

All matrix ranks and vector calculations in the paper are over $\mathbb F_2$.
Whenever we express a sum in a logical formula, the number of summands is fixed, so no counting quantifier is needed.
For every fixed $k$, the condition $\rho_G(X)\leq k$ is expressible by an FO formula with the free set variable $X$~\cite[Lemma~2.2 of the full version]{lowrankmso}.

As observed by Oum and Seymour, a graph class has bounded rankwidth if and only if it has bounded cliquewidth~\cite{rankwidth}.
We use rankwidth to bound the orders of cuts and we use cliquewidth when we apply the closure under MSO transductions from \cref{lem:transduction-preserves-cliquewidth}.

\subsection{The rank-one base case}

We prove by induction on the rankwidth bound that, for each fixed bound, one MSO transduction constructs cliquewidth decompositions of uniformly bounded width.
Oum proved that the graphs of rankwidth at most one are the distance-hereditary graphs, those in which every connected induced subgraph preserves the distances between its vertices~\cite{rwone}.
Each connected component has a split tree whose leaves are its vertices and whose internal nodes represent cliques and stars~\cite[Section~3]{gioanpaul}.
At a star node, one incident branch corresponds to the center of the star.
Two leaves are adjacent in the graph precisely when their path uses that branch at every star it passes through.
We construct the split trees by the MSO transduction of Campbell and K\"ohler~\cite[Corollary~7(3)]{laminar} and mark the center branch at each star so that we can recover adjacency.

\begin{restatable}{lemma}{rankonebase}\label{lem:rankonebase}
Graphs of rankwidth at most one have source-preserving MSO encodings by finitely labeled forests.
\end{restatable}

We give the proof in \cref{app:rankonebase}.
\medskip

We join the trees under a new marked root and recover the forest by deleting that root.
Applying \cref{lem:treecliquecompiler} gives a single decomposer for the base case.
For the induction step, fix~$k\geq 2$ and assume that one MSO decomposer accepts every graph of rankwidth less than~$k$.
We distinguish three cases according to the number of tangles of order~$k$.

\subsection{Tangles, robustness, and supports}

A family $\mathcal F\subseteq 2^V$ is \emph{laminar} if, for every $X,Y\in\mathcal F$, either $X\subseteq Y$, $Y\subseteq X$, or $X\cap Y=\varnothing$.
Thus, any two members are nested or disjoint, as illustrated in \cref{fig:laminarity}.
\input{figures/laminarity}

For the following definitions, let $\lambda$ be a connectivity function on $V$.
Fix $k\geq2$.
A \emph{tangle of order $k$} is a family $\mathcal T\subseteq 2^V$ of sets of order less than $k$ that contains exactly one side of every cut of order less than $k$.
We call the members of $\mathcal T$ \emph{big} and their complements \emph{small}.
Any three members of $\mathcal T$, allowing repetitions, must have nonempty intersection, and no singleton may belong to $\mathcal T$.
Equivalently, no union of at most three small sets covers $V$, and $V\setminus\{x\}$ is never small.
Throughout the proof, \emph{low-order} means of order strictly less than the order of the tangle.
Among low-order sets, any superset of a big set is big, since its complement is disjoint from that big set.
Similarly, if the intersection of two big sets has low order, it too is big, since the two big sets and the complement of their intersection have no common vertex.
We use the big sides, as do Grohe and Schweitzer~\cite{triplecovers}, because we will describe a tangle through the intersections of its big sets.
Taking complements gives the small-side convention used by Geelen, Gerards, and Whittle~\cite{ggwtangles}.
A tangle should be thought of as pointing towards one region that cannot be split into a few low-order pieces.
This intuition is illustrated in \cref{fig:tangle}.
\input{figures/tangle}

A set is \emph{weak} with respect to a tangle if it is contained in a small set, and it is \emph{strong} otherwise.
A weak set can have order at least $k$, even though it is contained in a small set.
Every superset of a strong set is again strong.
If both sides of a cut are strong, then its order is at least $k$.
For $k\geq2$, every singleton in a graph is small in an order-$k$ tangle, because its cut-rank is at most one.

A tangle is called \emph{robust} if no union of at most eight small sets covers the ground set~\cite{flowers}.
Equivalently, every intersection of at most eight big sets is nonempty.
For a robust tangle, no union of at most eight weak sets covers the ground set either, since each weak set is contained in a small set.
We use robustness in \cref{lem:coresplit} to rule out a cover of the ground set by six small sets.

\input{figures/weak-robust}

A nonempty set $A$ is \emph{titanic} if every partition of $A$ into at most three nonempty parts has a part of order at least $\lambda(A)$~\cite{tanglesorigin,ggwtangles}.
Equivalently, every cover of~$A$ by at most three subsets of~$A$ has a member of order at least $\lambda(A)$~\cite[Lemma~3.2]{hlinenyoum}.
\emph{Contracting} $A$ means replacing it by one element, called an \emph{atom}, and assigning each union of atoms the connectivity value of its original union.
Geelen, Gerards, Robertson, and Whittle proved the titanic contraction lemma for matroids~\cite[Lemma~2.1]{ggrw}.
Hlin\v{e}n\'y and Oum gave the connectivity-function version~\cite[Lemma~3.3]{hlinenyoum}.
In the formulation of Korhonen and Oum, if the original branchwidth is at most~$k$ and~$\lambda(A)\leq k$, then contracting a proper titanic set~$A$ leaves the branchwidth at most~$k$~\cite[Proposition~2.4]{titanic}.
In \cref{sec:robust-tangle} we will also need to contract certain sets of order~$k$ that need not be titanic.
Later, for titanic sets of order less than~$k$, we will use the tangle correspondence of Geelen, Gerards, and Whittle to compare the tangles before and after contraction~\cite[Lemmas~4.3 and~4.4]{ggwtangles}.
We illustrate the titanic condition and contraction in \cref{fig:titanic-contraction}.
\input{figures/titanic-contraction}

In the inner induction, we will need to recognize tangles that choose a big side inside one of the small replacement graphs.
For a fixed size bound~$L$, we call a set~$U\subseteq V$ with~$|U|\leq L$ a \emph{support}.
A \emph{support family} is a collection $\mathcal U$ of such sets.
A support $U$ \emph{captures} an order-$k$ tangle if there is a set $S\subseteq U$ with $\lambda(S)<k$ that is big in the tangle.
Thus, $U$ contains a big low-order set, although $U$ itself may have larger order.
A tangle is \emph{captured by $\mathcal U$} if some $U\in\mathcal U$ captures it, and is \emph{uncaptured} otherwise.
We represent the family by a bipartite incidence graph, with one auxiliary vertex for each support and edges from that vertex to the members of the support.
We call this auxiliary vertex the \emph{support index}, so its neighborhood is exactly the support it represents.
These definitions are illustrated in \cref{fig:support-capture}.
\input{figures/support-capture}

By tangle duality for connectivity functions, applied to cut-rank, a graph has rankwidth less than $k$ if and only if it has no order-$k$ cut-rank tangle~\cite[Theorem~3.2]{ggrwduality}, \cite[Theorem~3.1]{ggwtangles}.
The next observation uses the same restriction of cuts that Geelen, Gerards, and Whittle use to lift a tangle from a matroid minor~\cite[Lemma~5.1]{ggwtangles}.
For $W\subseteq V(G)$, we have \mbox{$\rho_{G[W]}(X\cap W)\leq\rho_G(X)$}, since the former cut matrix is a submatrix of the latter.
We can therefore orient each low-order cut of $G$ according to its restriction to $W$.

\begin{lemma}\label{lem:inducedtanglelift}
Let $G$ be a graph and $W\subseteq V(G)$, and let $Q$ be an order-$k$ tangle of $G[W]$ with respect to its cut-rank function $\rho_{G[W]}$.
For each set $X\subseteq V(G)$ with $\rho_G(X)<k$, declare~$X$ small if and only if $X\cap W$ is small in $Q$.
This defines an order-$k$ tangle of $G$ with respect to its cut-rank function $\rho_G$.
\end{lemma}
\begin{proof}
The two sides restrict to complementary sets in~$W$, and $Q$ chooses exactly one of them as big.
Any three big sides in~$G$ have a common vertex in~$W$, since their restrictions are big in~$Q$.
A singleton restricts to a singleton or the empty set, neither of which is big in~$Q$.
\end{proof}

\input{figures/cut-trace-lift}

We can now apply the outer induction when~$G$ has no order-$k$ tangle, since tangle duality gives~$\operatorname{rw}(G)<k$.
Suppose next that~$P$ is the only order-$k$ tangle of~$G$, and let~$W$ be a nonempty $P$-weak set.
An order-$k$ tangle of $G[W]$ would lift to a tangle that makes a $P$-small set containing $W$ big.
It would therefore differ from $P$, which is impossible.
Consequently, $G[W]$ has rankwidth less than~$k$.
The induction hypothesis gives a decomposition for every such weak piece.
In \cref{sec:robust-tangle} we combine these decompositions into one for~$G$.
When there are several tangles, the same lift can give a second tangle of~$G$, so we will need the inner induction to separate them.
For both constructions we first need to describe the tangles in~MSO.

\subsection{A tangle can be encoded inside any of its big sets}

Kalampakas applies flower theory to split-prime graphs of rankwidth at most two on at least nine vertices.
In this setting, there is a unique tangle of order two, and its small sets are the empty set and the singletons~\cite[Section~3]{ranktwo}.
For larger rankwidth, we need a description that also distinguishes different tangles.
We describe a tangle by a bounded set of vertices~$W$ and a list of subsets of~$W$.
The list consists of the \emph{traces} of the big sets on~$W$, that is, the intersections~$X\cap W$ for big sets~$X$.
For this description to determine the tangle, we must choose~$W$ so that a small set and a big set always have different traces.

\input{figures/localized-pair-witness}

A \emph{pair witness} is a set meeting the intersection of every two big sets (\cref{fig:localized-pair-witness}).
Indeed, if~$X$ is big and~$Y$ is small, the witness meets~$X\cap(V\setminus Y)$, so their traces differ.

Grohe and Schweitzer proved that a set of size bounded in terms of $k$ can meet every intersection of three big sets~\cite[Lemma~3.1]{triplecovers}.
Intersecting their triple cover with a prescribed big set $A$ already gives a pair witness inside $A$, since we can take $A$ as the third big set.
Seibt improved the triple-cover bound to $(3^k-1)/2$ \mbox{\cite[Theorem~6]{witnesses}}.
We adapt Seibt's argument to pairs to obtain the smaller witness bound below. 
We give the proof in \cref{app:localpairwitness}.

\begin{restatable}{lemma}{localpairwitness}
\label{lem:localpairwitness}
Let $\mathcal P$ be an order-$k$ tangle of a finite connectivity system, and let $A\in\mathcal P$ be big.
There is a pair witness $W\subseteq A$ with $|W|\leq2^k-1$.
\end{restatable}

For graph cut-rank, \cref{lem:localpairwitness} has the following logical consequence.
For each possible trace of a low-order set, we specify whether the set is big.
The pair-witness property ensures that this answer is consistent, even though many quite different sets may have the same trace.
We can therefore describe the tangle by finitely many vertex parameters and a finite list of choices.
We illustrate this description in \cref{fig:tangle-trace}.
\input{figures/tangle-trace}

\begin{corollary}
\label{cor:tangleparameters}
For fixed $k$, every order-$k$ cut-rank tangle of a graph can be represented by a bounded tuple of vertices inside any prescribed big set, together with a finite list specifying which traces on the tuple correspond to big sets.
Its orientations and the validity of the representation are MSO-definable.
\end{corollary}
\begin{proof}
Take a pair witness $W$ from the lemma.
Suppose that two low-order sets $X,Y$ have the same trace on $W$, with $X$ big and $Y$ small.
Then the two big sets $X$ and $V\setminus Y$ have no common vertex in $W$, contrary to the choice of $W$.
Thus, the trace determines the orientation.
The list specifies this orientation for every realized trace.
The choices for unrealized traces do not matter.
There are finitely many possible lists of choices because~$|W|$ is bounded in terms of $k$.
For each such list the orientation predicate is a fixed formula, and the tangle axioms can be checked by quantifying over at most three set variables.
A tangle may have several parameter descriptions.
To require that two descriptions give distinct tangles, we therefore ask for a low-order cut that they orient differently.
\end{proof}

\section{Combining the weak pieces around one tangle}
\label{sec:robust-tangle}

Suppose that for some order-$k$ tangle of~$G$, the graphs induced by its nonempty weak sets have uniformly bounded definable cliquewidth.
We show how to combine their decompositions into one tree from which we can recover~$G$.
When there is a unique order-$k$ tangle, the induction on rankwidth gives exactly this hypothesis.
We state the construction for an arbitrary tangle with this property because we will also use it in the terminal case of the inner induction.

\begin{theorem}
\label{thm:weakpieceencoding}
Fix $k\geq2$ and $m$.
There are a finite-copy MSO transduction $\mathsf T_{k,m}$ and an MSO interpretation $\delta_{k,m}$ with the following properties.
Let $G$ have rankwidth at most $k$, and suppose that some order-$k$ tangle $\mathcal T$ of $G$ satisfies
\[
 \operatorname{dcw}(G[W])\leq m
 \quad\text{for every nonempty $\mathcal T$-weak set }W.
\]
Then $G$ has an output under $\mathsf T_{k,m}$.
Every accepted output is a tree with a fixed finite set of unary labels and one distinguished leaf for each input vertex, with that vertex as its origin.
On these leaves, $\delta_{k,m}$ recovers $G$.
Every fixed finite collection of unary labels on the input can be kept.
\end{theorem}

\begin{figure}[H]
\centering
\color{black}
\begin{tikzpicture}[manuscript illustration]
  \node[box] (t) at (0.65,0.35) {$\mathcal T$};
  \node[box] (s) at (4.15,1.02) {at most eight small sets\\cover $V$};
  \node[key] (r) at (4.15,-0.32) {robust tangle};
  \node[box] (b) at (8.3,1.02) {ordered weak blocks};
  \node[key] (c) at (8.3,-0.32) {local port graphs};
  \draw[arr] (t) -- (s);
  \draw[arr] (s) -- (b);
  \draw[arr] (t) -- (r);
  \draw[arr] (r) -- (c);
\end{tikzpicture}
\caption{The two constructions in \cref{thm:weakpieceencoding}.
In the upper case, successive differences of at most eight small sets give weak blocks of bounded definable cliquewidth.
In the lower case, the core tree gives a local graph at each internal node, and the weak blocks are used within those graphs.}
\label{fig:weak-piece-cases}
\end{figure}
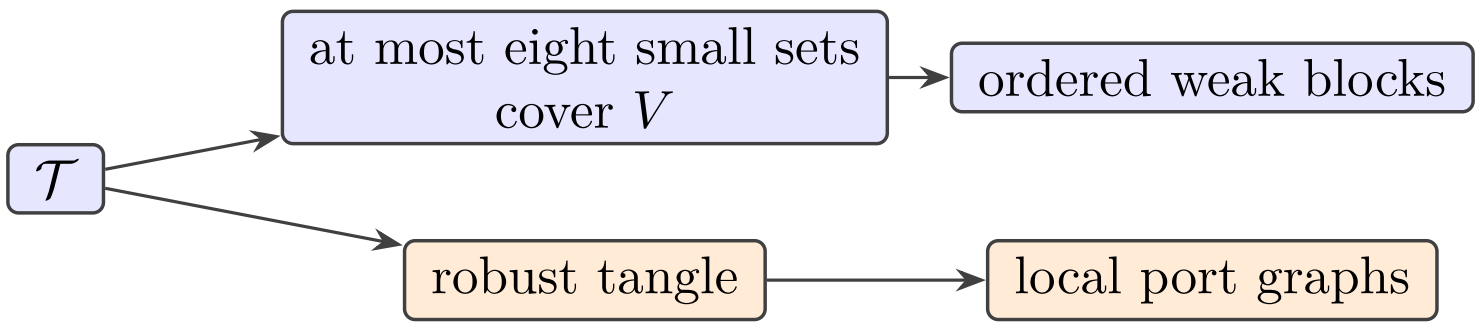

\noindent\textit{The nonrobust case.}
If the specified tangle $\mathcal T$ is nonrobust, at most eight of its small sets cover $V(G)$.
Taking successive differences and omitting empty sets gives an ordered partition into weak blocks.
Each block has definable cliquewidth at most $m$ by hypothesis.
The union of the first $i$ blocks is a union of at most eight of the small sets, so its cut-rank is at most $8(k-1)$ by subadditivity.
By \cref{lem:linearclosure}, there is one MSO decomposer for the nonrobust case, and its leaves keep the origins of the vertices.
Its expression tree is the required encoding.
The cover and its block order are used only to apply \cref{lem:linearclosure}.

Suppose now that the tangle is robust.
We use cuts of order exactly~$k$, where both sides may be strong.
Two cuts \emph{cross} if each side of either cut meets both sides of the other.
The cuts we need can cross, so we cannot expect them all to be displayed by edges of one tree.
We first recall how Clark and Whittle grouped such cuts into classes and represented them by a tree with flowers~\cite{flowers}.
From these classes we select sides that are nested or disjoint and take their inclusion tree.

At each node of this tree, we form a local graph by replacing certain sides of cuts with bounded sets of auxiliary vertices.
We need to show that the resulting graph has an ordering into weak blocks and bounded replacement blocks to which \cref{lem:linearclosure} applies.

\subsection{Closure, flowers, and the existing structure theorem}
\label{sec:flower-background}

As in the rankwidth-two argument of Kalampakas~\cite{ranktwo}, we use full closure and flower theory to organize separations around one tangle.
Here we work with tangles of arbitrary fixed order, so we use the general definitions and structural results of Clark and Whittle~\cite{flowers}.
We fix a tangle $\mathcal T$ of order $k$ in a finite connectivity system $(V,\lambda)$.
All references to weak and strong sets and to full closure are with respect to this tangle.
Following Clark and Whittle, we call $X$ \emph{$k$-separating} if $\lambda(X)\leq k$, and we call the cut $(X,V\setminus X)$ a \emph{$k$-separation} in this case.
This includes the cuts of order exactly~$k$, which the tangle itself does not orient.

Let $X$ be a strong $k$-separating set.
It is \emph{fully closed} if $\lambda(X\cup W)>k$ for every nonempty weak set $W\subseteq V\setminus X$.
Its \emph{full closure}, written $c(X)$, is the least fully closed strong $k$-separating set containing $X$.
Starting with $X$, we may add nonempty weak sets one at a time, each disjoint from the set reached so far, as long as the order stays at most $k$.
We call any such sequence of additions a \emph{partial $k$-sequence}.
The closure theorem of Clark and Whittle says that every maximal sequence reaches $c(X)$, and that $c$ is monotone and idempotent on its domain~\cite[Corollary~3.2 and Lemmas~3.3 and~3.6]{flowers}.
We illustrate this process in \cref{fig:closure-operation}.
\input{figures/closure-operation}

Adding a weak set to $X$ while keeping the order at most $k$ therefore preserves the full closure.
For arbitrary~$k$, one addition may have to include several vertices at once.

A separation $(X,V\setminus X)$ is \emph{strong} if both $X$ and $V\setminus X$ are strong with respect to the fixed tangle.
Every cut of order less than $k$ has a small side, so a strong separation of order at most $k$ necessarily has order exactly $k$.
A strong separation $(X,V\setminus X)$ of order at most~$k$ is \emph{non-sequential} if both $c(X)$ and \mbox{$c(V\setminus X)$} are proper subsets of $V$.
Otherwise, it is \emph{sequential}.
Two non-sequential separations are \emph{equivalent} when their unordered pairs of full closures agree.
We call each resulting equivalence class a \emph{separation class}.
To choose an \emph{orientation} of a separation, we distinguish one side as the first side.
For the orientation $(X,V\setminus X)$, we call $s=(A,B)=(c(X),c(V\setminus X))$ its \emph{oriented closure pair}.
We write $s^*=(B,A)$ for the reverse orientation.
A side \emph{represents}~$s$ if it is the first side of a separation with oriented closure pair~$s$.
Equality of the first full closure already determines the second.
The set~$V\setminus B$ represents~$s$ and satisfies~$c(V\setminus B)=A$.
We call it the \emph{core} of the orientation $s$ and write $F_s=V\setminus B$.
The reverse orientation has core $F_{s^*}=V\setminus A$.
The core $F_s$ is contained in every side representing $s$, since $V\setminus X\subseteq B$ implies $V\setminus B\subseteq X$.
These facts are the closure and equivalence lemmas of Clark and Whittle \mbox{\cite[Lemmas~3.7 and~3.8(iii)]{flowers}}.

\cref{fig:full-closure} shows how the two full closures distinguish the sequential and non-sequential cases, and identifies the cores of the two orientations.
\input{figures/full-closure}

Two separation classes are \emph{compatible} if we can choose one separation from each class and one side of each separation so that one chosen side is contained in the other.
A class is \emph{strong} if it is compatible with every non-sequential class.
Thus, whether a class is strong depends on its relation to the other classes, while whether a set is strong depends on the fixed tangle.
For oriented closure pairs, we use the order
\[
 (A,B)\preceq(C,D)\quad\Longleftrightarrow\quad
 A\subseteq C\text{ and }B\supseteq D.
\]
Two classes are compatible if and only if we can choose an orientation of each class such that the two chosen orientations are comparable in this order.
Indeed, comparable orientations have nested cores, while nested representatives give comparable closure pairs by monotonicity.
We illustrate compatibility and strong separation classes in \cref{fig:separation-classes}.
\input{figures/separation-classes}

To relate these definitions to the notation of Clark and Whittle, let $\mathcal S$ consist of the $k$-separating sets $X$ for which $V\setminus X$ is strong and $c(V\setminus X)\ne V$.
They call a cut a $(k,\mathcal S)$-separation when both sides belong to $\mathcal S$.
For this choice, these are exactly the non-sequential separations, and their general results apply to this family~\cite[Section~3]{flowers}.

A \emph{flower} is a partition $(P_1,\ldots,P_n)$ of $V$ into strong sets, called \emph{petals}, such that every petal and every union of two consecutive petals has order at most $k$.
Indices are read cyclically.
Submodularity then gives order at most $k$ for every union of consecutive petals.
A flower is an \emph{anemone} if every union of petals has order at most $k$, and it is a \emph{daisy} if a union of petals has this property if and only if its petals are consecutive.
Every flower has one of these forms~\cite[Corollary~4.1]{flowers}.
\cref{fig:flower-types} shows which unions of petals we can choose in the two cases.
\input{figures/flower-types}

A flower \emph{displays} a $k$-separation when one side is a union of petals.
Two flowers are \emph{equivalent} if they display the same non-sequential classes.
The \emph{order} of a flower is the least number of petals in an equivalent flower.
A flower is \emph{tight} if its number of petals equals its order.
Thus, an order-one flower displays no class, an order-two flower displays one class, and a flower of order at least three displays at least two different classes.
A petal is \emph{loose} if it is contained in the closure of a neighboring petal, where any ordering is allowed for an anemone.
Such a petal can be merged with its neighbor without losing a displayed class \mbox{\cite[Lemma~4.3]{flowers}}.
More generally, a \emph{concatenation} of a flower replaces groups of consecutive petals by their unions.
The resulting partition is again a flower.
For an anemone, we may first choose any order of the petals.
We will also use that, for a flower of order at least two, $V\setminus P\in\mathcal S$ for every petal $P$, and therefore $c(P)\ne V$~\cite[Lemma~4.4]{flowers}.
We illustrate a displayed separation and the merging of a loose petal in \cref{fig:flower-display}.
\input{figures/flower-display}

The tree of Clark and Whittle has two kinds of nodes~\cite[Section~6]{flowers}.
A \emph{bag node} carries a subset of $V$, and these bags partition $V$, with empty bags allowed.
A \emph{flower node} represents an anemone or a daisy whose petals are determined by the branches attached to that node.
At a daisy node, we also specify the cyclic order of these branches.

Deleting a tree edge splits the tree into two components.
Taking the union of the bags in each component gives the two sides of the separation displayed by that edge.
Similarly, deleting a flower node leaves several components, and the union of the bags in each component is one petal of the flower.
The bag nodes, flower node, and the cuts they display are illustrated in \cref{fig:partial-tree}.
\input{figures/partial-tree}

A tree satisfying the conditions below is called a \emph{partial $k$-tree}.
With our fixed choice of~$\mathcal S$, this is a partial $(k,\mathcal S)$-tree in the terminology of Clark and Whittle~\cite[Section~6]{flowers}.
A partial tree may still leave some non-sequential separations unresolved inside its bags, as allowed by~(P5).
\begin{enumerate}[label=(P\arabic*)]
\item Every tree edge displays a strong $k$-separation, and an edge between two bags displays a non-sequential separation.
\item Each non-bag node is an anemone or a daisy node, with the corresponding local petal order.
\item An anemone node has order at least three and has no loose petal.
\item A daisy node has order at least three and has no loose petal.
\item Every non-sequential separation is equivalent either to a displayed separation or to one with a side contained in a bag.
\end{enumerate}

The structure theorem for robust tangles states that a maximal such partial tree displays every non-sequential class, up to equivalence~\cite[Corollary~7.2]{flowers}.
Maximality here means that no partial tree displays a proper superset of the classes displayed by the chosen tree.
We use the existence of this tree to prove the local structure in the next subsection.
The transduction will construct the laminar family of strong cores directly, without selecting the flower tree or its cyclic orders.
\subsection{A class-normal core tree for one robust tangle}

We will modify the flower tree so that an edge displays each strong class.
Cutting these edges will then leave at most one flower in each component.
This is the class-normal form of Kalampakas for rankwidth two~\cite[Proposition~3.9]{ranktwo}, which we extend to a robust tangle of any fixed order~$k$.

Fix a robust tangle $\mathcal T$ of order $k$ in a finite connectivity system $(V,\lambda)$, and assume that every singleton is $\mathcal T$-weak.
This holds for graph cut-rank when $k\geq2$.
Since the cores are strong, this assumption keeps them distinct from the singleton sets that become leaves of the inclusion tree.

Choose one oriented strong class $r$ as an \emph{anchor} if one exists.
For each other strong class choose its unique orientation $s$ satisfying $s\preceq r$ or $s\preceq r^*$.
Compatibility with the anchor gives two comparable orientations.
If the orientation of the other class lies above the anchor orientation, reversing both orientations puts it below the opposite anchor orientation.
For uniqueness, write $r=(A,B)$ and $s=(C,D)$.
If both $s$ and $s^*$ lie below $r$, then $C,D\subseteq A$, contrary to $C\cup D=V$ and $A\ne V$.
The same argument applies below $r^*$.
If $s\preceq r$ and $s^*\preceq r^*$, then $C=A$ and~$D=B$.
The reversed comparison gives~$C=B$ and~$D=A$.
These last two cases identify the class with the anchor, which was excluded.
Keep $r$ itself in the chosen orientation.
The two anchor orientations and the resulting choice of cores are illustrated in \cref{fig:anchor-class}.

\input{figures/anchor-class}

The chosen cores are nested or disjoint, so we can use them to construct an inclusion tree.
We follow the laminarity argument of Kalampakas~\cite[Lemma~3.6]{ranktwo}, which uses only the order on closure pairs.

\begin{lemma}
\label{lem:generalcorelaminar}
The selected cores, together with $V$ and the singleton sets, form a laminar set system.
\end{lemma}
\input{figures/core-laminarity}
\begin{proof}
Cores below opposite anchor orientations are disjoint, because they are contained in the complements of the two anchor closures, whose union is $V$.
Consider selected orientations $s=(C,D)$ and $t=(F,H)$ below the same anchor orientation $(A,B)$.
They are compatible.
A direct comparison nests their cores.
If $s\preceq t^*$, then $D\supseteq F\supseteq V\setminus H$, so the cores $V\setminus D$ and $V\setminus H$ are disjoint.
The only remaining comparison $s^*\preceq t$ would imply $C,D\subseteq A$, since $C,F\subseteq A$ and $D\subseteq F$.
This contradicts $C\cup D=V$ and $A\ne V$.
\end{proof}

A flower may display a strong class through several petals together, whereas we want that class to appear on a single tree edge.
We can create such an edge by replacing the flower node by two flower nodes joined by a new edge, with the petals on one side of the chosen separation attached to the first node and the remaining petals attached to the second.
Kalampakas uses this splitting operation in the proof of the rank-two class-normal form \mbox{\cite[Lemma~3.8 and Proposition~3.9]{ranktwo}}.
We must show that the two new flowers and the edge between them still display all the classes of the old flower and that neither flower has a loose petal.
For arbitrary $k$, Clark and Whittle prove that a concatenation along such a separation has no loose petal when the original flower is tight and has order at least three~\cite[Lemma~4.7]{flowers}.
After a split, we only know that the original petals for the next split are not loose, so we give their uncrossing argument under this hypothesis.
\cref{fig:split-flower} illustrates this replacement.

\begin{lemma}
\label{lem:generalsplitflower}
Let a flower without loose petals display a non-sequential strong class as a union $U$ of petals.
If each side contains at least two petals, we can replace the flower node by two flower nodes joined by an edge displaying $(U,V\setminus U)$, without losing any displayed non-sequential class.
If a new flower has order two, we may replace its node by a bag node.
When the flower is a node of a partial tree that displays every non-sequential class, the replacement preserves the partial-tree axioms.
\end{lemma}
\input{figures/split-flower}
\begin{proof}
Arrange $U$ consecutively: this is already true for a daisy, and an anemone allows any petal order.
The two new partitions consist of the old petals on one side and one merged petal on the other side.
They are flowers by concatenation.

First check that concatenation creates no loose petal.
Let $W$ be one of the merged non-sequential sides, let $P$ be an adjacent old petal outside $W$, and choose the boundary petal $R$ inside $W$ next to $P$ in the old petal order.
If $P\subseteq c(W)$, take a maximal sequence of weak additions~$W=C_0\subsetneq\cdots\subsetneq C_t=c(W)$.
The two sets $C_j$ and $R\cup P$ have order at most $k$.
Their union contains the strong set $W$ and its complement contains the strong set~$V\setminus c(W)$, so its order is at least $k$.
Submodularity gives
\[
 \lambda\bigl(R\cup(P\cap C_j)\bigr)\leq k.
\]
Intersecting the weak additions with $P$ therefore absorbs all of $P$ into $R$, contrary to the old flower having no loose petal.
The reverse kind of new looseness, in which a merged petal is contained in the closure of an old neighbor, would already make its boundary old petal loose.
All other pairs are unchanged.
Both flowers therefore have no loose petal.

Now consider a different non-sequential class $q$ displayed by the old flower.
Since the class of $U$ is strong, we may orient the two classes so that the core $F$ of $q$ lies in the core on one side of the split, say on the side $U$.
Take a representative of $q$ displayed by the old flower and orient it so that its first side has full closure $c(F)$.
The flower displays two distinct classes and therefore has order at least three.
The localization lemma of Clark and Whittle now gives a representative of $q$ that is still displayed by the old flower and has a side contained in $U$~\cite[Lemma~4.8]{flowers}.
That side is a union of petals inside $U$, all of which are still separate petals at the new flower on this side.
The new flower thus displays this representative, and the new edge displays the class of $U$.

Each new flower displays at least this one class, so has order at least two.
If it has order two, every displayed non-sequential class is the class of $U$ and remains on the new edge.
Replace that center by an empty bag, and contract any resulting sequential bag-bag edge.
This last contraction merges the two endpoint bags and preserves every other displayed edge cut and every remaining flower partition.

We check the partial-tree axioms.
Each old edge that remains displays the same cut as before, and the new edge displays the strong non-sequential separation $(U,V\setminus U)$.
Condition~(P1) therefore holds after the sequential bag-bag contractions.
Give each remaining flower its induced cyclic order and its flower label, as required by (P2).
A concatenation of an anemone is an anemone.
In a concatenation of a daisy, a union of new petals is consecutive if and only if the corresponding old petals are consecutive, so the new flower is again a daisy.
This also holds when only three petals remain, by the definition of a daisy in~\cite{flowers}.
The remaining flowers have order at least three and no loose petals, proving (P3)-(P4).
Every non-sequential class displayed before the operation is still displayed afterwards.
We apply the operation to a tree displaying all such classes, so (P5) follows as well.
Empty bag nodes are allowed in the definition of a partial tree.
\end{proof}

We now apply the preceding operation until every strong class appears on an edge.
If we cut all edges displaying these classes, each of the remaining components contains at most one flower.
This gives us the local pieces in which we will later find the weak-block descriptions.
Several edges may still display the same class, but we will see that they form a path whose internal vertices are bags of degree two.
In the tree that tells us how the components are joined, we can therefore replace the whole path by one edge.

Kalampakas proves the four conclusions below for the order-two tangle of a split-prime graph on at least nine vertices~\cite[Proposition~3.9 and Lemma~3.11]{ranktwo}.
We follow that proof for a robust tangle of arbitrary order, using \cref{lem:generalsplitflower} at each split.

\begin{theorem}
\label{thm:generalclassnormal}
Assume that every singleton is $\mathcal T$-weak.
There is a maximal partial $k$-tree for $\mathcal T$ in which every strong non-sequential class is displayed by an edge.
Cut all edges displaying these classes and call the remaining connected subtrees \emph{components}.
Form a tree with one node for each component and one edge for each of the cut edges.
In this tree, the edges displaying any one class form a path, and all internal nodes of the path have degree two.
Replace each such path by one edge to obtain the \emph{component tree}.
Call the class of an edge incident with a component a \emph{boundary class} of that component.
To find its \emph{outward side}, take the first original edge on the corresponding path, starting at the component.
Deleting this edge from the partial tree gives two sides, and we take the one away from the component.
The \emph{outward core} is the core of this orientation.
Then the following hold.
\begin{enumerate}
\item The component tree is the inclusion tree of the selected cores, with the singleton leaves omitted.
\item Each component contains at most one flower node.
\item At a component containing a flower, each incident edge of the component tree corresponds to a petal $P$ containing its outward core $F$, with $F\subseteq P\subseteq c(F)$.
Every other petal $P$ has $c(V\setminus P)=V$ and contains no core of a non-sequential separation.
\item At a component without a flower, every non-sequential separation that keeps each incident outward core wholly on one side belongs to an incident boundary class.
\end{enumerate}
\end{theorem}
\input{figures/class-normal-tree}
\begin{proof}
\noindent
\textit{Making the strong classes into tree edges.}
Start with a maximal partial tree.
By the display theorem of Clark and Whittle, each non-sequential class is displayed at an edge or a flower~\cite[Corollary~7.2]{flowers}.
Apply \cref{lem:generalsplitflower} whenever a strong class is not yet displayed by an edge.
Previously introduced strong edges persist, so the process terminates.

We next consider the edges that display one fixed class.
The cores of its two orientations are nonempty.
Choose one element from each core.
Every edge of the class separates the bags containing these two elements, so all these edges lie on the path between the two bags.
Orient two such edges so that their sets $X,Y$ have the same closure $A$, and write $B$ for their common opposite closure.
Both sets contain~$V\setminus B$.
Since the edges lie on this path, we may take $X\subseteq Y$.
Each edge between them has a side $Z$ with $X\subseteq Z\subseteq Y$.
By (P1), its two sides are strong and have order at most $k$, so we may take their full closures.
Monotonicity in both orientations gives $c(Z)=A$ and $c(V\setminus Z)=B$.
Every edge between the two chosen edges therefore displays the same class.

Suppose there were a flower at a vertex between the two edges.
Consider the two petals that point along the path.
If a displayed non-sequential separation puts these petals on different sides, one side lies between $X$ and $Y$, so the separation belongs to the same class.
If a displayed separation keeps the two petals together, its other side $C$ lies in $Y\setminus X$.
Then
\[
 c(V\setminus C)\supseteq c(X)\cup(V\setminus Y)
 =A\cup(V\setminus Y)=V,
\]
so it is sequential.
This would leave the flower only one non-sequential class.
The two-petal flower for that class displays the same non-sequential classes, so the original flower has order at most two, contrary to its order being at least three.
It follows that every vertex between edges of this class is a bag.
Suppose a branch leaves the path at such a bag, and let $W$ be the side of its first edge pointing into the branch.
Then $W\subseteq Y\setminus X$.
The set $V\setminus W$ contains~$X$ and $V\setminus Y$, so its full closure contains both $c(X)$ and $V\setminus Y$, and thus is $V$.
By (P1), the neighbor across this edge cannot be a bag, since a bag-to-bag edge displays a non-sequential separation.
If the neighbor is a flower, take a non-sequential separation that it displays and choose the side $Z$ that omits the petal pointing back to the path.
Then~$Z\subseteq W$, so $c(V\setminus Z)=V$, again a contradiction.

Thus, every bag between two edges of the class has degree two, and both its incident edges belong to that class.
Cutting the strong edges leaves each such bag in a component of its own.
Paths belonging to different classes can therefore meet only at their endpoints.
We can therefore form the component tree by replacing each such path by one edge, as shown in \cref{fig:class-normal-duplicates}.
At either end, we still have the original component and its incident boundary edge in the partial tree.
If the boundary edge is incident with a flower node, we can therefore still use the petals at that node.
\input{figures/class-normal-duplicates}
\smallskip

\noindent
\textit{Identifying the component tree.}
Each boundary edge in the original partial tree displays some representative of its class.
We must therefore check that replacing a representative by its core does not change the nesting pattern.
For this comparison, choose one original edge from each path of a strong class.
Contracting all other edges gives the same abstract tree as suppressing the intermediate nodes above.
We show that the selected sides of the chosen edges have the same nesting and disjointness relations as their cores.
Since the cuts come from tree edges, two such sides are nested, disjoint, or have union~$V$.
For completeness, write $s=(A,B)$ and \mbox{$t=(C,D)$}, and let $X,Y$ be the corresponding sides of their cuts, so that $c(X)=A$, $c(V\setminus X)=B$, and similarly for $Y$.
If $s\preceq t$, both $X,Y$ contain the nonempty core $F_s$, so they meet.
If $Y\subseteq X$, monotonicity gives the reverse closure comparison, identifying the two classes.
If $X\cup Y=V$, then $D\subseteq A\subseteq C$, which forces $C=V$ because $C\cup D=V$.
Both possibilities are excluded, so $X\subseteq Y$.
If $s\preceq t^*$, nesting $X\subseteq Y$ would give $B\supseteq C\cup D=V$, and reverse nesting would give $D=V$.
The case $X\cup Y=V$ gives the reverse comparison~$t^*\preceq s$, again identifying the two classes.
It follows that $X$ and $Y$ are disjoint.
These cases cover every selected pair: below the same anchor this was checked in \cref{lem:generalcorelaminar}, and below opposite anchor orientations transitivity gives~$s\preceq r\preceq t^*$.
If there is no strong class, the component tree consists of one node and the inclusion tree has just its root after the element leaves are omitted.
Otherwise, root the component tree at the endpoint of the new anchor edge outside the chosen anchor side.
Every selected side points away from this root, so the nesting just proved identifies the component tree with the inclusion tree of the selected cores, after omitting the singleton leaves.
This proves the first assertion.
\smallskip

\noindent
\textit{The structure left inside a component.}
We first show that a component contains at most one flower.
Every edge displaying a non-sequential separation displays a strong class, since its cut is nested with every displayed separation and every class has a displayed representative.
If two flowers survived in a component, let $P$ be the petal of the first pointing toward the second.
Its edge is sequential.
By~\cite[Lemma~4.4]{flowers}, we have $c(P)\ne V$, so $c(V\setminus P)=V$.
The second flower displays a non-sequential side $Z\subseteq P$, but then monotonicity gives $c(V\setminus Z)=V$, a contradiction.
This proves the second assertion.
The excluded configuration is shown in \cref{fig:class-normal-two-flowers}.
\input{figures/class-normal-two-flowers}

The local containments used in the last two assertions are illustrated in \cref{fig:class-normal-local}.
\input{figures/class-normal-local}

At the flower, a non-sequential petal edge is one of the original boundary edges.
Its petal~$P$ represents the outward orientation, so the outward core $F$ satisfies $F\subseteq P\subseteq c(F)$.
This core is the same for every edge along the path representing the class.
If the edge for a petal~$P$ displays a sequential separation, the same petal argument gives $c(V\setminus P)=V$.
If the core $F$ of a non-sequential separation lay in $P$, then $V\setminus P\subseteq V\setminus F$ would give $c(V\setminus F)=V$, a contradiction.
Consequently, there is no boundary edge farther along this branch, since its outward core would lie in $P$.
Each incident boundary edge therefore gives a distinct petal, and the remaining petals have the stated property.

Finally, at a component without a flower, consider a non-sequential separation that keeps every incident outward core whole.
We follow the proof of Kalampakas's Lemma~4.5 for rankwidth two~\cite{ranktwo}.
An equivalent separation is displayed somewhere in the partial tree.
It cannot be displayed inside the component, which has neither a flower nor a remaining non-sequential edge.
In the original partial tree, let $P$ be the outward side of the first boundary edge on the way to this display, and orient the displayed separation so that its side $R$ lies in $P$.
Since $V\setminus P\subseteq V\setminus R$, monotonicity gives $A=V\setminus c(V\setminus R)\subseteq V\setminus c(V\setminus P)=F$, where $A$ and $F$ are the corresponding cores.
Let $X_0$ be the side of our original separation with core $A$.
Since $X_0$ contains $A$ and keeps $F$ whole, it contains $F$, and we have $c(A)=c(X_0)$.
It follows that
\[
 c(A)\subseteq c(F)\subseteq c(X_0)=c(A).
\]
The first full closure determines the orientation, so the original separation belongs to the boundary class of $F$.
This proves the last assertion.
\end{proof}

We can now construct the core inclusion tree by an MSO transduction.
For fixed $k$, \cref{cor:tangleparameters} describes the tangle by a bounded tuple of vertices and a finite list specifying which intersections with that tuple correspond to big sets.
We can test weakness by asking for a small set containing the given set.
For a strong $k$-separating set, we can define its full closure as the least fully closed strong $k$-separating superset.
To test whether two classes are compatible, or whether a class is strong, we then need only quantify over vertex sets.
One guessed anchor set orients all strong classes.
\cref{lem:generalcorelaminar} and the counting-free laminar-tree transduction~\cite[Theorem~2]{laminar} therefore produce the core inclusion tree with the original vertices kept as leaves.
The partial tree, its flower labels, and its petal orders are needed only to prove the local block-layout theorem and are never selected by the transduction.

\input{figures/definable-core-tree}
\subsection{Contracting disjoint cores with fully closed complements}

Fix a tangle $\mathcal T$ of order $k$ in $(V,\lambda)$.
Our aim is to contract a side $A$ of a separation of order $k$ when the other side is fully closed.
If some branch decomposition of width at most $k$ already displays $A$, we can simply replace that branch by one leaf.
To find such a decomposition, we adapt a modification of branch decompositions used by Hall, Oxley, Semple, and Whittle for 3-connected connectivity functions of branchwidth three~\cite[Theorem~4.1]{hosw}.

\begin{lemma}\label{lem:strongsurgery}
Let $(A,B)$ be a strong separation of order $k$, and let $T$ be a branch decomposition of width at most $k$.
Suppose an edge of $T$ displays $(X,Y)$ and both $X\cap A$ and $X\cap B$ are strong.
There is a branch decomposition of width at most $k$ that displays $X$ and every nonempty member of $\{Y\cap A,Y\cap B\}$.
When all three are nonempty, deleting one internal vertex gives three branches with precisely these leaf sets.
Moreover, every set displayed inside the original $X$-branch remains displayed.
\end{lemma}
\begin{proof}
The tree operation is shown in \cref{fig:strong-branch-surgery}.
\input{figures/strong-branch-surgery}
Keep the $X$-branch.
Take two copies of the $Y$-branch, keeping in one only the leaf labels in $Y\cap A$ and in the other only those in $Y\cap B$, and join the three branches at a new vertex.
Suppress unlabeled leaves and degree-two vertices afterwards.

We now check the cuts in the new branches.
Restricting a branch may change its connectivity value, so this check does not follow merely from deleting leaf labels.
Consider an old edge in the $Y$-branch, with sides $Z,W$ where $X\subseteq Z$.
In the copy restricted to $A$, its new set is $W\cap A$, whose complement is $Z\cup B$.
The set $Z\cap B$ contains the strong set $X\cap B$, and its complement contains the strong set $A$.
It follows that $\lambda(Z\cap B)\geq k$.
Submodularity gives
\[
 \lambda(Z\cup B)
 \leq \lambda(Z)+\lambda(B)-\lambda(Z\cap B)
 \leq\lambda(Z)\leq k.
\]
The copy restricted to $B$ is treated with $A$ and $B$ interchanged.
The cuts in the unchanged $X$-branch do not change.
This also checks the three cuts at the new joining vertex.
\end{proof}

The second surgery follows the construction of Hall, Oxley, Semple, and Whittle \mbox{\cite[Lemma 4.2]{hosw}}.
It becomes particularly useful once we have found one strong displayed set on each side of the cut we want to display.
We keep the branch displaying each of these sets and join two copies of the tree along the desired cut.

\begin{lemma}\label{lem:strongseeds}
Let $(A,B)$ be a strong separation of order $k$.
Suppose a width-$k$ branch decomposition simultaneously displays a strong set $C\subseteq A$ and a strong set $D\subseteq B$.
Then it can be changed into a width-$k$ branch decomposition displaying $(A,B)$.
\end{lemma}
\begin{proof}
If one edge already displays $(C,D)$, then $C=A$ and $D=B$ and there is nothing to prove.
Otherwise, the branches displaying $C$ and $D$ are disjoint.
Take a copy of the component obtained by removing the $D$-branch and keep only its $A$-labels, and a copy of the component obtained by removing the $C$-branch and keep only its $B$-labels.
Join the two copies at their attachment vertices, as in~\cite[Lemma~4.2]{hosw}.
The new joining edge displays~$(A,B)$.

\input{figures/seeds-join}

For completeness, consider a cut in the $A$-copy outside the unchanged $C$-branch.
It comes from an old cut $(X,Y)$ with $D\subseteq Y$, and its new set is $X\cap A$.
The set $Y\cap B$ contains~$D$, and its complement contains $A$, so both sides are strong and $\lambda(Y\cap B)\geq k$.
Equivalently $\lambda(X\cup A)\geq k$.
Submodularity therefore gives
\[
 \lambda(X\cap A)
 \leq \lambda(X)+\lambda(A)-\lambda(X\cup A)
 \leq\lambda(X)\leq k.
\]
Cuts inside the $C$-branch still have their original orders.
The $B$-copy is symmetric, using $C$.
Deleting unlabeled leaves and suppressing degree-two vertices preserves the bound.
\end{proof}

In the width-three proof, Hall, Oxley, Semple, and Whittle use the sizes of intersections to find displayed sets on opposite sides~\cite[proof of Theorem~4.1]{hosw}.
Here we use full closure and robustness to find the two strong sets required by \cref{lem:strongseeds}.
If a side $X$ of order at most $k$ meets the core in a nonempty weak set, the fact that the opposite side is fully closed tells us that the part of $X$ outside the core is small.
Thus, a cut that fails to give a strong intersection also gives a short cover by small sets.
Robustness will rule out too many such failures.

\begin{lemma}
\label{lem:weakcorner}
Let $(A,B)$ be a strong separation of order $k$, with $B$ fully closed.
If $\lambda(X)\leq k$ and $\varnothing\ne A\cap X$ is weak, then $\lambda(X\setminus A)<k$, the set $X\setminus A$ is small, and $X$ is covered by two small sets.
\end{lemma}
\begin{proof}
Apply full closure of $B$ to the nonempty weak set $W=A\cap X$.
By symmetry, $\lambda(A\setminus X)=\lambda(B\cup W)>k$.
Posimodularity, which follows from symmetry and submodularity, gives
\[
 k+\lambda(X)=\lambda(A)+\lambda(X)
 \geq\lambda(A\setminus X)+\lambda(X\setminus A).
\]
It follows that $\lambda(X\setminus A)<\lambda(X)\leq k$.
Its complement contains the strong set $A$, so the tangle orients $X\setminus A$ as small.
The other part $A\cap X$ is contained in one small set by weakness.
\end{proof}

\input{figures/two-small-corners}

Consider an arbitrary decomposition of the required width.
We want an edge that cuts the core into two strong parts.
If no such edge exists, we may direct every edge towards its strong intersection with the core.
At a vertex where all three edges point inwards, each of the three branches meets the core in a weak set.
The preceding lemma then covers the three branches by six small sets, contrary to robustness.

\begin{lemma}
\label{lem:coresplit}
Let $(A,B)$ be a strong separation of order $k$ with $B$ fully closed, and let $\mathcal T$ be robust.
Assume every singleton in $A$ is weak.
Every width-$k$ branch decomposition has an edge whose two sides both meet $A$ in a strong set.
\end{lemma}
\begin{proof}
Suppose no edge has this property.
No edge can have two weak intersections with~$A$.
If one were empty, the other would be~$A$ and hence strong.
If neither were empty, \cref{lem:weakcorner} would cover each side by two small sets, contradicting robustness.
Thus, every edge has exactly one side with strong intersection with $A$.
Orient the edge towards that set.

\begin{figure}[H]
\centering
\color{black}
\begin{tikzpicture}[manuscript illustration]
  \node[dot] (v) at (3,0) {};
  \node[box] (x) at (1.35,1.0) {$X_1$};
  \draw[arr] (x) -- (v);
  \node[box] (y) at (4.65,1.0) {$X_2$};
  \draw[arr] (y) -- (v);
  \node[box] (z) at (3,-1.15) {$X_3$};
  \draw[arr] (z) -- (v);
  \node[note] (n7) at (8.75,.78) {each branch is covered by two small sets\\in the argument for $A$};
  \node[note] (n8) at (8.75,-.63) {each branch meets $B$ in a weak set\\in the argument for $B$};
\end{tikzpicture}
\caption{The same sink argument is used in \cref{lem:coresplit,lem:oppositeseed}.
We direct each edge towards the side whose intersection with $A$, or with $B$ in the second argument, is strong.
At a sink, the branches $X_1,X_2,X_3$ each meet that set in a weak set.
For the core $A$, this gives a cover of $V$ by six small sets.
For the opposite set~$B$, it covers $B$ by three weak sets.}
\label{fig:strong-sink}
\end{figure}
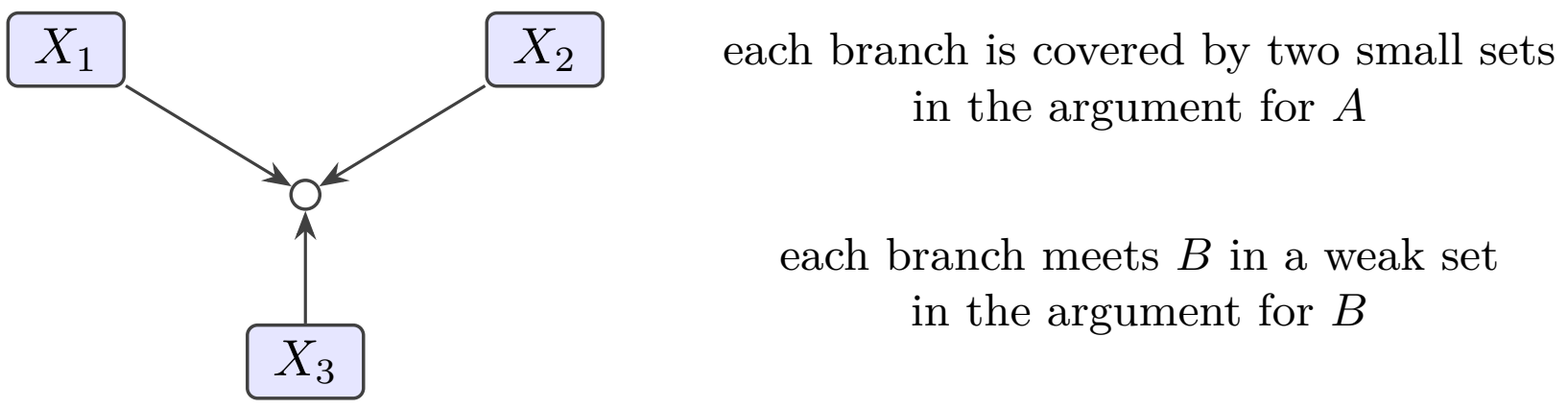

Every leaf edge points away from its leaf.
Indeed, its intersection with $A$ is either empty or a weak singleton.
A finite directed tree has a sink, which is consequently an internal degree-three vertex.
Let $X_1,X_2,X_3$ be the three branches there.
Each $X_i\cap A$ is weak.
None can be empty: otherwise $A$ would be partitioned into at most two nonempty weak intersections, and an edge at the sink would have weak intersections on both sides, already ruled out.
Applying \cref{lem:weakcorner} to all three branches covers $V$ by six small sets, again a contradiction.
\end{proof}

Once a strong set on one side has been displayed, we also need a strong set on the opposite side.
For this second step, full closure is no longer needed.
For comparison, Hall, Oxley, Semple, and Whittle find an edge with at least two marked leaves on each side whenever at least four leaves are marked~\cite[Lemma~3.3]{hosw}.
Here the same tree argument works with strong intersections whenever three weak sets cannot cover the opposite side.

\begin{lemma}
\label{lem:oppositeseed}
Let $B$ be a set not covered by three weak sets.
In every branch decomposition, either a leaf in $B$ is itself strong, or some edge has strong intersections with $B$ on both sides.
\end{lemma}
\begin{proof}
If there is no such edge and no such leaf, orient every edge towards its unique strong intersection with $B$.
A unique strong side exists, since two weak sides would cover $B$ by two weak sets.
All leaf edges point away from their leaf.
The three incoming branches are shown in \cref{fig:strong-sink}.
An internal sink partitions $B$ into three weak intersections, a contradiction.
\end{proof}

We can now prove that replacing a core by one atom does not destroy the width bound.
For a titanic set $A$, Geelen, Gerards, Robertson, and Whittle prove the contraction result in the matroid setting~\cite[Lemma~2.1]{ggrw}, while Korhonen and Oum prove it for arbitrary connectivity functions~\cite[Proposition~2.4]{titanic}.
Kalampakas obtains a corresponding display in the rank-two graph setting from the width-three theorem when both sides have at least four atoms~\cite[proof of Lemma~4.4]{ranktwo}.
For the non-titanic case at order $k$, full closure and robustness give us a strong set on each side, and the two changes to the tree then give a decomposition displaying the core.

\begin{theorem}
\label{thm:corecontraction}
Let $\mathcal T$ be a robust order-$k$ tangle in a connectivity system of branchwidth at most $k$.
Let $(A,B)$ be a strong separation of order $k$, and suppose $B$ is fully closed.
Then contracting $A$ preserves branchwidth at most $k$.
In fact, if $A$ is not titanic, some width-$k$ branch decomposition displays $(A,B)$.
\end{theorem}
\begin{proof}
If $A$ is titanic, the contraction bound follows directly from~\cite[Proposition~2.4]{titanic}, since~$\lambda(A)=k$.

Otherwise, $A$ has a partition into at most three nonempty sets of order less than $k$.
In this case every singleton in $A$ is weak.
Indeed, a part containing a strong singleton would itself be strong, while its complement contains the strong set $B$, forcing that part to have order at least $k$.
Each part is small, because its complement contains the strong set $B$.
Consequently $B$ cannot be covered by five weak sets, since together with the small parts of~$A$ they would violate robustness.

\input{figures/core-display-contract}

Take a width-$k$ branch decomposition.
By \cref{lem:coresplit}, some edge $(X,Y)$ has both $A$-intersections strong.
At least one of $X\cap B,Y\cap B$ is strong, since otherwise $B$ is covered by two weak sets.
Say $X\cap B$ is strong.
\cref{lem:strongsurgery} now gives a width-$k$ decomposition displaying the strong set $C=Y\cap A$.

If this decomposition already has a strong singleton in $B$, take that displayed leaf as~$D$ and apply \cref{lem:strongseeds}.
Otherwise, \cref{lem:oppositeseed} gives an edge with both $B$-intersections strong.
An edge in the branch displaying $C$ cuts off a subset of $C$ disjoint from $B$.
The edge given by the lemma therefore lies outside this branch.
Let $Z$ be its side containing $C$.
Both~$Z\cap A$ and $Z\cap B$ are strong.
Apply \cref{lem:strongsurgery} again, keeping the $Z$-branch and splitting the other branch.
It displays a strong set $D\subseteq B$ and preserves the already displayed set $C$.
\cref{lem:strongseeds} now displays $(A,B)$ at width $k$.
Contracting its $A$-branch proves the required bound.
\end{proof}

A node of the component tree may have several outward cores.
To contract them one at a time, we must check that the complement of each remaining core is still fully closed after the earlier contractions.

Kalampakas likewise contracts the boundary blocks one at a time at rank two, allowing an exceptional local graph with at most twelve vertices~\cite[Lemma~4.4]{ranktwo}.
Here \cref{thm:corecontraction} applies at each step because the tangle on the remaining atoms still has the required properties.

\begin{corollary}
\label{cor:simultaneouscores}
Let $(V,\lambda)$ have branchwidth at most $k$ and let $\mathcal T$ be a robust order-$k$ tangle.
Let $A_1,\ldots,A_m$ be pairwise disjoint strong order-$k$ sets, each with a strong fully closed complement.
Replacing all $A_i$ by atoms yields a connectivity system of branchwidth at most $k$.
\end{corollary}
\begin{proof}
Contract the sets successively.
On the current atom set, restrict $\mathcal T$ to atom unions.
This gives an order-$k$ tangle, since the orientation and covering axioms follow by lifting the sets of atoms to the original system.
The cut around each new atom has order~$k$, so the restricted tangle has no singleton big set consisting of a new atom.
Original singleton atoms still satisfy the singleton axiom.
Robustness also lifts directly.

Each uncontracted $A_j$ remains strong, as does its complement: a weak atom union lifts to a weak original set.
Its complement remains fully closed.
Indeed, a nonempty atom union disjoint from that complement consists of original singleton atoms in $A_j$.
If it is weak in the restricted tangle, it lifts to a weak set in $\mathcal T$, and adding it to the complement of $A_j$ therefore raises the order above $k$.
The theorem therefore applies at each step, preserving the width bound.
\end{proof}

\input{figures/disjoint-core-atoms}

The proof uses robustness and the fact that each core has a fully closed complement.
It applies even when there are other tangles of order $k$ or when the separation classes of the cores are not strong.

We can therefore replace each of the disjoint boundary cores by one element and still have a branch decomposition of width at most $k$.
The remaining elements of~$V$ stay as singleton atoms, and a cut of the new system has the value of the corresponding union of original parts (\cref{fig:core-contraction-result}).
We next replace the atoms for the cores by bounded graphs so that we can recover adjacencies across the cuts.

\input{figures/core-contraction-result}
\subsection{Local port graphs as linear sums of weak pieces}

Let $G$ have rankwidth at most $k\geq2$, and let $\mathcal T$ be a robust order-$k$ tangle.
Consider one of the components represented in the component tree of \cref{thm:generalclassnormal}.
Write $F_1,\ldots,F_m$ for its disjoint outward boundary cores and $K=V(G)\setminus\bigcup_iF_i$.
An original vertex belongs to $K$ if it lies outside all these outward cores, even if its bag was suppressed when we formed the component tree.
Each $F_i$ is strong, has rank $k$, and has a strong fully closed complement.
We use the following three properties from \cref{thm:generalclassnormal}.
\begin{enumerate}
\item A component has either no flower or one flower.
\item In the flower case, each boundary core $F_i$ belongs to a petal $P_i$ satisfying $F_i\subseteq P_i\subseteq c(F_i)$.
Every remaining petal $P$ contains no outward boundary core and satisfies $c(V(G)\setminus P)=V(G)$.
\item In the flower-free case, every non-sequential cut of order at most $k$ that keeps all boundary cores whole is equivalent to one of the boundary cuts.
\end{enumerate}

In \cref{lem:localweakblocks}, we assume these component properties and prove the required bound on the cut-rank.
For split-prime graphs of rankwidth two, Kalampakas orders the individual vertices of each local graph so that the three vertices representing any one boundary occur consecutively~\cite[Lemma~4.3 and Proposition~4.8]{ranktwo}.
We use the same ordering argument at order $k$, allowing a whole weak set to form one block and measuring cut-rank after each completed block.

We keep the vertices of~$K$ with their induced edges and call them the \emph{ordinary vertices}.
In each~$F_i$, choose~$k$ vertices whose rows form a basis of the adjacency matrix across the cut around~$F_i$.
We replace~$F_i$ by a set~$T_i$ of new vertices, called \emph{ports}, one for each nonzero vector of~$\mathbb F_2^k$.
The vector specifies a sum of basis rows, which gives the neighborhood of its port in~$K$.
For ports in different sets, we take the row and column combinations specified by their vectors in the adjacency matrix between the two sets of basis vertices.
We may choose the edges inside each~$T_i$ arbitrarily.
Call the resulting graph~$H$ the \emph{port graph}.

Suppose that a cut of~$H$ keeps each port set~$T_i$ on one side.
Replacing each~$T_i$ by~$F_i$ gives the corresponding cut of~$G$.
In the cut matrix of~$G$, we may restrict each core to its basis vertices without changing the rank, since their rows or columns span the others.
In the cut matrix of~$H$, we may likewise restrict each port set to the ports for the individual basis rows.
The two remaining matrices agree entry by entry, so the original cuts have the same rank~\cite[Lemma~4.2]{ranktwo}.

\input{figures/boundary-basis-ports}

We now seek an order of the blocks of~$H$ in which every prefix ending at a block boundary has bounded cut-rank.
By \cref{lem:linearclosure}, it suffices to prove that such an order exists.

\begin{lemma}\label{lem:localweakblocks}
Under the stated component properties, $H$ admits an ordered partition into blocks of the following two kinds:
\begin{enumerate}
\item a port set $T_i$;
\item a subset of $K$ that is $\mathcal T$-weak in the original graph.
\end{enumerate}
Every union of a completed initial segment of blocks has cut-rank at most $2k$ in $H$.
\end{lemma}
\begin{proof}
There are two sources for the required order.
At a flower, we use the petal order and then fill in each petal by a weak closure sequence.
Without a flower, a branch decomposition of the contracted system provides three sequences that can be placed one after another.

\input{figures/weak-block-layout}

Suppose first that the component has a flower.
If a petal $P_i$ contains the boundary core~$F_i$, we can use full closure to partition $P_i\setminus F_i$ into a sequence of weak blocks, with every union of $F_i$ and a completed prefix having rank at most $k$.
To see this, take a maximal partial $k$-sequence for $F_i$ and intersect its blocks with $P_i$.
Each intersection remains weak.
A partial union $C$ intersects $P_i$ in the required prefix.
The union $C\cup P_i$ contains the strong core $F_i$, and its complement contains the strong opposite core $V(G)\setminus c(F_i)$.
The union therefore has rank at least $k$.
Submodularity applied to $C$ and $P_i$ now bounds the rank of their intersection by $k$.

In the port graph, start this petal with the whole block $T_i$ and then use that weak-block sequence.
The equality of cut-ranks just proved applies after each complete block.
For every other petal~$P$, its complement has full closure~$V(G)$.
Take a weak closure sequence from that complement to~$V(G)$ and read its blocks in reverse.
Each completed prefix is the complement of an intermediate set in the closure sequence, so it has rank at most~$k$.
Since that petal contains no boundary core, all its blocks lie in $K$, and the same identity applies.
Concatenate the petal sequences in flower order.
A completed prefix is the union of a consecutive set of complete petals, of rank at most $k$, and a completed prefix in one petal, of rank at most $k$.
Subadditivity gives $2k$.

Suppose now that the component is flower-free.
Contract every $F_i$ to one atom and keep the vertices of $K$ as singleton atoms.
By \cref{cor:simultaneouscores}, there is a branch decomposition~$D$ of the resulting atom connectivity function $\mu$ of width at most $k$.
If there are at most two atoms, list them as separate blocks.
Each is an interior weak singleton or one boundary atom, and every completed prefix has order at most~$k$.
We may therefore assume that $D$ has an internal vertex.

We show that every edge of $D$ has a side from which all atoms can be reached by additions of two permitted kinds: a weak set of interior vertices, or one boundary atom.
All intermediate atom unions have $\mu$-value at most $k$.

Let $X$ be the original lift of a side of an edge cut.
If $X$ is strong and $c(X)=V(G)$, then~$X$ contains every boundary core.
Indeed, if~$X$ missed a core~$F_i$, then it would lie in~$V(G)\setminus F_i$, and monotonicity would force the full closure of $X$ into the fully closed proper set $V(G)\setminus F_i$.
A closure sequence from $X$ therefore adds only weak subsets of $K$ and transfers unchanged to the atom system.
The same argument applies if the opposite side is strong and its full closure is $V(G)$.
If $X$ is weak, it contains no strong boundary core.
Since~$X$ is a union of atoms, it consists entirely of interior vertices.
Its complement thus reaches the whole ground set by adding $X$ in one step.
The same argument applies with the two sides exchanged.

The remaining case is a strong non-sequential cut.
By the component property, its class is that of an incident boundary, say the one with core $F_i$.
Orient its lift so that $F_i\subseteq X$ and $c(X)=c(F_i)$.
For $j\ne i$, the fully closed set $V(G)\setminus F_j$ contains $F_i$ and therefore its full closure.
It follows that $X$ contains no other boundary core, and $X=F_i\cup Y$ for some $Y\subseteq K$.
Equivalence gives
\[
 c(V(G)\setminus X)=V(G)\setminus F_i.
\]
A closure sequence from the left-hand set adds weak subsets of $Y$.
Transfer these additions to the atom system and then add the single atom $F_i$, reducing the connectivity from $k$ to zero.
This proves the assertion for every edge.

Orient each edge towards a set with such a completion.
Leaf edges may always point away from their leaf, since the remaining atom is either a weak original singleton or one boundary atom and can be added in one step.
At an internal sink with branches $A_1,A_2,A_3$, reverse the three completion sequences.
This partitions each $A_i$ into permitted blocks whose completed prefixes have connectivity at most $k$: these prefixes are the complements of intermediate sets in the completion starting at the complement of $A_i$.
Concatenate the first two branch orderings and the reverse of the third.
A completed prefix in the first branch has rank at most $k$.
In the second branch, it is $A_1$ together with one rank-at-most-$k$ prefix, so subadditivity gives at most $2k$.
In the third branch, its complement is a prefix of the third branch's original ordering, so its rank is again at most $k$.
Finally replace each boundary atom by the whole set $T_i$ and use the equality of cut-ranks for cuts that keep each port set whole.
\end{proof}

\input{figures/whole-marker-block}

The same prefix bound holds in every induced subgraph of~$H$.
We intersect each block with the new vertex set and omit the empty blocks.
Every completed-prefix cut matrix is a submatrix of the old one, weak blocks remain weak, and port blocks remain of bounded size.
It is also preserved on adding one marker adjacent only to the ports of each $T_i$, by putting that marker in the same block.
No new marker edge crosses a completed-block boundary.

We can now apply the linear closure lemma to the weak-block layouts constructed above.
When the tangle is unique, \cref{lem:inducedtanglelift} shows that every graph induced by a weak set has smaller rankwidth.
The induction hypothesis therefore deals with every possibly large weak block, while the port blocks have bounded size from the start.
Since every completed prefix has bounded cut-rank, \cref{lem:linearclosure} gives a uniform decomposer for the local graph without requiring the block order as input.
For rank two, Kalampakas already obtains a linear layout of bounded width after restricting to the local graph~\cite[Corollary~4.9]{ranktwo}.

\begin{corollary}
Suppose in addition that $\mathcal T$ is the unique order-$k$ tangle of $G$, and assume that graphs of rankwidth less than $k$ have bounded definable cliquewidth.
The local port graphs have uniformly bounded definable cliquewidth.
Definable cliquewidth remains uniformly bounded if, at a node of the rooted tree, we keep only its original vertices and the port sets for its children, and add one marker for each of these port sets.
\end{corollary}
\begin{proof}
Every weak block induces its original graph from $G[K]$ and has rankwidth less than~$k$ by \cref{lem:inducedtanglelift} and the uniqueness of the tangle.
Each port block, including its optional marker, has at most $2^k$ vertices.
Apply \cref{lem:linearclosure} to the ordering from \cref{lem:localweakblocks}, whose completed-prefix ranks are at most $2k$.
The order witnesses this bound, so the transduction can act on the graph alone.
\end{proof}
\subsection{Assembling the local trees around the tangle}

For split-prime graphs of rankwidth two, Kalampakas uses a tree of cores and compatible bases at their boundaries to evaluate each recognizable property~\cite[Section~5]{ranktwo}.
Here we use this idea at order~$k$ to construct one tree from which we can recover all adjacencies.
The core tree and its finite boundary data are chosen by MSO.
The weak-block orders are used only to justify a uniform bound for the local decomposers.

\begin{proof}[Completion of the proof of \cref{thm:weakpieceencoding}]
\noindent\textit{The robust core tree.}
Suppose now that $\mathcal T$ is robust.
Encode it by a bounded tuple and its finite list of choices, and check its tangle axioms by an MSO guard.
By \cref{cor:tangleparameters}, the chosen tuple and list of choices give an MSO formula testing whether a set is big in this tangle.
We check robustness by universally quantifying sets $X_1,\ldots,X_8$ and requiring that, if all eight are big, then $X_1\cap\cdots\cap X_8$ is nonempty.
These variables may denote equal sets, so this also checks intersections of fewer than eight big sets.
This guard ensures that the chosen tangle satisfies the hypothesis of the robust-tangle structure theorem used below.
The hypothesis of \cref{thm:weakpieceencoding} guarantees a choice of~$\mathcal T$ with the required bound on its weak pieces.

Use full $\mathcal T$-closure, with additions of arbitrary weak sets.
It is the least fully closed strong $k$-separating superset, hence an MSO-definable set operation.
Define the strong non-sequential classes.
If there is at least one such class, choose an anchor orientation and form the selected cores.
The selected cores are laminar by \cref{lem:generalcorelaminar}, and \cref{thm:generalclassnormal} describes their local pieces.
The counting-free laminar-tree theorem~\cite[Theorem~2]{laminar} gives a rooted tree $D$ with the original vertices as leaves.
If there is no strong class, $D$ has just one internal root and those leaves.

We call the set~$S_u$ of original vertices at leaves below an internal node~$u$ its \emph{territory}.
Write~$I_u$ for the vertices at its leaf children and~$\operatorname{Ch}(u)$ for its internal children.
Then
\[
 S_u=I_u\mathbin{\dot\cup}
       \mathop{\dot\bigcup}_{v\in\operatorname{Ch}(u)}S_v.
\]
The root has territory~$V(G)$.
Every other internal territory is a strong non-sequential core of rank~$k$.
At a nonroot node $u$, the canonical boundary in the parent direction is
\[
 F_u^{\mathrm{out}}=V(G)\setminus
       c(S_u).
\]
This is the core of the opposite orientation of the same class.
It is disjoint from $S_u$, and its complement is fully closed.
This core and the child territories $S_v$ are exactly the outward cores at the corresponding node of the component tree.
The parent-side core can be smaller than~$V(G)\setminus S_u$, since~$c(S_u)$ may strictly contain~$S_u$.

\smallskip
\noindent\textit{Choosing all boundary bases at once.}
We extend the two-selector construction of Kalampakas~\cite[Lemma~5.1]{ranktwo} to $k$ selectors.
At every nonroot internal node~$u$, choose vertices in~$S_u$ whose rows form an ordered basis of the cut matrix at~$S_u$, and write
\[
 R_u=(r_1(u),\ldots,r_k(u)).
\]
These choices can be made coherent with $k$ selector predicates on the tree nodes.
Whenever a parent representative lies in a child territory, keep it in its slot in the child basis.
The inherited rows remain independent, since the cut matrix at the child still contains all the columns outside the parent territory.
Extend the inherited rows to a basis in the unused slots.
For each slot, mark the child leading to its representative at every nonroot internal node.
Starting at such a node and following marked children leads to its representative leaf, since an inherited representative keeps its slot.
One unary predicate per slot therefore suffices for the whole tree.
At the root we use the empty basis.
One fixed MSO guard verifies all paths and basis conditions.

\smallskip
\noindent\textit{The local graphs and their width bound.}
Put $\Lambda=\mathbb F_2^k\setminus\{0\}$.
At node $u$, create
\[
 V(H_u)=I_u\mathbin{\dot\cup}
    \mathop{\dot\bigcup}_{v\in\operatorname{Ch}(u)}
       T_v,\qquad T_v=\{t_v^a:a\in\Lambda\}.
\]
Keep the edges induced on $I_u$, choose any fixed internal graph on each $T_v$, and define the remaining adjacencies by
\[
 E(t_v^a,x)=\sum_{p=1}^k a_pE_G(r_p(v),x),
 \qquad
 E(t_v^a,t_z^b)=\sum_{p,q=1}^k a_pb_qE_G(r_p(v),r_q(z))
 \quad(v\ne z).
\]
The arithmetic is over $\mathbb F_2$, so these are fixed Boolean formulas.
Each port $t_v^a$ is a copy of the child node $v$ and carries a unary label for its vector $a$.
The zero vector represents an empty outside neighborhood, and the reconstruction formulas below handle it by an empty sum.

To prove the needed local bound, first take the port graph of the whole component, including the ports in the parent direction when there is one.
Its disjoint canonical boundary cores all have fully closed complements.
By \cref{cor:simultaneouscores}, the connectivity system on the atoms has a branch decomposition of width at most $k$.
By \cref{thm:generalclassnormal,lem:localweakblocks}, we can order the atoms in blocks so that each completed prefix has rank at most~$2k$ and every block is either a weak set of ordinary vertices or one boundary atom.
Replacing atoms by their finite port groups preserves this bound, since row and column spanning preserve the rank of every cut that keeps the groups intact.

The rooted graph $H_u$ is the induced subgraph obtained by deleting the parent port group and all original vertices outside $S_u$.
At a nonroot node, the original vertices deleted in this step lie in $c(S_u)\setminus S_u$.
Use the coherent bases chosen above for all child territories in the port graph of the whole component.
The choice of basis for the parent direction has no effect on~$H_u$, since we delete its port group.
Intersect each block of the ordering with~$V(H_u)$ and omit the empty intersections.
Prefix cut-rank cannot increase under induced restriction, and subsets of weak sets remain weak.
Consequently, $H_u$ has a block ordering of prefix rank at most $2k$ whose blocks are either induced graphs on weak original sets or finite port groups.

For each child $v$, add a marker $a_v$ adjacent exactly to all of $T_v$, and call the resulting graph $J_u$.
We use separate unary labels for the markers and the original vertices.
Place $a_v$ in the same block as $T_v$.
The marker contributes no cross-edge at a completed block boundary, so the prefix-rank bound remains $2k$.
Every port block with its marker has at most $2^k$ vertices.
Every nonempty weak block has definable cliquewidth at most $m$ by hypothesis.
Linear closure of definable cliquewidth therefore gives a uniform origin-preserving MSO decomposer for all the graphs $J_u$.
The weak blocks, flower labels, petal orders, and block orders serve only to prove the uniform local bound.
The local decomposer acts on~$J_u$ alone.

The domains of all $J_u$ are disjoint and have total size linear in $|V(G)|$.
The children of~$u$ identify its domain: each leaf child gives one ordinary vertex and each internal child gives its ports and marker.
We can therefore run the decomposer in parallel on these domains as in \cref{sec:sparse-encodings}, keeping all fixed unary labels.
Obtain an expression tree $P_u$ with its source leaves identified with $V(J_u)$.
The usual decomposer filter ensures that every accepted local output decodes the correct local graph.

\smallskip
\noindent\textit{The finite labels needed to recover the original graph.}
For $x\in S_u$ and nonroot internal $u$, write $\ell_u(x)\in\mathbb F_2^k$ for the coordinate vector of its outside row in the basis $R_u$.
Keep $\ell_u(x)$ on every original leaf $x\in I_u$.
If $v$ is an internal child of a nonroot node $u$, keep on its marker the matrix
\[
 M_v[p,q]=\ell_u(r_p(v))_q.
\]
For $x\in S_v$, restrict its row at~$v$ to the columns outside~$S_u$ and express each remaining basis row in~$R_u$.
This gives
\[
 \ell_u(x)=\ell_v(x)M_v\qquad(x\in S_v).
 \tag{UT}
\]
No transition into the empty root boundary is required.
All vectors and matrices range over finite sets depending only on $k$, and their correct values are defined by fixed formulas before the original adjacency is discarded.

\smallskip
\noindent\textit{The output tree and its inverse.}
Mark the root $p_u$ of each $P_u$.
For each internal child $v$ of~$u$, join the marker leaf $a_v$ in $P_u$ to $p_v$ by a path of length two, using a new copy of $v$ with the label \textsc{Link} as its middle vertex.
These paths join disjoint trees according to the internal-node tree of $D$, so the resulting graph is a tree (\cref{fig:local-tree-link}).
Original vertices remain designated leaves.
Deleting all \textsc{Link} vertices recovers the local trees as connected components, each with its unique marked root.
Their MSO decoders can therefore be relativized to those components.

\input{figures/local-tree-link}

The links recover the parent relation on local roots and hence the core hierarchy.
Decoding~$J_u$ recovers the marker $a_v$, and among its neighbors the label for $a$ identifies the port $t_v^a$.
For an original vertex $x$, its owner is the unique internal node $w$ with $x\in I_w$.
For every nonroot ancestor $v$ of $w$, repeated use of (UT) computes $\ell_v(x)$ from $\ell_w(x)$.
A run along this path uses only $2^k$ states, so it is definable with finitely many monadic state sets.
This is MSO, without any counting quantifier.

Let $u$ be the least common ancestor of the owners of distinct original vertices $x,y$.
If both belong to $I_u$, use their decoded edge in $H_u$.
For each~$q\in[k]$, let~$e_q$ be the vector with a single one in position~$q$, so at each internal child~$v$ the port~$t_v^{e_q}$ represents its $q$th basis row.
If~$x\in I_u$ and~$y\in S_v$ for a child~$v$, use
\[
 E_G(x,y)=\sum_{q=1}^k\ell_v(y)_q E_{H_u}(x,t_v^{e_q}).
\]
If $x\in S_v$ and $y\in S_z$ for distinct children, use
\[
 E_G(x,y)=\sum_{p,q=1}^k
       \ell_v(x)_p\ell_z(y)_q
       E_{H_u}(t_v^{e_p},t_z^{e_q}).
\]
These are the two row-basis expansions of the relevant cut matrix.
They also cover zero coordinate vectors.
Each sum has fixed length, so the equations define one MSO decoder on the output tree.

Finally, every choice is guarded while the original input is still available: tangle and anchor parameters, the core system, selectors, local decomposers, and all finite labels.
Composition of finite-copy transductions remains finite-copy.
Every accepted output is therefore a correct tree encoding, and at least one output exists for every input in the theorem.
A wrong choice of the robust tangle may cause some local decomposer to reject, but it cannot produce an incorrect accepted output.
The good tangle promised by the hypothesis provides a successful choice.
Take the finite union of this construction and the nonrobust-case expression decomposer, using a root label to distinguish their two fixed decoders.
This proves \cref{thm:weakpieceencoding}.
\end{proof}

We return to the unique-tangle case of the outer induction.

\begin{corollary}
\label{thm:uniquetangleencoding}
Fix $k\geq2$ and assume that graphs of rankwidth less than $k$ have uniformly bounded definable cliquewidth.
Then graphs of rankwidth at most $k$ having exactly one order-$k$ tangle have source-preserving MSO encodings by finitely labeled trees.
The same class has bounded definable cliquewidth.
Every fixed finite collection of unary labels can be kept.
\end{corollary}
\begin{proof}
If $\mathcal T$ is the unique order-$k$ tangle, every $\mathcal T$-weak induced graph has rankwidth less than $k$ by \cref{lem:inducedtanglelift} and tangle duality.
The inductive hypothesis gives one uniform bound~$m$ for their definable cliquewidth.
Apply \cref{thm:weakpieceencoding}.
By \cref{lem:treecliquecompiler}, we can turn this tree encoding into a cliquewidth decomposition.
\end{proof}

\begin{figure}[H]
\centering
\color{black}
\begin{tikzpicture}[manuscript illustration]

\draw[rounded corners=6pt] (-2.8,-1.4) rectangle (2.8,1.4);
\node[anchor=south east,inner sep=0pt] at (2.45,-1.05) {$V(G)$};
\draw[fill=orange!15] (-.3,0) ellipse (1.9 and 1.05);
\node at (.95,.5) {$S$};
\draw[fill=blue!18] (-.75,0) ellipse (.95 and .6);
\node at (-.75,0) {$W$};
\node[note] at (6,.8) {$\mathcal T$ makes $S$ small};
\node[note] at (6,0) {a tangle on $G[W]$ would lift\\to one making $S$ big};
\node at (6,-.9) {$\operatorname{rw}(G[W])<k$};
\end{tikzpicture}
\caption{The weak set $W$ is contained in a small set $S$ of the unique order-$k$ tangle $\mathcal T$ of~$G$.
An order-$k$ tangle of $G[W]$ would lift to one that makes $S$ big, since $S\cap W=W$.
This would give a second tangle of order $k$ in $G$, so $G[W]$ has rankwidth less than $k$.}
\label{fig:unique-weak-lift}
\end{figure}
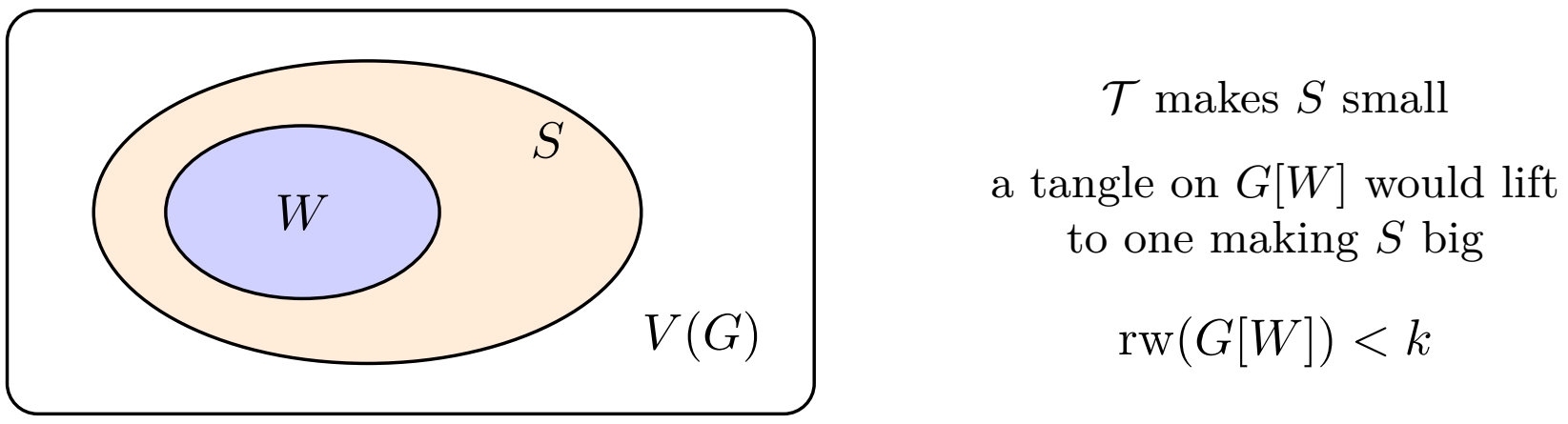

This completes the outer induction step for graphs with exactly one order-$k$ tangle.
It remains to handle graphs with several such tangles.

\input{figures/unique-tangle-result}

\section{The induction at fixed rankwidth}
\label{sec:fixed-rankwidth}

Suppose now that the input graph has several order-$k$ tangles.
We will separate them by cuts of order less than~$k$ and replace the regions beyond these cuts by graphs of bounded size.
Some tangles in the resulting local graph may have a big side contained entirely in one of these replacements.
We therefore keep the vertex set of each replacement as a support.
Recall that a support \emph{captures} a tangle if it contains one of its big sides (\cref{sec:tangles}).
We need to separate only the uncaptured tangles.
We first state the induction for a graph with supports and deal with the cases in which we can finish the construction.
In \cref{sec:boundaries} we construct the local graphs, and in \cref{sec:supports} we show why the induction applies to them.

At each step we have a graph~$H$ of rankwidth at most~$k$ and a family of supports, each with at most~$L$ vertices.
The enriched input also has a separate vertex for each support, called its index, and a relation joining each index to the members of its support.
By the \emph{graph reduct} we mean~$H$ itself.
The graph, the support incidences, and any trees already constructed must all come from one fixed MSO transduction of a bounded-cliquewidth class.
This ensures, in particular, that the incidence graph on its own has bounded cliquewidth.
The number of support indices over all local graphs is linear in the size of the preceding enriched input, since the transduction makes only a fixed number of copies of each element.

The parameter~$r$ is a lower bound on the order of every cut that distinguishes two uncaptured tangles.
On the original graph we take the support family to be empty and put~$r=0$, since every cut has nonnegative order.
The inner induction is on~$k-r$.
By \cref{cor:ownerthreshold}, we can increase $r$ by one in every nonroot local graph after a separating stage.
When $r=k$, at most one uncaptured tangle remains, and \cref{thm:terminalsupport} applies.

\begin{lemma}\label{lem:recursiveencoding}
Fix $k\geq2$ and $0\leq r\leq k$, and assume that all graphs of rankwidth less than $k$ have uniformly bounded definable cliquewidth.
Let $\mathscr H$ be a class of enriched graphs jointly obtained by a fixed MSO transduction from a bounded-cliquewidth graph class, whose graph reducts have rankwidth at most $k$ and whose supports have uniformly bounded size.
Suppose that distinct uncaptured order-$k$ tangles have minimum distinguishing order at least~$r$.
Then there is a fixed MSO transduction on the enriched inputs that gives source-preserving encodings of their graph reducts by uniformly sparse graphs.
In the same output, we keep the support indices, their incidences with the distinguished copies of the graph vertices, and any fixed unary labels.
Along each branch of the recursion, the construction uses at most~$k-r$ separating stages before it reaches the terminal case.
\end{lemma}

The distinguished copy of a graph vertex has that vertex as its origin.
The enriched input also contains support indices, so an auxiliary vertex of the output may be a copy of an index.
The bounds on the number of copies, the labels, and the degeneracy are uniform over the input class.

\subsection{Coloring overlapping supports}

We call a set \emph{rainbow} when its vertices receive different colors.
If all supports are rainbow, restricting the graph to two colors leaves at most two vertices in each support.
We will use this observation to exclude tangles from the induced graph.

We first recall a simple consequence of the tangle axioms.
Every set of fewer than $k$ vertices is small in an order-$k$ graph tangle.
Indeed, enumerate the set and add its vertices one by one.
Each prefix has cut-rank less than $k$, and a union of two small sets is small whenever its order is less than $k$, since otherwise the two sides and the small complement of their union would cover the graph.
This proves the assertion by induction from singletons.
We will also need the case of two vertices when $k=2$.
Here we must assume separately that the pair has order less than $k$, and then its two singleton sets and its complement give the same argument.

Suppose now that every order-$k$ tangle of~$H$ is captured.
A tangle in a one- or two-color induced subgraph would lift to~$H$, where some support contains one of its big sides.
The intersection of that side with the induced subgraph would still be big, although it has at most two vertices.

\begin{lemma}\label{lem:rainbowregions}
Let $\mathcal U$ be a family of vertex sets in a graph $H$.
Suppose a coloring with $q$ colors is injective on every $U\in\mathcal U$, and every order-$k$ tangle of $H$ has a big set $S$ of order less than~$k$ contained in some $U\in\mathcal U$.
Then every nonempty union of fewer than $k$ color classes induces a graph of rankwidth less than $k$.
For $k\geq2$, the same holds for every nonempty union of at most two color classes.
\end{lemma}
\begin{proof}
Suppose that the induced graph on a union $W$ of $p<k$ color classes has an order-$k$ tangle, and lift this tangle to $H$ by \cref{lem:inducedtanglelift}.
By assumption, the lifted tangle has a big low-order set $S\subseteq U$ for some support $U$.
Its trace $S\cap W$ is big in the induced tangle, while the rainbow property gives
\[
               |S\cap W|\leq |U\cap W|\leq p<k.
\]
This contradicts the preceding small-set observation.
Hence, no such induced tangle exists, and tangle duality gives rankwidth less than $k$.

For the additional assertion, only the case $k=p=2$ is new.
Although cardinality alone does not give the required strict rank bound, the trace $S\cap W$ has order less than $k$ by restriction of the original cut.
Its complement and the singletons of its vertices would therefore cover $W$ by at most three small sets, which is again impossible.
\end{proof}

\begin{figure}[H]
\centering
\color{black}
\begin{tikzpicture}[manuscript illustration]

\draw[fill=blue!10] (0,0) ellipse (2.2 and .95);
\node at (0,.58) {$U$};
\foreach \x/\c/\l in {-1.2/blue/1,-.4/orange/2,.4/violet/3,1.2/green/4}
 \node[circle,draw,fill=\c!25,inner sep=3pt] at (\x,0) {$\l$};
\draw[-{Stealth[length=2mm]},thick] (2.6,0)--(4,0);
\node[note] at (3.3,.55) {keep colors $1,2$};
\draw[fill=blue!10] (6,0) ellipse (1.5 and .95);
\node at (6,.58) {$U\cap W$};
\node[circle,draw,fill=blue!25,inner sep=3pt] at (5.5,0) {$1$};
\node[circle,draw,fill=orange!25,inner sep=3pt] at (6.5,0) {$2$};
\end{tikzpicture}
\caption{Keeping colors $1$ and $2$ leaves at most two vertices of a rainbow support, and therefore at most two vertices of any big low-order set contained in that support.
For $k=2$ one also uses that the restricted cut still has order less than two.
Such a trace cannot be big in the induced tangle.}
\label{fig:rainbow-trace}
\end{figure}
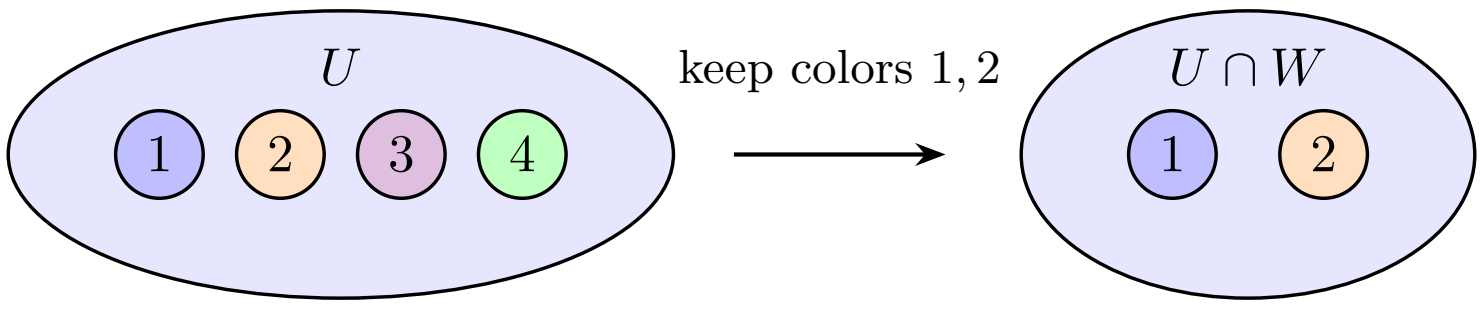

For overlapping supports, a bound on their sizes alone gives no bound on the number of colors.
For example, if every pair of vertices is a support, all the vertices need different colors.
In our setting we can use the theorem of Gurski and Wanke to bound the treewidth of the incidence graph~\cite{gw}.
We then turn a tree decomposition of this graph into one in which every support lies in a bag, as in the proof of Samer and Szeider~\cite[Lemma~2(3)]{incidenceprimal}.
The decomposition proves that a fixed palette suffices, so the transduction can simply guess the coloring and check it.

\begin{lemma}
\label{lem:supportincidence}
Suppose $L\geq1$, every member of $\mathcal U$ has size at most $L$, and the incidence graph between vertices of $H$ and indices for $\mathcal U$ has cliquewidth at most $c$.
Then a number of colors depending only on $L,c$ suffices to make every member of $\mathcal U$ rainbow.
Such a coloring can be guessed by unary predicates with a first-order guard over the incidence structure.
\end{lemma}
\begin{proof}
Every support index has at most $L$ neighbors, and the original vertices form an independent set in the incidence graph.
Thus, the incidence graph is $L$-degenerate, and together with the cliquewidth bound $c$ this gives a treewidth bound $w=w(c,L)$ by the theorem of Gurski and Wanke~\cite{gw}.

Form the \emph{conflict graph} on the vertices of $H$, joining two distinct vertices when some support contains both.
Take a width-$w$ tree decomposition of the incidence graph, and in every bag replace each support index by all its neighbors.
Original graph vertices in the bag are kept.
Since $L\geq1$, the resulting bags have size at most $L(w+1)$.

\input{figures/support-clique}

These bags form a tree decomposition of the conflict graph.
Every conflict edge is covered by a bag that contained its support index.
For a fixed original vertex, the added bag subtree of each incident support index meets the original bag subtree at a bag containing their incidence edge, so all its occurrences remain connected.
The conflict graph consequently has treewidth at most $L(w+1)-1$ and admits a proper coloring with $L(w+1)$ colors.
Such a coloring is injective on each support.
A transduction can guess these finitely many color classes and check, by a first-order formula, that no two distinct members of one support have the same color.
\end{proof}

In \cref{lem:recursiveencoding}, the incidence graph is obtained by a fixed MSO transduction from a class of bounded cliquewidth, so the lemma gives one palette for all inputs.

\subsection{At most one uncaptured tangle}

We first handle the situation in which the supports have accounted for all tangles except possibly one.
If there is no exception, the rainbow coloring reduces the problem to graphs of smaller rankwidth.
If one tangle remains, the same coloring gives decomposers for its weak pieces, and the one-tangle construction can combine them.
This will be the stopping case of the inner induction, where we use \cref{thm:finite-cover-encoding,thm:weakpieceencoding} for the two cases.

\begin{theorem}
\label{thm:terminalsupport}
Fix $k\geq2$ and $m$.
Assume that graphs of rankwidth less than $k$ have bounded definable cliquewidth.
Consider the graphs $H$ of rankwidth at most $k$ that admit an~$m$-coloring in which every order-$k$ tangle, with at most one exception, has a rainbow big set.
This class has bounded definable cliquewidth, with MSO decomposers preserving the origins of vertices and any fixed unary labels.
\end{theorem}

\begin{figure}[H]
\centering
\color{black}
\begin{tikzpicture}[manuscript illustration]
  \node[box] (a) at (0.78,0.42) {rainbow big sets};
  \node[box] (n) at (5.02,1.03) {every tangle has one};
  \node[key] (p) at (5.02,-0.28) {one exceptional tangle $P$};
  \node[box] (c) at (9.65,1.03) {use the color cover};
  \node[key] (w) at (9.65,-0.28) {encode the $P$-weak pieces};
  \draw[arr] (a) -- (n);
  \draw[arr] (n) -- (c);
  \draw[arr] (a) -- (p);
  \draw[arr] (p) -- (w);
\end{tikzpicture}
\caption{The two stopping cases.
When every tangle has a rainbow big low-order set, the colors suffice on the whole graph.
With one exception, we first use those colors on its weak pieces, and the one-tangle construction then assembles their decompositions.}
\label{fig:terminal-cases}
\end{figure}
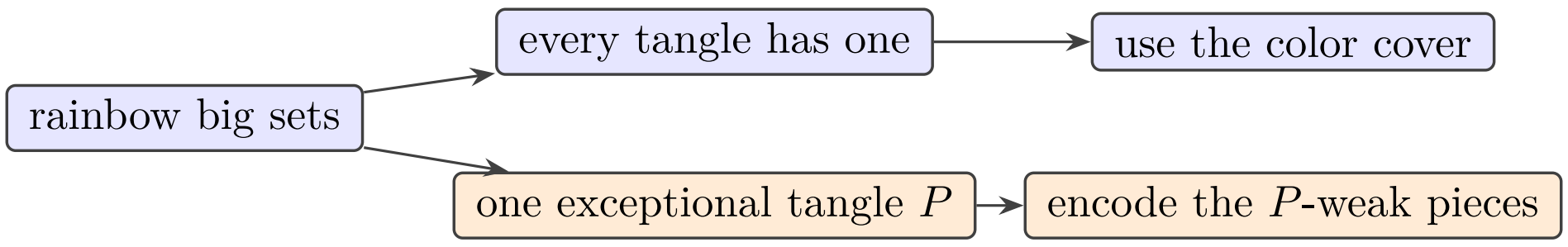

In particular, the theorem applies when all but at most one tangle are captured by a family of bounded supports, provided the supports admit a common coloring with $m$ colors that is injective on each support.
The family itself need not be given to the decomposer; it proves the existence of the coloring used in the theorem.

Suppose that~$P$ is the exceptional tangle.
The coloring argument still works inside each~$P$-weak set, because a tangle of an induced subgraph there would lift to a tangle different from~$P$.

\begin{lemma}
\label{lem:weakrainbowblocks}
Let $k\geq2$, fix a vertex coloring $c$ of $H$, let $P$ be an order-$k$ tangle of $H$, and suppose that every other order-$k$ tangle has a rainbow big low-order set.
If $W$ is $P$-weak, then for every set $I$ of one or two colors that occur in $W$,
\[
 \operatorname{rw}\bigl(H[W\cap c^{-1}(I)]\bigr)<k.
\]
\end{lemma}
\begin{proof}
Suppose an order-$k$ tangle $Q$ exists in the indicated induced graph, and lift it to a tangle $\widehat Q$ of $H$ using \cref{lem:inducedtanglelift}.
A $P$-small set containing $W$ is big in $\widehat Q$, so $\widehat Q\ne P$.
Choose a rainbow big set $S$ of $\widehat Q$ with $\rho_H(S)<k$.
By the definition of the induced-tangle lift, $S\cap W\cap c^{-1}(I)$ is big in $Q$.
It has at most two vertices, and its cut-rank in the induced graph is at most $\rho_H(S)<k$, contradicting the small-set observation used in \cref{lem:rainbowregions}.
Tangle duality proves the width bound.
\end{proof}

To apply \cref{thm:weakpieceencoding} to $P$, we need a decomposer that works on the weak block itself.
It must not need access to the ambient tangle or to the support family.
We obtain this stronger form by using those objects only to prove that a suitable finite coloring exists, while the transduction guesses the coloring on its own input and checks the resulting local encodings.

\begin{lemma}
\label{lem:rainbowblockdcw}
Fix $k\geq2$ and $m$, and assume that graphs of rankwidth less than $k$ have bounded definable cliquewidth.
Let $H$ have rankwidth at most $k$, let $c\colon V(H)\to[m]$, and let~$P$ satisfy the assumptions of \cref{lem:weakrainbowblocks} for this coloring.
Then every nonempty $P$-weak set~$W$ induces a graph of definable cliquewidth bounded in terms of $k,m$.
\end{lemma}
\begin{proof}
By \cref{lem:weakrainbowblocks}, each nonempty subgraph of~$H[W]$ induced by one or two colors has rankwidth less than~$k$.
By the induction hypothesis, one decomposer works on all these subgraphs.
Its expression trees give source-preserving encodings, so \cref{thm:finite-cover-encoding} gives a sparse encoding of~$H[W]$.
The graphs $H[W]$ themselves have rankwidth at most $k$, since they are induced subgraphs of $H$.
The encoded graphs consequently have bounded cliquewidth by closure under MSO transductions, and bounded treewidth because they are uniformly sparse, by Gurski and Wanke~\cite{gw}.
By \cref{prop:boundedtwencoder}, which uses the decomposition theorem of Boja\'nczyk and Pilipczuk~\cite{treewidthlogic}, we can encode each sparse graph by a labeled forest.
We join its components under a new marked root and compose the forest interpretation with the inverse of the sparse encoding.
The resulting interpretation recovers~$H[W]$ on its designated leaves, so \cref{lem:treecliquecompiler} gives the required origin-preserving decomposer.

The decomposer takes only~$H[W]$ as input.
It guesses the colors there and checks the local encodings, and the restriction of~$c$ to~$W$ gives one choice that succeeds.
\end{proof}

\begin{proof}[Proof of \cref{thm:terminalsupport}]
If every order-$k$ tangle has a rainbow big low-order set, then every one- or two-color induced subgraph of $H$ has rankwidth less than $k$.
Indeed, an order-$k$ tangle in such a subgraph would lift to a tangle of $H$ with a rainbow big set.
The intersection of this set with the subgraph would be big in the tangle of the subgraph, although it has at most two vertices and has order less than $k$.
We can therefore apply the same finite-cover argument as in the proof of \cref{lem:rainbowblockdcw}, this time to all of $H$.
This also covers the case that~$H$ has no order-$k$ tangle.

Otherwise, let~$P$ be the one exceptional tangle.
By \cref{lem:rainbowblockdcw}, every nonempty $P$-weak induced graph has definable cliquewidth at most a fixed bound $m'$, depending only on $k,m$.
The weak-piece encoding criterion, \cref{thm:weakpieceencoding}, applies directly, whether or not $P$ is robust.
It gives a source-preserving labeled-tree encoding of $H$, and \cref{lem:treecliquecompiler} gives the required decomposer.

In the second case, we may choose the $m$ colors and the bounded parameters for $P$, and check in MSO that every other order-$k$ tangle has a rainbow big set.
In both cases, we can check the proposed decomposition against the input, and the argument above gives at least one successful choice.
\end{proof}

This completes the construction when at most one tangle remains uncaptured.
We turn to the case in which at least two uncaptured tangles must still be separated.

\input{figures/terminal-support-result}

\section{Finite graphs at low-order boundaries}
\label{sec:boundaries}
When at least two order-$k$ tangles remain uncaptured, we choose one as an anchor and separate each of the others from it by a cut of minimum order.
The sides we choose form an inclusion tree.
At each internal node we will keep the original vertices at its leaf children and replace the regions in the other branches by graphs of bounded size.

The replacements must preserve the rankwidth bound, the tangles that make all the replaced regions small, and the minimum order distinguishing any two of these tangles.
We first show how to do this for a titanic piece by replacing it with a clique of ports.
We then partition each region into titanic pieces and show how to define all the replacements along the tree by one MSO transduction.
\subsection{The closest sets towards one tangle}

Fix an order-$k$ tangle~$R$, which we call the \emph{anchor tangle}.
For another order-$k$ tangle $P$, a \emph{$P$-$R$ set} is a set that is big in $P$ and small in~$R$.
We first minimize the order $\lambda(X)$ among all $P$-$R$ sets $X$ (in the graph case this is the cut-rank).
Among the sets of this minimum order, choose an inclusion-minimal one~$X_P$.
We call~$X_P$ the \emph{closest set} of~$P$ with respect to the anchor~$R$.
The next lemma shows that intersections preserve this minimum order and still distinguish the two tangles.
Consequently, $X_P$ is uniquely determined and is contained in every other minimum-order $P$-$R$ set.
The role of the anchor and the choice of~$X_P$ are illustrated in \cref{fig:anchor-tangle}.
\input{figures/anchor-tangle}

Grohe and Schweitzer call $X_P$ the \emph{leftmost minimum separation} between $P$ and $R$ and prove that it is contained in every minimum $P$-$R$ set~\cite[Lemma~2.8]{computingtangles}.
We use their uncrossing argument first for this assertion and then to show that the sets obtained by keeping $R$ fixed and varying $P$ form a laminar family.

\begin{lemma}\label{lem:closestunique}
Fix a tangle $R$ of order $k$.
For every tangle $P\neq R$ of order $k$, the set $X_P$ is uniquely determined and is contained in every minimum-order $P$-$R$ set.
The sets~$X_P$ for all order-$k$ tangles~$P\ne R$ form a laminar family.
\end{lemma}
\begin{proof}
Let $a$ be the minimum distinguishing order, and let $X,Y$ be two $P$-$R$ sets of order~$a$.
If $X\cap Y$ has order at most $a$, the tangle axioms make it big in $P$ and small in $R$.
If instead its order were greater than $a$, submodularity would give $\lambda(X\cup Y)<a$.
The union would also be big in $P$ and small in $R$, contradicting minimum order.
Thus, the intersection is another minimum set.
Repeated intersections prove uniqueness and the containment assertion.

For laminarity, take $X=X_P$ and $Y=X_Q$, put $a=\lambda(X)$ and $b=\lambda(Y)$, and assume~$a\leq b$.
Suppose that $X\cap Y$, $X\setminus Y$, and $Y\setminus~X$ are all nonempty.
If $Q$ made $X$ big, then $X$ would distinguish $Q$ from $R$, so $a=b$ and the first assertion would give $Y\subseteq X$, a contradiction.
It follows that $Q$ makes $X$ small.

The order of $Y\setminus X$ must be greater than $b$.
Otherwise, the big sets $Y$ and $V\setminus X$ force their intersection~$Y\setminus X$ to be big in $Q$, while it is small in $R$ as a low-order subset of $Y$.
It would be a competing set properly contained in $Y$.
Posimodularity now gives $\lambda(X\setminus Y)<a$.

Moreover $\lambda(X\cup Y)\geq b$, since a smaller-order union would distinguish $Q$ from $R$.
Submodularity therefore gives $\lambda(X\cap Y)\leq a$.
Both $X\setminus Y$ and $X\cap Y$ are small in $R$.
The first cannot be big in $P$, since its order is smaller than $a$.
Together with the fact that $X$ is big in $P$, the tangle axioms then force $X\cap Y$ to be big in $P$.
This is a competing minimum set properly contained in $X$, giving the required contradiction.
\end{proof}

\input{figures/minimum-intersection}

All parts of this definition are MSO properties, once the anchor is represented by its finite parameters.
The minimum order belongs to $\{0,\ldots,k-1\}$, and containment in every competing minimum set can be tested with a set quantifier.
We may restrict the other tangles by any fixed MSO condition on their finite descriptions.
We now want to construct a tree whose nodes represent the sets in this laminar family.
After adding $V(G)$ and all singleton sets, its \emph{inclusion tree} has one node for each set, with $V(G)$ as its root and an edge from each other set to its smallest strict superset in the family.
The leaves represent the singleton sets and can therefore be identified with the original graph vertices.

Campbell and K\"ohler proved that such an inclusion tree can be constructed by an MSO transduction from a laminar family given by a predicate on sets~\cite[Theorem~2]{laminar}.
In particular, if membership in the family is defined by a fixed MSO formula on a graph, we can substitute this formula for their set predicate and obtain an MSO transduction on the graph itself.
Their construction represents the tree nodes by a fixed number of copies of the original vertices.
Here we apply the theorem to the family of closest sets, together with~$V(G)$ and all singleton sets.
This family is laminar by \cref{lem:closestunique}, and its membership predicate is MSO-definable by the preceding argument.

The leaves below the node for~$X_P$ are exactly the vertices of~$X_P$.
Several tangles may have the same closest set, and the tree puts them at the same node.
We will show in \cref{lem:ownertangleclasses} that any cut distinguishing two such tangles has order strictly greater than the order of their common closest set.

To obtain a graph at an internal node~$u$, let~$A$ be the set of leaves below~$u$, which we call its \emph{territory}.
The vertices at the leaf children of~$u$ are the vertices \emph{owned} by~$u$.
For each internal child, the set of leaves below that child is one \emph{direction} at~$u$.
If~$u$ has a parent, there is one further direction, consisting of the vertices outside~$A$.
The local graph at~$u$ keeps the vertices owned by~$u$ and replaces each direction by a bounded graph.
We now show what these replacements must preserve for the tangles that regard all the directions at~$u$ as small.

\subsection{Faithful transfer through a low-order titanic contraction}

We describe contraction in terms of the sides a tangle regards as small.
Let $(V,\lambda)$ be a finite connectivity system.
If $A\subseteq V$ is nonempty, let $V/A$ be the atom set obtained by replacing~$A$ by one atom $a$, and let
\[
 \lambda_A(Z)=\lambda(\widehat Z)
\]
for the union $\widehat Z\subseteq V$ of the atoms in $Z$.
Thus, the cuts of $V/A$ are exactly the cuts of $V$ that do not split $A$, which we call the \emph{$A$-saturated cuts}.

A cut of the original ground set may split the set that we want to contract, and such a cut is no longer available after contraction.
The next lemma shows that we can move the whole titanic set to one side of the cut without increasing the order of the cut.
We call this operation \emph{rounding the cut}.
This is the submodularity argument in the proof of the tangle lifting lemma of Geelen, Gerards, and Whittle \mbox{\cite[Lemma~4.4]{ggwtangles}}, also used by Korhonen and Oum for branchwidth \mbox{\cite[Proposition~2.4]{titanic}}.

\begin{lemma}\label{lem:titanicround}
Let $A$ be titanic.
For every $X\subseteq V$, at least one of $X\setminus A$ and $X\cup A$ has order at most $\lambda(X)$.
\end{lemma}
\begin{proof}
Apply titanicity to the nonempty members of $\{A\cap X,A\setminus X\}$.
At least one of these two sets has order at least $\lambda(A)$.
If $\lambda(A\cap X)\geq\lambda(A)$, submodularity gives
\[
 \lambda(X\cup A)\leq\lambda(X)+\lambda(A)-\lambda(X\cap A)
 \leq\lambda(X).
\]
If $\lambda(A\setminus X)\geq\lambda(A)$, posimodularity instead gives
\[
 \lambda(X\setminus A)\leq\lambda(X)+\lambda(A)-\lambda(A\setminus X)
 \leq\lambda(X).
\]
\end{proof}

\input{figures/round-cut}

The order of a cut is only one part of the information that we need to preserve.
We also need to know which side is small in a tangle.
The following theorem says that contraction preserves exactly those tangles that regard the contracted set as small.
Geelen, Gerards, and Whittle proved the restriction to these cuts in \cite[Lemma~4.3]{ggwtangles} and the extension by rounding in \cite[Lemma~4.4]{ggwtangles}.
The latter generalizes a result of Robertson and Seymour for hypergraphs~\cite[(8.3)]{tanglesorigin}.
We give their argument in our notation to make explicit why the extension is unique and how to read the orientation of a cut from the contracted tangle.

\begin{theorem}
\label{thm:titanictanglebijection}
Let $A$ be titanic and let $\lambda(A)<k$.
Restriction to $A$-saturated cuts gives a bijection between
\begin{itemize}
\item order-$k$ tangles of $(V,\lambda)$ that orient $A$ as small, and
\item order-$k$ tangles of $(V/A,\lambda_A)$.
\end{itemize}
For each original set $X\subseteq V$ of order less than $k$, the extension makes $X$ small if and only if one of $X\setminus A$ and $X\cup A$ has order less than $k$ and its image in $V/A$ is small in the contracted tangle.
\end{theorem}
\begin{proof}
An original tangle that makes $A$ small restricts to a tangle on $V/A$.
The orientation axiom and the axiom that no three small sets cover the ground set restrict immediately.
The singleton axiom for the new atom says that its complement is not small, and this follows because $A$ is small and has order less than $k$.
The other singleton axioms are unchanged.

Conversely, let $\mathcal Q$ be an order-$k$ tangle on $V/A$, viewed as orientations of the $A$-saturated cuts of $V$.
The singleton $\{a\}$ has order $\lambda(A)<k$, so it is small in $\mathcal Q$ by the singleton axiom.

For a low-order set $X\subseteq V$, call either of
\[
 X^-:=X\setminus A,\qquad X^+:=X\cup A
\]
a \emph{valid rounding} if its order is less than $k$.
At least one is valid by \cref{lem:titanicround}.
When both are valid, they receive the same orientation from $\mathcal Q$.
One direction follows from monotonicity of low-order tangle orientations.
For the other direction, if $X^-$ were small and $X^+$ big, the three big sets
\[
 V\setminus X^-,\qquad X^+,\qquad V\setminus A
\]
would have empty intersection, contrary to the tangle axiom.

Define $X$ to be small for $\mathcal P$ if a valid rounding is small for $\mathcal Q$.
This is well-defined by the preceding paragraph.
Complementary cuts receive complementary orientations, since complements interchange the two possible roundings.

\begin{figure}[H]
\centering
\color{black}
\begin{tikzpicture}[manuscript illustration]
  \node[box] (p) at (0,0.45) {$\mathcal P$ on $V$\\$A$ small};
  \node[key] (q) at (4.4,0.45) {$\mathcal Q$ on $V/A$\\atom $a$ small};
  \node[box] (xx) at (8.85,0.45) {orientation of $X$\\from a valid rounding};
  \draw[arr] (p) -- (q);
  \draw[arr] (q) -- (xx);
  \node[note] (n5) at (2.12,1.26) {restrict to saturated cuts};
  \node[note] (n6) at (6.67,-0.43) {extend from the atoms};
\end{tikzpicture}
\caption{When both roundings of $X$ have order below $k$, their images in $V/A$ differ only by the atom $a$.
Since $a$ is small, the contracted tangle gives both images the same orientation.
This makes extension unique.
For disjoint titanic sets these maps compose one atom at a time, as in \cref{cor:disjointtangletransfer}.}
\label{fig:tangle-restriction-extension}
\end{figure}
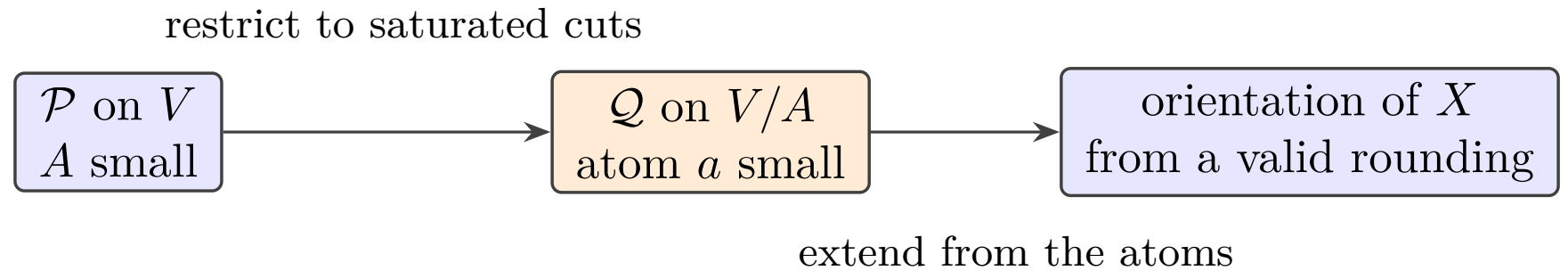

Suppose that three $\mathcal P$-small sets $X_1,X_2,X_3$ cover $V$.
Their intersections with $A$ cover~$A$.
By the cover form of titanicity from \cref{sec:tangles}, some $i$ satisfies
\[
 \lambda(A\cap X_i)\geq\lambda(A).
\]
Submodularity then makes the upward rounding $X_i\cup A$ valid, with order at most $\lambda(X_i)<k$.
Choose this rounding for $X_i$ and any valid rounding for the other two sets.
All three rounded sets are $\mathcal Q$-small.
They still cover $V\setminus A$, since rounding changes no point outside $A$, and the chosen upward rounding covers all of $A$.
Thus, they cover $V$, a contradiction.

For the singleton axiom, consider $V\setminus\{v\}$ when its order is less than $k$.
If $v\notin A$, it is already an $A$-saturated complement of an original singleton, so it is big for $\mathcal Q$.
If $v\in A$, its upward rounding is $V$, which is big.
Hence, it is never $\mathcal P$-small.
This proves that $\mathcal P$ is a tangle, and its restriction is $\mathcal Q$.

Finally, any original tangle making $A$ small must orient a low-order set $X$ in the same way as any valid rounding.
This follows from the same nested-set and three-big-set argument, because the difference between the two sets is contained in the small set $A$.
Therefore, the extension is unique.
\end{proof}

At a node of our decomposition tree, several directions will be replaced at the same time.
Their sets are disjoint, which is useful because contracting one of them leaves the connectivity values inside the others unchanged.
We may therefore apply \cref{thm:titanictanglebijection} successively, with a unique tangle correspondence at every step.

\begin{corollary}
\label{cor:disjointtangletransfer}
Let $A_1,\ldots,A_s$ be pairwise disjoint titanic sets with $\lambda(A_i)<k$.
The order-$k$ tangles of the connectivity system obtained by contracting all $A_i$ are in bijection with the original order-$k$ tangles that orient every $A_i$ as small.
\end{corollary}
\begin{proof}
Contract the sets one at a time.
Each remaining $A_i$ is still titanic, since the contraction leaves the order of every subset of $A_i$ unchanged.
Composing the bijections from \cref{thm:titanictanglebijection} gives the result.
\end{proof}
\subsection{A titanic finite port gadget}

We now show how to replace a titanic boundary piece by a finite graph.
There is a related construction for matroids, where Hlin\v{e}n\'y and Oum first reduce each part by deletions and contractions that preserve the connectivity of cuts outside it.
They then attach a uniform matroid to each remaining nonsingleton part, which makes the part titanic and again preserves these cut values~\cite[Corollary~4.2 and Lemma~4.3]{hlinenyoum}.
Kalampakas uses three ports at a rank-two cut, one for each nonzero boundary vector, and proves that the replacement preserves the ranks of all cuts that keep each group of ports together~\cite[Construction~4.1 and Lemma~4.2]{ranktwo}.
If we make these three ports a triangle, they illustrate why titanicity is tested on partitions into three parts.
In every partition of the triangle into two nonempty parts, two ports lie together and their rows are independent on the exterior, but the partition into three singletons has cut-rank one on every part.
For the tangle correspondence, we need the replacement itself to be titanic.
We therefore include the zero vector as well and call the clique on all these ports the \emph{affine clique gadget}.
The clique gives one extra constant coordinate at every proper internal cut, and this is what makes the counting argument in the next proof work.

\begin{lemma}
\label{lem:affineporttitanic}
Let $A$ be a rank-$r$ boundary block.
Replace it by ports
\[
 T=\{t^v:v\in\mathbb F_2^r\},
\]
where the external row of $t^v$ is the linear combination of a fixed basis of the original boundary rows with coefficient vector $v$, and make $T$ a clique.
Assume that the external columns still span a space of dimension $r$, as they do when we replace disjoint blocks simultaneously using their cut bases.
Then $\rho(T)=r$, and $T$ is titanic for the cut-rank function of the resulting graph.
\end{lemma}
\begin{proof}
All basis rows occur among the ports, so $\rho(T)=r$.
If $r=0$, titanicity is immediate from nonnegativity.
Consider a partition of $T$ into at most three nonempty parts.
If one part is all of $T$, its rank is $r$.
Otherwise every part is proper.

\input{figures/affine-clique-rows}

For a nonempty proper part $P\subset T$, the external columns of the whole block realize the vector rows $v$ faithfully.
A port in $T\setminus P$ gives one further cut column that is one at every row of $P$, because $T$ is a clique.
Therefore,
\[
 \rho(P)\ \geq\
 \operatorname{rank}_{\mathbb F_2}\{(v,1):t^v\in P\}.
\]
If $r=1$, this rank is at least one and we are done.
If $r\geq2$ and $\rho(P)<r$, the displayed vectors lie in a linear subspace of dimension at most $r-1$.
Its intersection with the affine hyperplane whose last coordinate is one has at most $2^{r-2}$ elements.
Hence, $|P|\leq2^{r-2}$.
Three parts all of rank less than $r$ could then cover at most $3\cdot2^{r-2}<2^r$ ports, a contradiction.
\end{proof}

A cut that keeps each port group together has the same rank as before.
The zero port contributes only a zero row or column across such a cut, and the edges inside a clique do not cross it.
Thus, an original titanic boundary block and its affine clique port replacement contract to the same atom connectivity function.
For any tangle order greater than the rank of the block, \cref{thm:titanictanglebijection} therefore identifies the tangles on the two sides that make the corresponding block small.

The replacement graph must not create an additional tangle hidden inside one port group.
To exclude this possibility, we use the affine structure of its labels once more.
Repeatedly dividing an affine space into two halves gives a tree of low-order cuts, and the tangle axioms then force the whole group to be small.

\begin{lemma}
\label{lem:affineportnoghosts}
In the construction of \cref{lem:affineporttitanic}, every tangle of order $q>r$ selects the whole port set $T$ as small.
\end{lemma}
\begin{proof}
If $r=0$, the port set is a single isolated vertex and the assertion follows from the singleton axiom.
Suppose $r\geq1$.
For a proper affine subspace $U\subsetneq\mathbb F_2^r$ of dimension $d$, put~$T_U=\{t^v:v\in U\}$.
Its cut rows are obtained from the vectors $(v,1)$ by a fixed linear map: external columns are linear functions of $v$, and the columns of $T\setminus T_U$ are constant one.
Consequently,
\[
 \rho(T_U)\leq d+1\leq r<q.
\]
Recursively bisect the affine space into two affine hyperplanes, continuing down to singletons.
This gives a binary partition tree of $T$ in which every node-set has cut-rank at most $r$, including the root since $\rho(T)=r$.
Since $q>r\geq1$, every singleton has order less than~$q$ and is small by the singleton axiom.
Whenever two small sets have union of order less than~$q$, their union is small as well: otherwise those two sides and the small complement of their union would cover the ground set.
Induction up the binary tree therefore makes $T$ small.
\end{proof}

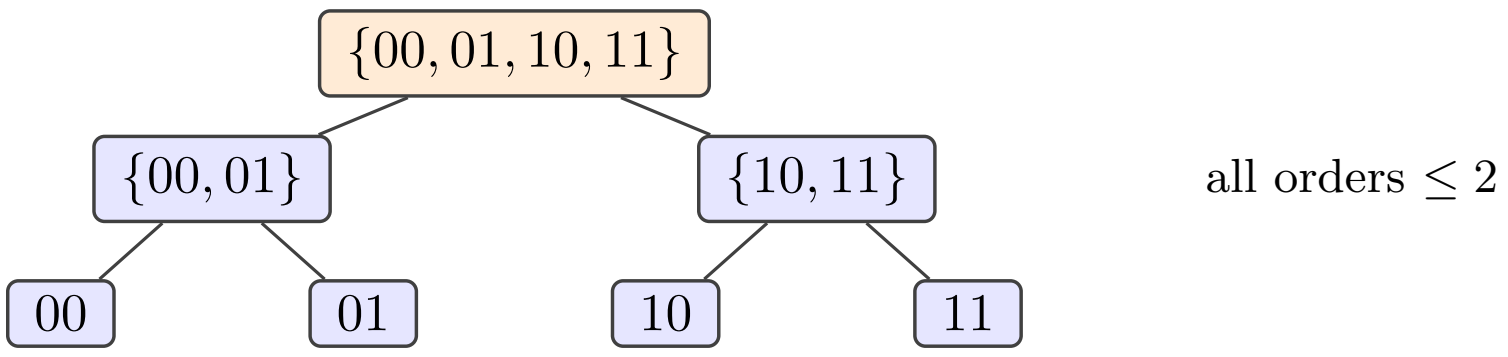
\begin{figure}[H]
\centering
\color{black}
\begin{tikzpicture}[manuscript illustration]
  \node[key] (t) at (4,0.98) {$\{00,01,10,11\}$};
  \node[box] (l) at (2,0.15) {$\{00,01\}$};
  \node[box] (r) at (6,0.15) {$\{10,11\}$};
  \draw[edge] (l) -- (t);
  \draw[edge] (t) -- (r);
  \node[box] (a0) at (1,-0.74) {$00$};
  \draw[edge] (l) -- (a0);
  \node[box] (a1) at (3,-0.74) {$01$};
  \draw[edge] (l) -- (a1);
  \node[box] (a2) at (5,-0.74) {$10$};
  \draw[edge] (r) -- (a2);
  \node[box] (a3) at (7,-0.74) {$11$};
  \draw[edge] (r) -- (a3);
  \node[note] (n13) at (9.54,0.15) {all orders $\leq 2$};
\end{tikzpicture}
\caption{Bisect a rank-two port set by the first coordinate and then by the second.
In a tangle of order greater than two, each singleton is small, hence so is each pair and then the whole port set.
Attaching this same tree in place of an atom leaf gives the internal cuts used in \cref{cor:affinewidth}.}
\label{fig:affine-bisection}
\end{figure}

We can now compare the original graph and its finite replacement through the same contracted connectivity system.
We apply \cref{cor:disjointtangletransfer} to both graphs and use \cref{lem:affineportnoghosts} on the port side.

\begin{corollary}\label{cor:affinetangletransfer}
Suppose disjoint original boundary blocks are titanic and have rank less than $q$.
Replace them simultaneously by affine clique port sets.
The order-$q$ tangles of the resulting port graph are in bijection with the order-$q$ tangles of the original graph that select every replaced boundary block as small.
\end{corollary}
\begin{proof}
Contract the original blocks and the port blocks, respectively.
The simultaneous block identity gives the same atom connectivity function.
Titanic tangle transfer identifies the tangles on both sides with the tangles of this atom system.
On the port side, every tangle makes every port block small by \cref{lem:affineportnoghosts}.
\end{proof}

\begin{figure}[H]
\centering
\color{black}
\begin{tikzpicture}[manuscript illustration]
  \node[box] (g) at (0.65,0.9) {original blocks};
  \node[box] (h) at (8.0,0.9) {affine port cliques};
  \node[key] (a) at (4.32,-0.12) {common atom system};
  \draw[arr] (g) -- (a);
  \draw[arr] (h) -- (a);
  \node[note] (n5) at (4.3,1.13) {$\lambda_{\rm atoms}(Z)=\lambda(\widehat Z)$};
\end{tikzpicture}
\caption{Each cut of the atom system keeps every block whole.
Contracting on either side gives it the same order.
On the left, we restrict to tangles making every replaced block small.
On the right, every tangle of the specified order already makes each port clique small.}
\label{fig:common-atoms}
\end{figure}
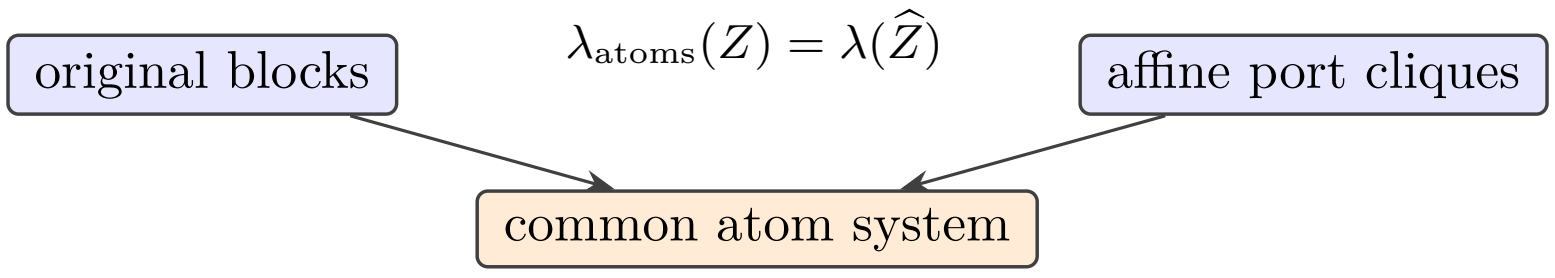

The same affine partition tree also explains why the replacement does not increase the width.
We begin with a decomposition in which each replaced block is a single atom, and then expand that leaf by the partition tree of its ports.
This is the usual way of joining branch decompositions at an atom~\cite[Section~2]{titanic}, with the affine partition tree giving the bound on the cuts inside the gadget.

\begin{corollary}\label{cor:affinewidth}
If the common atom connectivity function has branchwidth at most $k$ and every replaced block has rank at most $k$, the affine clique port graph has rankwidth at most $k$.
\end{corollary}
\begin{proof}
Take a width-$k$ branch decomposition of the atom function.
The tree in \cref{fig:affine-bisection} shows how to expand a leaf of rank two.
Replace each port-atom leaf by the affine binary partition tree from the preceding proof.
Old cuts keep whole port sets together and therefore have their original atom rank.
A new cut inside a rank-$r$ port set has rank at most $r\leq k$.
Thus, the expanded tree has width at most $k$.
\end{proof}
\subsection{Defining finite boundary gadgets without choosing nested partitions}

A low-order set need not itself be titanic, so we cannot always replace it by a single port group.
We first divide it into titanic pieces.
By the definition of titanicity, each piece that fails this condition splits into at most three nonempty parts of smaller order.

\begin{lemma}
\label{lem:titanicpartition}
Every nonempty set $A$ of order $a$ has a partition into titanic sets $F_1,\ldots,F_m$ of orders $b_i\leq a$ such that
\[
 \sum_{i=1}^m2^{b_i}\leq3^a.
\]
\end{lemma}
\begin{proof}
If $A$ is titanic, use the one-part partition; $2^a\leq3^a$.
Otherwise there is a partition into two or three nonempty parts, each of order less than $a$.
Apply induction to its nonempty parts and sum the bounds, obtaining at most $3\cdot3^{a-1}=3^a$.
Every nonempty set of order zero is titanic, so the induction terminates.
\end{proof}

Fix a nonempty set~$A$ of rank~$a$ and choose vertices~$R_A=(r_1,\ldots,r_a)$ in~$A$ whose rows across the cut form a basis.
We call this the \emph{inward basis} of~$A$.
A \emph{gadget type} consists of the following finite data.
\begin{enumerate}
\item the number $m$ of titanic pieces and their ranks $b_i\leq a$, with $\sum_i2^{b_i}\leq3^a$;
\item a binary matrix $M_i$ of size $b_i\times a$ for each piece;
\item a binary matrix $B_{ij}$ of size $b_i\times b_j$ for each $i<j$.
\end{enumerate}
A realization consists of a partition of $A$ into titanic sets $F_1,\ldots,F_m$ of the prescribed ranks and an ordered cut basis $R_i=(r_{i,1},\ldots,r_{i,b_i})$ in each $F_i$, with the prescribed matrices.
The matrix $M_i$ expresses those rows on the exterior of $A$ in the fixed basis $R_A$, while $B_{ij}[p,q]=E_G(r_{i,p},r_{j,q})$.

For one side, MSO can quantify a partition and all the required basis vertices.
However, this does not by itself give a simultaneous choice for every set in a deeply nested family with a fixed list of unary parameters.
We avoid this problem by remembering only the finite matrices that a choice produces.
For fixed $a$, the list of types is finite, so we can test them one by one using an MSO formula and choose the first type that has witnesses.

\begin{lemma}
\label{lem:finitegadgettype}
For each fixed type $\tau$, the property that $\tau$ is realized at $(A,R_A)$ is MSO-definable.
Some type is always realized.
Hence, the first realizable type in a fixed finite enumeration is definable without keeping a partition or a basis for any piece.
\end{lemma}
\begin{proof}
Existentially quantify the fixed number of sets $F_i$ and their fixed-size basis tuples.
Require the sets to partition $A$ into nonempty pieces of the prescribed cut-ranks.
For each $i$, require the $b_i$ chosen rows to be independent on the complement of $F_i$, so that they form a cut basis of $F_i$.
To check titanicity, require every partition of $F_i$ into at most three nonempty parts to have a part of rank at least $b_i$.
All rank predicates have a fixed bound and are MSO-definable by finite row-basis tests.

For the prescribed profile matrix, require for every $z\notin A$ that
\[
 E_G(r_{i,p},z)=\sum_{j=1}^a M_i[p,j]E_G(r_j,z).
\]
Require the finitely many basis adjacencies given by $B_{ij}$.
These are fixed Boolean formulas.
\cref{lem:titanicpartition} and the existence of cut bases provide a realization of some type.
Taking the first type is a finite Boolean combination of the realizability formulas.
No particular witnesses remain in the output.
\end{proof}

The selected type defines at most $3^a$ \emph{virtual ports}, indexed by pairs $(i,v)$ with $v\in\mathbb F_2^{b_i}$.
Make the ports of one piece a clique, and put between different pieces the adjacency $v^{\mathsf T}B_{ij}w$.
The external profile of port $(i,v)$ in the fixed $A$-basis is the row vector
\[
 c(i,v)=v^{\mathsf T}M_i\in\mathbb F_2^a.
\]
Thus, its adjacency to an exterior original vertex $x$ is $\sum_jc(i,v)_jE_G(r_j,x)$.
Between ports with profiles $c,d$ from two disjoint sets $A,A'$, adjacency is
\[
 c\,A_G[R_A,R_{A'}]\,d^{\mathsf T}.
\]
Thus, the selected types and the bases of the cuts around~$A$ and~$A'$ determine the edges between their ports.

The ports have now been defined from finite data, but we must still show that the resulting graph has the promised structural properties.
For this proof alone, we may choose witnesses for the selected types.
When the sets are disjoint, these witnesses can be chosen separately, and the displayed bilinear formula makes their replacements agree on every edge between different sets.

\begin{proposition}
\label{prop:virtualtitanicports}
Let $A_1,\ldots,A_s$ be pairwise disjoint nonempty sets in a graph of rankwidth at most $k$, each of cut-rank less than $k$.
The resulting virtual port graph has rankwidth at most $k$.
Fix any realization of the chosen type for each $A_j$.
The order-$k$ tangles of the port graph are in bijection with the original order-$k$ tangles that make every piece of these realizations small.
In particular, every original tangle that makes each $A_j$ small has a corresponding tangle in the port graph.
Every other order-$k$ tangle of the port graph makes the whole set of ports replacing some $A_j$ big.
\end{proposition}
\begin{proof}
Choose any realization of each selected finite type.
These choices are independent because the sets are disjoint.
Their internal piece matrices agree with the selected types.
Across two sides, the row expansions first in the $A$-basis and then in the $A'$-basis give exactly the displayed bilinear formula.
Consequently, the virtual graph is the simultaneous replacement by affine clique gadgets for these realizations.

Contract all titanic pieces.
Contracting the disjoint pieces preserves branchwidth at most $k$ by the titanic partition lemma of Hlin\v{e}n\'y and Oum~\cite[Lemma~3.4]{hlinenyoum}, applied to the partition completed by singleton parts.
Expanding each atom by its affine clique gadget preserves this width by \cref{cor:affinewidth}.
The tangle correspondence now follows from \cref{cor:affinetangletransfer}.
Orientations agree on unions of whole pieces, so the port set replacing $A_j$ is big if and only if $A_j$ is big in the corresponding original tangle.
Each of these port sets has size at most $3^{k-1}$.
\end{proof}

\begin{figure}[H]
\centering
\color{black}
\begin{tikzpicture}[manuscript illustration]
  \node[box] (p) at (0,0.3) {$(i,v)$};
  \node[key] (c) at (2.65,0.3) {$c=v^{\mathsf T}M_i$};
  \node[box] (b) at (5.61,0.3) {$A_G[R_A,R_{A^{\prime}}]$};
  \node[key] (d) at (8.6,0.3) {$d=w^{\mathsf T}M_j^{\prime}$};
  \node[box] (q) at (11.0,0.3) {$(j,w)$};
  \draw[edge] (p) -- (c);
  \draw[edge] (c) -- (b);
  \draw[edge] (b) -- (d);
  \draw[edge] (d) -- (q);
  \node[note] (n9) at (5.62,-0.4) {edge value $=$ $c\,A_G[R_A,R_{A^{\prime}}]\,d^{\mathsf T}$};
\end{tikzpicture}
\caption{The two ports are in different directions, and each profile is written in the inward basis of its own direction.
Expanding the original cut matrix on both sides gives the displayed entry.
Thus, any disjoint realizations of the selected types yield the same cross edges.}
\label{fig:virtual-port-edge}
\end{figure}
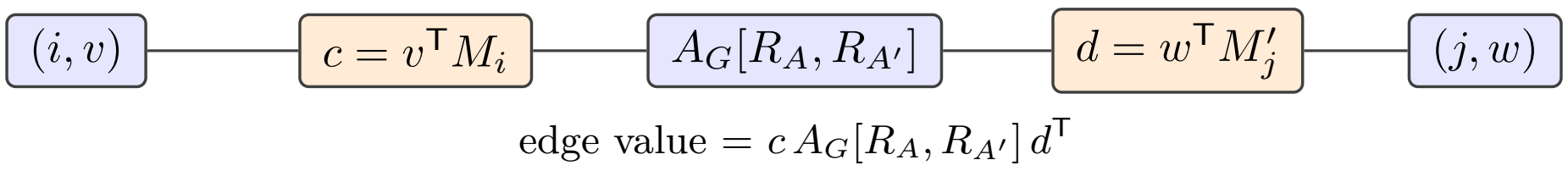

The recursion will measure progress by the minimum order of a cut that distinguishes two tangles.
A bijection between tangles would not by itself preserve this number.
The rounding in \cref{lem:titanicround,thm:titanictanglebijection} lets us pass from either graph to a distinguishing cut of no larger order on the common atoms.

\begin{corollary}
\label{cor:gadgetdistinguishingorder}
Fix realizations as in \cref{prop:virtualtitanicports}.
For two distinct order-$k$ tangles of the port graph, the minimum distinguishing order equals that of the corresponding tangles in the original graph.
\end{corollary}
\begin{proof}
Round a distinguishing cut successively across all replaced titanic pieces.
A rank-nonincreasing upward or downward rounding always exists.
Since each piece is small in both tangles, each valid rounding preserves its orientation in both: small low-order unions remain small, and removing a small piece from a big low-order set remains big.
The final saturated cut transfers to the common atom function with the same rank.
Conversely, perform the same operation on a distinguishing cut in the virtual graph, using its titanic affine port sets, all of which are small in every order-$k$ tangle.
Transfer the saturated cut back.
These two inequalities prove equality of the minimum orders.
\end{proof}

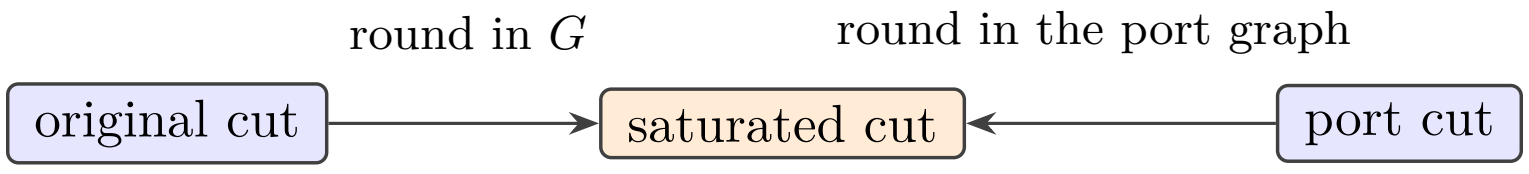
\begin{figure}[H]
\centering
\color{black}
\begin{tikzpicture}[manuscript illustration]
  \node[box] (g) at (0.7,0) {original cut};
  \node[key] (a) at (4.77,0) {saturated cut};
  \node[box] (p) at (8.85,0) {port cut};
  \draw[arr] (g) -- (a);
  \draw[arr] (p) -- (a);
  \node[note] (n5) at (2.69,0.6) {round in $G$};
  \node[note] (n6) at (6.83,0.6) {round in the port graph};
\end{tikzpicture}
\caption{Start with a minimum distinguishing cut on either side.
Rounding preserves the disagreement and never increases the order of the cut.
The middle cut has the same order in both graphs.
Applying the argument from both ends gives equality of the two minimum orders.}
\label{fig:distinguishing-rounding}
\end{figure}

The port graph is determined by its types, and every choice of realizations gives a correspondence with the stated properties.
On an MSO-defined laminar tree, the child node of each edge can provide the copies for the ports at both ends of that edge.
Every nested set selects its own first realizable type by a fixed formula.
No global family of nested witness partitions is selected.

To recover the adjacencies of a vertex in~$A$, we need to express its row on the exterior of~$A$ as a linear combination of port rows.
The virtual port profiles need not contain every standard unit vector, but they span $\mathbb F_2^a$: the original rows of $A$ are spanned by the cut bases of all its pieces.
In each finite gadget type, choose the first ordered basis among its profile rows.
Its inverse matrix is finite data.
For a vertex~$x\in A$, multiplication of its row coordinates in~$R_A$ by this inverse gives its coefficients in the selected port basis.
This gives the exact outside-adjacency decoder and avoids adding extra ports that might change the width or tangle properties.

For the parent direction, we use column profiles, so there is no need to choose coherent bases in the complements of the laminar sets.
We give the formulas for this construction.
Let~$A$ be the territory of a nonroot node and let~$B=V(G)\setminus A$ be the direction towards its parent.
We already have the inward basis~$R_A=(r_1,\ldots,r_a)$ on the other side of this cut.
For $x\in A$, write $\ell_A(x)\in\mathbb F_2^a$ for its row coordinates in $R_A$, and for $y\in B$, put
\[
 p_A(y)=(E_G(r_1,y),\ldots,E_G(r_a,y)).
\]
With both profiles written as rows, we have
\[
 E_G(x,y)=\ell_A(x)p_A(y)^{\mathsf T}.
\]
Since~$a<k$, the displayed product is a sum of bounded length and can be evaluated by an MSO formula.

For the parent direction $B$, a finite type again specifies titanic pieces $F_i$ of ranks $b_i\leq a$, their internal cross-basis matrices $B_{ij}$, and matrices $P_i$ of size $b_i\times a$.
In the realizability formula, quantify a partition of $B$ and a cut basis $R_i=(r_{i,1},\ldots,r_{i,b_i})$ inside each piece, as before, but replace the external row-coordinate condition by
\[
 P_i[p,j]=E_G(r_j,r_{i,p})
 \qquad(1\leq p\leq b_i,\ 1\leq j\leq a).
\]
Thus, row $p$ of $P_i$ is the column profile $p_A(r_{i,p})$.
All other tests, including titanicity and exact cut-basis rank in the full graph, remain unchanged.
Some such type is realized, and taking the first realized type again removes every partition and basis witness from the output.

The parent port $(i,v)$ has profile $h=v^{\mathsf T}P_i\in\mathbb F_2^a$ towards $A$.
Its adjacency to a local interior vertex $x\in A$ is
\[
 \ell_A(x)h^{\mathsf T}.
\]
For a child territory $C\subseteq A$ with inward basis $R_C$, let $N_{C,A}$ be the matrix whose row indexed by $s\in R_C$ is $\ell_A(s)$.
A child port with external row coordinates $c$ in the basis $R_C$ is adjacent to the parent port if and only if
\[
 cN_{C,A}h^{\mathsf T}=1.
\]
Indeed, expanding first in the child cut basis and then in the parent piece basis gives this equality.
Two child directions use the previously displayed bilinear formula.
Edges within each gadget are given by its type, and two interior vertices keep their original adjacency.

\begin{figure}[H]
\centering
\color{black}
\begin{tikzpicture}[manuscript illustration]
\node[box] (c) at (0,0) {child profile $c$};
  \node[box] (ca) at (3.1,0) {$cN_{C,A}$};
  \node[key] (h) at (11.45,0) {parent profile $h$};
  \node[key] (test) at (6.95,0) {edge if and only if $cN_{C,A}h^{\mathsf T}=1$};
  \draw[arr] (c)--node[above,note] {$N_{C,A}$} (ca);
  \draw[arr] (ca)--(test);
  \draw[arr] (h)--(test);
\end{tikzpicture}
\caption{We express the child profile $c$ in the inward basis $R_A$ by multiplying it by $N_{C,A}$.
The parent port gives its own column profile $h$ on $R_A$.
The two ports are adjacent if and only if $(cN_{C,A})h^{\mathsf T}=1$.}
\label{fig:parent-column-profile}
\end{figure}
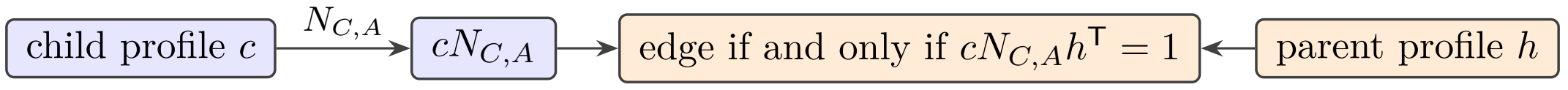

The rows of all $P_i$ together span $\mathbb F_2^a$: the piece bases span the rows of $B$ restricted to $A$, and the column profiles of the vertices of $B$ have rank $a$.
Consequently, a canonical basis among the finitely many parent port profiles also provides any needed reverse coordinate conversion.
The matrices $N_{C,A}$ are computed from the coherent inward bases alone.
Thus, all nested local graphs, including their parent-direction gadgets, are simultaneously defined without selecting any actual basis in any complement $V(G)\setminus A$.
\cref{fig:localports} shows the local replacement and its relation to the laminar tree.
\input{figures/port}

Given the laminar tree and the inward bases at its nodes, we have constructed all the local graphs by one MSO transduction, including the replacements in the parent directions.
In \cref{lem:coherentgeneralbases}, we will see how to choose these bases along the whole tree at once.
Each boundary is represented by a graph of bounded size, and every local graph still has rankwidth at most $k$.
Every tangle that regards all the replaced directions as small survives, and the minimum order distinguishing two such tangles is the same before and after the replacement.
A tangle of a local graph may still regard a whole boundary graph as big, even though it regards each of its port groups as small.
We keep the vertices of each whole boundary graph as a support so that the next stage can recognize these tangles.

\input{figures/local-replacements-result}

\section{Bounded supports and recursive progress}
\label{sec:supports}
The new supports account for any local tangle that makes a whole replacement big.
We must also account for tangles captured by a support in the graph before the replacement.
We first determine at which nodes such a support is needed and what its members become there.
We then show why the remaining uncaptured tangles are harder to distinguish.

\subsection{Transporting bounded tangle supports through local port graphs}

Let~$G$ have rankwidth at most~$k$.
We describe the argument for any rooted tree~$D$ in which each vertex of~$G$ is assigned to one node, its \emph{owner}.
For the inclusion tree of \cref{sec:boundaries}, we take the tree on its internal nodes and assign each original vertex to the parent of its leaf.
The \emph{interior}~$I_v$ of a node~$v$ is the set of vertices it owns.
For each edge~$e$ at~$v$, let~$B_{v,e}$ be the vertices whose owners lie in the component of~$D-v$ reached along~$e$.
Assume every boundary set $B_{v,e}$ has cut-rank less than~$k$.
A tangle \emph{owns} $v$ if it regards every incident direction~$B_{v,e}$ as small.
Omit any empty boundary sets.
For each remaining boundary set, choose a partition into titanic pieces of order less than~$k$ as in \cref{lem:titanicpartition}.
Replacing all these pieces by their affine clique ports gives the local graph~$H_v$.
Let $L$ bound the size of each whole boundary gadget.

Let~$U$ be a support consisting of at most~$b$ original vertices.
An empty support captures no tangle and needs no transported copy.

To preserve capture, we need a copy of~$U$ only at nodes where a tangle captured by~$U$ can occur.
Suppose~$U$ contains a big set of a tangle owning~$v$, but no member of~$U$ is owned by~$v$.
Then~$U$ must meet at least three directions at~$v$, since the union of two small directions cannot contain a big set.
Such a node is a branch vertex of the smallest subtree joining the owners of the members of~$U$ (\cref{fig:supportlocations}).

\input{figures/support}

\begin{lemma}
\label{lem:supportlocations}
Suppose a tangle owns $v$ and is captured by $U$.
Then either $U\cap I_v\ne\varnothing$, or $U$ meets at least three boundary directions at $v$.
There are at most~$2b-2$ such nodes when~$b\geq2$, and at most one when~$b=1$.
\end{lemma}
\begin{proof}
Let $S\subseteq U$ be a big low-order set.
If $U$ has no interior vertex and meets at most two directions, their small sets $B_1,B_2$, together with the small set $V(G)\setminus S$, cover the whole graph.
This contradicts the tangle axiom, with the empty set used for any missing direction.

In \cref{fig:supportlocations}, the support has three owner nodes and just one further node where three occupied directions meet.
Let $Q$ be the set of owner nodes of the vertices in $U$, so $|Q|\leq b$.
Nodes with an interior member belong to $Q$.
Every other relevant node is a vertex of degree at least three in the minimal subtree connecting $Q$.
If $|Q|\geq2$, there are at most $|Q|-2$ such branch vertices, since every leaf of the connecting subtree belongs to $Q$.
Adding the at most $|Q|$ owner nodes gives the claimed bound.
If $|Q|\leq1$, there is no branch vertex, and only the possible owner node is relevant.
\end{proof}

We call a node \emph{relevant for $U$} if it satisfies either condition in \cref{lem:supportlocations}.
At each such node define the \emph{transported support}
\[
 U_v=(U\cap I_v)\ \cup\!
 \bigcup_{\substack{e\text{ incident with }v\\U\cap B_{v,e}\ne\varnothing}}
 T_{v,e},
\]
where~$T_{v,e}$ is the vertex set of the graph replacing~$B_{v,e}$.
Each interior member of~$U$ contributes one vertex to~$U_v$.
Each occupied direction contributes at most~$L$ vertices, and we can charge it to a different member of~$U$ outside the interior.
Thus, $|U_v|\leq |U|\max(1,L)\leq b\max(1,L)$.

At a relevant node, we replace each occupied direction by its whole finite gadget.
To capture the local tangle, it suffices that the transported support contain a big set, so we may include the whole gadget even when only some of its ports are needed.
We apply the rounding argument of \cref{lem:titanicround,thm:titanictanglebijection} to show that a big set survives inside the transported support.

\begin{lemma}
\label{lem:supportcapture}
Suppose an original tangle $\mathcal T$ owns $v$ and is captured by $U$.
Then $U_v$ captures the corresponding tangle of $H_v$ from \cref{prop:virtualtitanicports}.
\end{lemma}
\begin{proof}
Choose a big low-order $S\subseteq U$.
Round $S$ across the disjoint titanic pieces of all boundary sets.
For a titanic piece $F$ of rank $r$, at least one of $S\cup F$ and $S\setminus F$ has rank at most $\rho_G(S)$, by the titanic two-part condition and submodularity or posimodularity.
If~$F$ is disjoint from $S$, or already contained in it, leave $S$ unchanged.
Otherwise, choose a rank-nonincreasing rounding.

Every boundary piece is small in $\mathcal T$, since the whole boundary set is small.
Enlarging a big side preserves its orientation whenever the enlarged side still has low order.
If removing~$F$ made the side small, this smaller side, the set~$F$, and the old complement would be three small sets covering the graph.
All rounded sets have order less than $k$.

After all steps, the resulting set $S^*$ keeps every titanic piece whole.
A piece can be added only if it originally met $S$, since all pieces are disjoint.
Thus, its boundary direction meets~$U$.
The local port image of $S^*$ is contained in $U_v$.
The simultaneous block identity preserves its cut-rank, and faithful tangle transfer preserves its big orientation.
It is therefore the required big low-order subset of $U_v$.
\end{proof}

\input{figures/support-rounding}

The preceding bounds are combinatorial, and we still have to turn the relevant pairs of a support and a tree node into actual output vertices.
By \cref{lem:supportincidence}, we can choose a coloring that makes each support rainbow and then name its members by their colors.
A relevant node is either the owner of one member or the least common ancestor of the owners of two members, so a fixed number of copies of the support index is enough.

\begin{lemma}
\label{lem:supportcopies}
Suppose $D$, its owner map, and a linear-size family of bounded supports with their incidence relation are jointly MSO-defined from a bounded-cliquewidth input class.
Suppose the same transduction also gives the local graphs and identifies the direction to which each port belongs.
Then a fixed-copy MSO transduction gives all relevant pairs $(U,v)$ and their transported supports $U_v$, with linear total size.
\end{lemma}
\begin{proof}
The incidence graph has bounded cliquewidth because it is obtained by a fixed MSO transduction from the input class.
By \cref{lem:supportincidence}, we may guess a coloring with a fixed palette and check that every support is rainbow.

For a support $U$, each color identifies at most one member, and therefore at most one owner node.
Every relevant node is either one of these owners or the least common ancestor of two of them.
Indeed, if~$v$ owns no member of~$U$ and at least three directions at~$v$ meet~$U$, then at least two of them are child directions.
Choosing one support member in each of these two directions makes~$v$ the least common ancestor of their owners.
Represent $(U,v)$ by a copy of the index of $U$, using one copy for each color and one for each ordered pair of colors.
Keep the first index that gives $v$, and keep it only when $v$ has an interior member of~$U$ or at least three occupied directions.
These are fixed MSO tests.
This gives a bounded number of copies per support, regardless of the height of $D$.

Once we know $v$, the definition of $U_v$ uses only membership in $U$, the owner of each member, and the ports belonging to each direction at $v$.
Our boundary construction gives these port incidences together with the local graphs, using \cref{lem:finitegadgettype} and the inward bases of \cref{lem:coherentgeneralbases}.
Each transported support has at most~$b\max(1,L)$ members by the bound above.
Since each old support gives only a bounded number of copies, all copies and their incidences have linear total size.
\end{proof}

Consequently, when every newly created boundary gadget is itself taken as a support, an uncaptured local tangle can only come from an uncaptured original tangle owning that node.
Titanic tangle transfer makes this correspondence injective for each fixed simultaneous realization of the finite gadget types.
We may fix any realizations for this argument, since the types already determine the local graph.

\subsection{Owner tangles and the increase of distinguishing order}

Let $H$ be a graph of rankwidth at most $k\geq2$, and let $\mathcal U$ be a support family with size bound $L$.
Write $\mathcal P$ for the order-$k$ tangles that are uncaptured by $\mathcal U$.

Suppose that $\mathcal P$ has at least two members.
Fix an anchor $P\in\mathcal P$.
For each $Q\in\mathcal P\setminus\{P\}$, let $X_Q$ be the inclusion-minimal minimum-order set that is small in $P$ and big in $Q$.
By \cref{lem:closestunique}, these sets are uniquely determined and form a laminar family.
Add $V(H)$ and all singletons, and let $D$ be their inclusion tree.
For an internal node with territory $A$, let $I_A$ be the original vertices immediately below it.
Its \emph{directions} are its maximal proper closest-set children, together with $V(H)\setminus A$ when $A\ne V(H)$.
The directions are pairwise disjoint, have cut-rank less than $k$, and cover $V(H)\setminus I_A$.

If~$A\subsetneq B$ are closest sets, the cut around~$B$ also separates the anchor from every tangle with closest set~$A$.
This gives the following comparison of their orders.

\begin{lemma}
\label{lem:closestchainorders}
If $A\subsetneq B$ are closest sets, then $\rho_H(A)\leq\rho_H(B)$.
\end{lemma}
\begin{proof}
Choose $Q$ with $X_Q=A$.
Since $Q$ makes $A$ big, it also makes the low-order superset~$B$ big.
The anchor makes $B$ small, so $B$ is a competing $Q$-$P$ set.
Minimality of the order of $A$ proves the claim.
\end{proof}

We call the uncaptured tangles that make all directions at a node small its \emph{owner class}.
We now identify this class from the tree and compare the orientations chosen by its members.

\begin{lemma}
\label{lem:ownertangleclasses}
At the root, the anchor is the only uncaptured tangle making every direction small.
At a nonroot node~$A$, the uncaptured tangles making every direction small are precisely those with closest set~$A$.

If $X_Q=X_R=A$ and $Q\ne R$, then $Q,R$ agree on every separation of order at most~$\rho_H(A)$.
\end{lemma}
\begin{proof}
If $X_Q=A$, then $Q$ makes $A$ big and its complement small.
A proper child $B$ has order at most that of $A$ by \cref{lem:closestchainorders}.
If $Q$ made $B$ big, $B$ would be a smaller-order competing set or a proper contained minimum set, both impossible.
Thus, $Q$ makes every child direction small.

\input{figures/owner-directions}

Conversely, suppose $A$ is a nonroot node, $Q\ne P$ makes all directions at $A$ small, and write $B=X_Q$.
If $B\subsetneq A$, it lies in a child direction, which must then be big in $Q$, a contradiction.
If $A,B$ are disjoint, they cannot both be big in $Q$.
If $A\subsetneq B$, then $Q$ makes $A$ big because it makes its complement small.
Both $A$ and $B$ are small in $P$, and minimal order together with \cref{lem:closestchainorders} makes their orders equal.
This contradicts inclusion-minimality of~$B$.
Hence $A=B$.
At the root, every nonanchor $Q$ makes the direction containing $X_Q$ big, while $P$ makes every direction small.
At any nonroot node $A$, the anchor makes $A$ small and therefore makes its parent direction $V(H)\setminus A$ big.

For the last assertion put $a=\rho_H(A)$, and suppose a separation $Y$ of order $b\leq a$ distinguishes $Q$ and $R$.
Orient it so that $P$ makes $Y$ small, and rename $Q,R$ so that $Q$ makes $Y$ big.
It distinguishes $P,Q$, so $b\geq a$ and therefore $b=a$.
The unique inclusion-minimal minimum $Q$-$P$ set is contained in every such set, hence $A\subseteq Y$.
But $R$ makes $A$ big and $Y$ small, a contradiction.
\end{proof}

For each direction $B$ at $A$, we partition $B$ into a bounded number of titanic pieces of order less than~$k$, contract the pieces, and replace each atom by its affine clique gadget.
Every such group is titanic, has the same boundary rank as the piece it replaces, and has a binary partition tree of cuts of at most this rank.
The resulting local graph is denoted $H_A$.
Its untouched vertices are $I_A$, and the whole replacement of a direction $B$ is denoted $Z_B$.
There is a constant $g=g(k)$ with $|Z_B|\leq g$.

The canonical finite type specifies this local graph without selecting the original partitions globally.
In the proofs below, fix any realizing partition and basis witnesses separately for each direction.
Directions at one node are disjoint, so these witnesses can be realized simultaneously.
No assertion depends on a logical selection of these witnesses.

We apply \cref{prop:virtualtitanicports,cor:gadgetdistinguishingorder} at~$A$ to compare the tangles of~$H_A$ with those of~$H$.
For the recursion, it matters whether a tangle regards the whole set~$Z_B$ as small, even when each of its port groups is small.

\begin{lemma}
\label{lem:faithfulownertransfer}
The graph $H_A$ has rankwidth at most $k$.
For each fixed choice of the realizations at $A$, its order-$k$ tangles are in bijection with those order-$k$ tangles of $H$ that make every piece in these realizations small.
Under this bijection, tangles of $H_A$ that make every $Z_B$ small correspond to tangles of $H$ that make every direction $B$ small.
If $Q_1,Q_2$ are distinct tangles of $H_A$ and $P_1,P_2$ are their corresponding tangles of $H$, then the minimum order distinguishing $Q_1,Q_2$ equals the minimum order distinguishing $P_1,P_2$.
\end{lemma}
\begin{proof}
By \cref{prop:virtualtitanicports}, contracting the disjoint titanic pieces preserves the width bound, and we can expand the resulting atoms to obtain the graph $H_A$.
Expanding a piece of rank~$a<k$ using the affine binary partition tree creates internal cuts of rank at most~$a<k$, while cuts respecting whole groups keep their ranks.
Thus, the expanded graph has rankwidth at most $k$.

Apply the titanic tangle-bijection theorem on the original side and on the expanded side.
Every order-$k$ tangle on the expanded side makes each affine clique port group small by \cref{lem:affineportnoghosts}.
Both contractions yield the same atom connectivity function, by the simultaneous row-and-column spanning identity.
This gives the stated bijection.
A whole direction and its whole gadget are corresponding atom unions, so their orientations agree.

If a tangle makes the whole set~$Z_B$ big, then~$Z_B$ itself captures it.

For the order statement, round a distinguishing cut successively around the titanic pieces on the side where it is given.
Each rounding can be chosen not to increase the order and preserves the orientation in both tangles, since the pieces are small in both.
It yields a distinguishing atom cut of no larger order.
Its lift on the other side has the same order and still distinguishes the corresponding tangles.
Applying the argument in both directions gives the equality from \cref{cor:gadgetdistinguishingorder}.
\end{proof}

Recall that $A$ is relevant for an original support $U$ if $U$ meets $I_A$ or meets at least three directions at~$A$.
At each such node, we keep the transported support
\[
 U_A=(U\cap I_A)\ \cup\!
       \bigcup_{B\cap U\ne\varnothing} Z_B.
 \tag{OS}
\]
Also keep every $Z_B$ as a new support.
The transported supports in (OS) have size at most~$L(1+g)$.

The new support~$Z_B$ captures any tangle that makes it big.
The transported support~$U_A$ accounts for a tangle captured by~$U$ before the replacement.
We now apply these two observations to an uncaptured tangle of~$H_A$.

\begin{lemma}
\label{lem:ownersupporttransfer}
Every uncaptured order-$k$ tangle of $H_A$, with respect to these new supports, corresponds injectively to an old uncaptured tangle making every direction at $A$ small.
\end{lemma}
\begin{proof}
A tangle not captured by a whole $Z_B$ makes $Z_B$ small, because $\rho_{H_A}(Z_B)=\rho_H(B)<k$.
Its preimage $Q$ therefore makes every direction small.

Suppose $Q$ was captured by a big low-order $S\subseteq U$ for an old support $U$.
If $U$ misses $I_A$ and meets at most two directions, those small directions together with the small set~$V(H)\setminus S$ cover the ground set.
This contradicts the tangle axiom.
Hence, the node~$A$ is relevant for~$U$.

By \cref{lem:supportcapture}, we may round $S$ successively around all original titanic pieces without increasing its order or changing its big orientation.
Pieces disjoint from $S$ can be left on the outside, so an upward rounding is needed only in a direction meeting $S$.
The final atom union has order at most $\rho_H(S)<k$ and lifts to a big set $S'$ in $H_A$.
Its interior vertices lie in~$U\cap I_A$, and every added port lies in a direction gadget whose original direction meets~$U$.
Therefore $S'\subseteq U_A$.
The local tangle would be captured by~$U_A$, contrary to our assumption.
Thus,~$Q$ is uncaptured in the original graph.
Injectivity is part of \cref{lem:faithfulownertransfer}.
\end{proof}

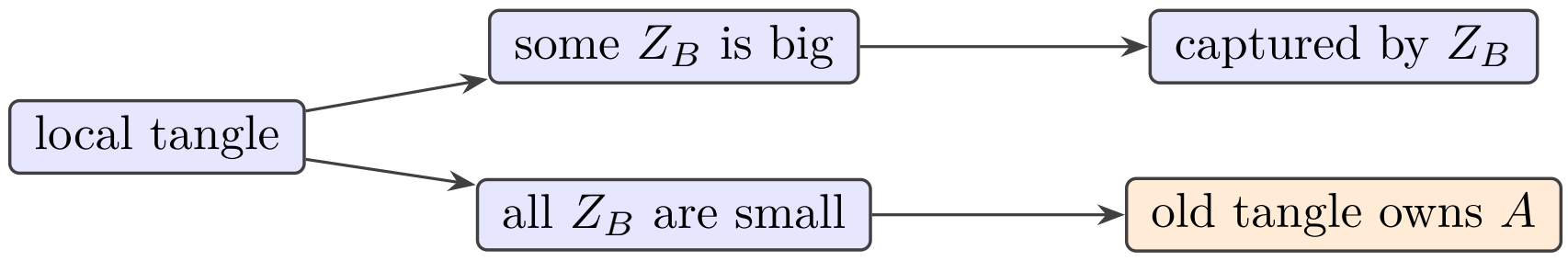
\begin{figure}[H]
\centering
\color{black}
\begin{tikzpicture}[manuscript illustration]
  \node[box] (q) at (0.4,0.4) {local tangle};
  \node[box] (a) at (4.12,1.05) {some $Z_B$ is big};
  \node[box] (b) at (4.12,-0.16) {all $Z_B$ are small};
  \node[box] (s) at (8.94,1.05) {captured by $Z_B$};
  \node[key] (o) at (8.94,-0.16) {old tangle owns $A$};
  \draw[arr] (q) -- (a);
  \draw[arr] (a) -- (s);
  \draw[arr] (q) -- (b);
  \draw[arr] (b) -- (o);
\end{tikzpicture}
\caption{Fix a realization of the local boundary types.
An uncaptured local tangle must follow the lower branch.
If the corresponding old tangle were captured by $U$, its transported support $U_A$ would capture the local tangle as well.
Thus, the corresponding original tangle is uncaptured and owns $A$.}
\label{fig:owner-support-alternative}
\end{figure}

We can now see why the lower bound on the distinguishing order increases.
Two uncaptured tangles at a nonroot node~$A$ both come from tangles with closest set~$A$.
By \cref{lem:ownertangleclasses}, these original tangles agree on every cut of order at most~$\rho_H(A)$.

\begin{corollary}
\label{cor:ownerthreshold}
Suppose distinct old uncaptured tangles have minimum distinguishing order at least $r$.
The root local graph has at most one new uncaptured tangle.
At every nonroot local graph, distinct new uncaptured tangles have minimum distinguishing order at least~$r+1$.
\end{corollary}
\begin{proof}
At each node, \cref{lem:ownersupporttransfer,lem:ownertangleclasses} give an injection from its uncaptured local tangles into the owner class of that node.
At the root that class has size one.
At a nonroot node $A$, the order $\rho_H(A)$ is the minimum order distinguishing the anchor from any tangle with closest set $A$, so $\rho_H(A)\geq r$.
Two different members of its owner class can only be distinguished at order strictly greater than $\rho_H(A)$.
Transfer preserves the minimum distinguishing order by \cref{lem:faithfulownertransfer}.
Orders are integers, giving the bound $r+1$.
\end{proof}

\input{figures/order-increment}

The recursion on the lower bound $r$ therefore has depth at most $k$.
This argument applies to every uncaptured anchor tangle, and \cref{thm:terminalsupport} handles the case in which at most one uncaptured tangle remains.

\section{The induction and the final encoding}
\label{sec:induction}
We return to the sparse encoding in \cref{lem:recursiveencoding}.
If at most one order-$k$ tangle remains uncaptured, we use the terminal construction.
Otherwise we form the local graphs of the inclusion tree.
By \cref{cor:ownerthreshold}, the root local graph is terminal, and at every other internal node we can use the induction hypothesis with lower bound~$r+1$ (\cref{fig:induction}).
\input{figures/induction}

Once we have a sparse encoding of each local graph, we still have to combine these encodings so that we can recover adjacency between vertices owned by different nodes.

\subsection{Choosing the boundary bases}

To recover an edge between vertices in different local graphs, we must carry their boundary coordinates towards a common ancestor.
Choosing a new basis independently at every node would be difficult to describe with unary parameters.
Instead, we keep each previously chosen representative in the same slot whenever it lies in the child territory, and only fill the remaining slots with new rows.
For cuts of rank two, Kalampakas selects the two basis vertices in this way using two colors~\cite[Lemma~5.1]{ranktwo}.
We use the same proof for variable ranks up to $k$, marking at each node which of the $k$ positions are occupied.

\begin{lemma}
\label{lem:coherentgeneralbases}
Let $D$ be a rooted tree whose leaves are the vertices of a graph $G$, and let~$S_u$ be the leaves below $u$.
Suppose $\rho_G(S_u)\leq k$ for every node.
One can choose $k$ unary selector predicates on the nonroot nodes and a \emph{set of active slots} $J_u\subseteq[k]$ at every node so that together they define an ordered cut basis $R_u=(r_j(u):j\in J_u)\subseteq S_u$.
If a parent representative $r_j(u)$ belongs to a child territory $S_v$, then it remains the representative $r_j(v)$ in the same slot.
From $G$ and $D$ with these labels, we can define the representatives and check the choices in MSO.
\end{lemma}
\begin{proof}
The root has its empty cut basis.
Proceed down the tree.
At a child $v$ of $u$, keep every parent representative that lies in $S_v$, in its existing slot.
These selected rows are independent on $V(G)\setminus S_u$, so they remain independent on the larger column set $V(G)\setminus S_v$.
Extend them to a cut basis for $S_v$ and give the added representatives unused slots.
At most~$k$~slots are needed.
A cut of rank zero has no active slot.
At a leaf of cut-rank one, exactly one slot is active, and its representative is the vertex at that leaf.

\begin{figure}[H]
\centering
\color{black}
\begin{tikzpicture}[manuscript illustration]
  \node[box] (u) at (3,1.39) {$u$};
  \node[box] (v) at (1.65,0.43) {$v$};
  \node[box] (w) at (4.35,0.43) {$w$};
  \node[key] (x) at (0.7,-0.58) {$x$};
  \node[box] (y) at (2.6,-0.58) {$y$};
  \node[box] (z) at (4.35,-0.58) {$z$};
  \draw[edge,densely dotted] (u.north) -- ++(0,.4);
  \draw[edge] (u) -- (w);
  \draw[edge] (w) -- (z);
  \draw[edge] (v) -- (y);
  \draw[edge,very thick,violet] (u) -- (v);
  \draw[edge,very thick,violet] (v) -- (x);
  \node[note] (n11) at (7.4,0.88) {slot $j$ at $u$ and $v$};
  \node[note] (n12) at (7.4,0.15) {$r_j(u)=r_j(v)=x$};
\end{tikzpicture}
\caption{The picture shows part of the tree below a nonroot node $u$.
Following the marked child for slot $j$ leads from $u$ to $v$ and then to the leaf $x$.
The rows inherited at $v$ remain linearly independent, since the columns outside $S_v$ include all the columns outside $S_u$.
We extend these rows to a basis at $v$ using the free slots.}
\label{fig:coherent-slot}
\end{figure}
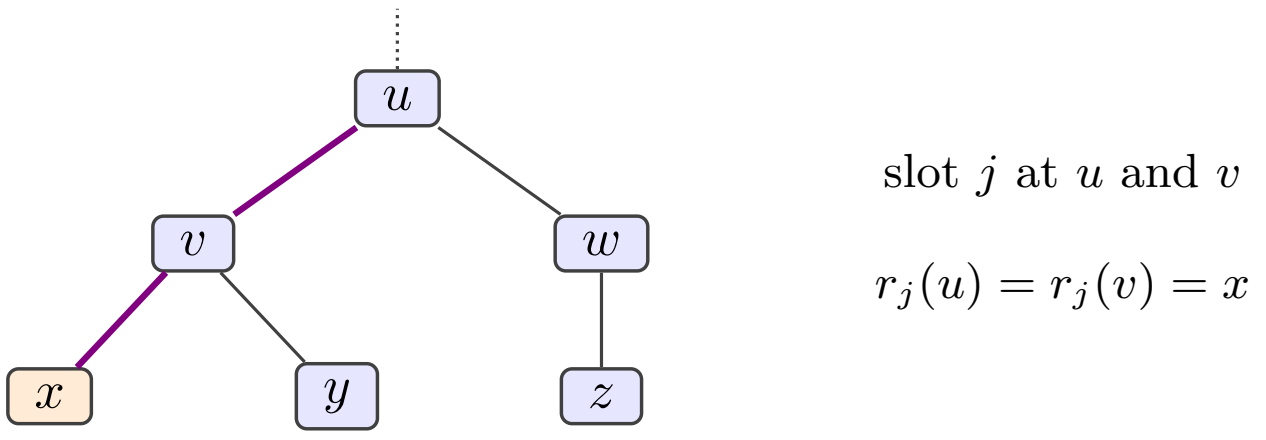

For each active slot $j$ at an internal node $u$, mark its unique child containing $r_j(u)$ with selector predicate $C_j$.
Each child has only one parent, so these marks never conflict.
Following $C_j$-marked children from $u$ reaches exactly the original leaf $r_j(u)$, because the representative is inherited at every step.
Thus, $x$ represents slot $j$ at $u$ if and only if $x$ is a leaf below $u$, $j\in J_u$, and every node strictly below $u$ on the path from $u$ to $x$ is marked $C_j$.
When $u$ is itself a leaf, it is the representative of slot $j$ if and only if $j$ is active there.

A guard requires exactly one marked child for each active slot of an internal node, no marked child for an inactive slot, and that every marked child keeps that slot active.
It also requires the resulting representatives to be a basis of the relevant cut.
Since $k$ is fixed and we can define tree paths in MSO, we can check these conditions by one MSO formula.
The construction above proves that the guard has a solution.
Thus, $k$ selector predicates and a finite set of labels for the active slots suffice to choose the bases along the whole tree.
\end{proof}

\subsection{Proof of the recursive encoding lemma}
\begin{proof}[Proof of \cref{lem:recursiveencoding}]
Let~$H$ be the graph in the enriched input.
We take cuts only of~$V(H)$ and compute their ranks using adjacency in~$H$.
The support indices and their incidences are used to test capture.

We first explain how to test whether a tangle is captured, since this test is needed both to choose the anchor and to recognize the terminal case.
By \cref{cor:tangleparameters}, a valid tuple and list of choices give an MSO test for whether a vertex set is big in the represented tangle.
To test capture, we existentially choose a support index $u$ and a set $S$ of graph vertices, require every vertex of $S$ to be incident with $u$, and check that $\rho_H(S)<k$ and that $S$ is big.
The incidence requirement says that $S$ is contained in the support represented by $u$, and the fixed cut-rank bound is MSO-definable.
Negating this formula tests whether the represented tangle is uncaptured.
We can also test whether at least two uncaptured tangles exist by choosing two valid parameter descriptions and requiring a low-order set that is big in one and small in the other.
This last requirement distinguishes the tangles even when one tangle has several parameter descriptions.

\emph{The terminal case.} We use downward induction on $r$, with the outer rankwidth bound~$k$ fixed throughout this lemma.
If $r=k$, there is at most one uncaptured tangle, because distinct order-$k$ tangles disagree on some cut of order less than $k$.
More generally, whenever at most one uncaptured tangle remains, rainbow-color the bounded supports and apply the terminal support theorem.
The support incidence graph is a fixed MSO projection of the enriched input and therefore has bounded cliquewidth.
By \cref{lem:supportincidence}, unary parameters suffice to choose the coloring from a fixed palette and a first-order guard checks it.

The terminal theorem gives an origin-preserving cliquewidth expression, hence a labeled tree.
We mark its root to tell the inverse that this is a terminal output.
For each support, we keep a separately marked copy of its index and join it to the distinguished copies of its members.
Deleting these marked indices gives the expression tree, and their neighborhoods give the support incidences.
Each such index has bounded degree, and the terminal encoding keeps all unary labels on the distinguished leaves.
This adds a uniformly sparse relation and is still a joint fixed MSO transduction of the enriched input.

\emph{One separating stage.} Suppose now that there are at least two uncaptured tangles and~$r<k$.
Choose an uncaptured anchor tangle by its bounded witness tuple and finite list of choices.
For every other uncaptured tangle, take its closest set with respect to this anchor.
They form a laminar family, and the laminar-tree transduction gives their inclusion tree $D$, with all original vertices kept.
We use the capture test above to require that the chosen anchor is uncaptured and to restrict the other tangle descriptions in the closest-set construction.

Choose coherent inward row bases along $D$.
At each internal node $v$, replace all its incident directions by the finite gadgets for the chosen types.
The parent direction uses the column profiles in the inward basis, so no second family of outside basis choices is needed.
The resulting local graphs $H_v$ have pairwise disjoint domains and total size linear in the input size.
Each has rankwidth at most $k$.
All their graphs, finite gadget labels, and ownership relations are produced together by one fixed MSO transduction.

Add every whole boundary gadget as a support.
For each nonempty old support, take the smallest subtree of~$D$ joining the owners of its members.
Transport the support to these owners and to the branch vertices of that subtree.
There are at most twice as many such nodes as members of the old support.
At a transported occurrence, keep the local members and the whole gadgets in directions meeting the old support.
These sets have bounded size.
Fix any realization of the chosen boundary types at~$v$.
By \cref{lem:ownersupporttransfer}, every uncaptured tangle of $H_v$ then corresponds to an uncaptured tangle of the original graph that makes every direction at $v$ small.

As in \cref{lem:supportcopies}, we use a coloring that is rainbow on every old support to construct these occurrences.
One color names a member and hence its owner, while a pair of colors names the least common ancestor of two owners.
This lets us represent every occurrence by a copy of the support index.

At the root, at most one uncaptured local tangle remains.
Its new supports have bounded size, and their incidence graph is obtained by a fixed transduction of the joint structure.
We can therefore choose a rainbow coloring for these supports by \cref{lem:supportincidence} and use the terminal encoding at the root, with the marked support indices attached to the expression tree as above.
At a nonroot node with territory $A$, any two distinct uncaptured local tangles correspond to distinct tangles of the original graph with the same closest set $A$.
Every cut distinguishing these two original tangles has order strictly greater than the cut-rank of $A$, which is at least $r$.
By \cref{lem:faithfulownertransfer}, the minimum distinguishing order is the same for the two local tangles.
Hence, the new threshold is at least $r+1$, and the induction hypothesis applies to all these local graphs.

\emph{Parallel recursion.} To apply the induction hypothesis, we first check that the individual nonroot domains satisfy its assumption on the enriched inputs.
We can mark one nonroot internal node of~$D$ with a unary predicate and take its local graph together with its support indices and their incidence relation.
This is a fixed MSO transduction of the joint structure, which was itself obtained from a class of bounded cliquewidth.
The bound on support size depends only on the previous bounds, and the threshold is at least $r+1$ in every such domain.
The induction hypothesis therefore gives one transduction for all the nonroot domains.
We run it in parallel by relativizing its formulas and guards to each domain, as in \cref{sec:sparse-encodings}.
The domains are disjoint, so a unary parameter on their union can make the required choices independently in each of them.
Thus, each recursive call decreases~$k-r$, independently of the depth of~$D$.

\emph{Assembling the sparse graph.}
Keep the internal nodes of~$D$ and the edges between them, and mark the root.
We call each such node a \emph{hub} and attach its local encoding there, as shown in \cref{fig:sparse-membership}.
Each original vertex $x$ already has a distinguished copy in the local encoding at its owner $u$, so we can omit the singleton leaves of $D$.
Before running the local encoders, we label~$x$ with the coordinates of its row across the cut around~$u$ in the inward basis at~$u$.
We also label each port with its index in the corresponding gadget.
The encoders keep these labels on their distinguished copies.
Below we use these coordinates to start the calculation in \cref{eq:generaltransition} and the port indices to identify the ports where the paths meet.
For every output vertex of the local encoder at $v$, add a marked path of length two from that vertex to the hub $v$.
A single extra copy of the vertex gives the middle of its path, and all these paths together form a subdivided star forest.
To recover the domain of the inverse at~$v$, take the endpoints of the marked paths from~$v$.
This also works when the local encoding is disconnected.
For each edge of the tree, use one copy of its child hub to represent the edge.
Join this copy to the two hubs and to the distinguished copies of the ports belonging to this edge at either end.
There are only boundedly many such ports at each end.
Unary labels distinguish the different kinds of auxiliary vertices, so these edges can be read separately from the edges within the local outputs.
On each hub~$v$, we mark the set $J_v$ of active slots in its inward basis.
On the copy representing the edge from~$v$ to its parent, we keep the two gadget types and the matrix that expresses the restricted child basis rows in the parent basis (\cref{eq:generaltransition}).
All of this uses a fixed number of unary labels.
At this stage, a unary predicate marks precisely the distinguished copies of the vertices owned by the hubs.
These copies form the domain of the interpretation of the current graph, including any vertices that were ports at the preceding stage.
The ports introduced at this stage instead carry their gadget labels for the local inverse.
We distinguish the source, root, and auxiliary labels of the different stages.
There are only boundedly many stages, so this uses a fixed set of labels and makes each port recognizable as a source vertex of the domain in which it was encoded.
Finally, keep the support indices of the current input on a separate layer.
For each member of a support, join its distinguished copy to the support index through a marked vertex.
If the member has color $i$ in the rainbow coloring, we use the $i$th copy of its support index for this vertex.
This represents the given support system, in addition to any transported support occurrences inside local domains.
The tree edges and membership paths form two forests, and the remaining incidence relations have bounded degree on one side.
Their union with the disjoint local encodings therefore has bounded degeneracy, by the edge bound in \cref{sec:sparse-encodings}.

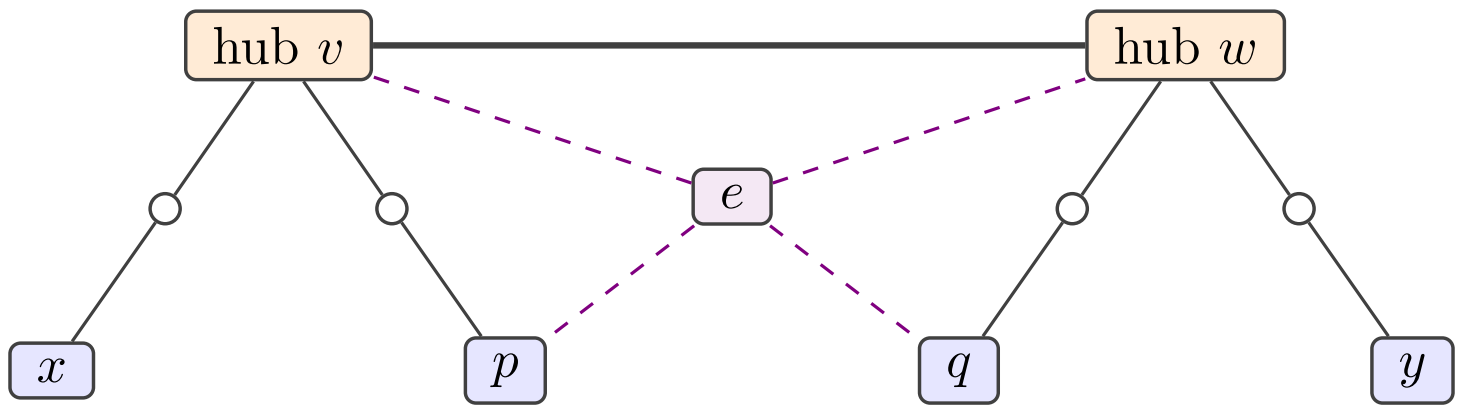
\begin{figure}[H]
\centering
\color{black}
\begin{tikzpicture}[manuscript illustration]
\node[key] (v) at (1.5,1.55) {hub $v$};
  \node[key] (w) at (7.5,1.55) {hub $w$};
  \draw[edge,very thick] (v) -- (w);
  \node[box,fill=violet!9] (e) at (4.5,.55) {$e$};
  \foreach \x/\lab/\owner/\mid in {0/x/v/.75,3/p/v/2.25,6/q/w/6.75,9/y/w/8.25}{
    \node[dot] (m\lab) at (\mid,.47) {};
    \node[box] (\lab) at (\x,-.6) {$\lab$};
    \draw[edge] (\owner) -- (m\lab) -- (\lab);
  }
  \foreach \endpoint in {v,w,p,q}
    \draw[edge,violet,dashed] (e) -- (\endpoint);
\end{tikzpicture}
\caption{Two adjacent hubs and part of their local encodings.
The paths through the white circles identify the domain of each local inverse, even if the local encoding is disconnected.
The vertex $e$ represents the tree edge between the hubs.
Its dashed edges lead to the two hubs and to the ports for this edge in their local graphs.
Here we show one such port at each end, $p$ and $q$.}
\label{fig:sparse-membership}
\end{figure}

\emph{Recovering the input graph.}
If the root has the terminal label for the current stage, we evaluate the expression tree obtained by deleting the support indices of that stage.
At a separating stage, the marked hub gives the root of the tree of domains.
The membership paths identify each local domain, and we apply the local inverse in that domain to recover its graph.
To recover the edges between vertices with different owners, we follow their rows up the tree of hubs.
Write $S_u$ for the territory of a node~$u$, $I_u$ for the original vertices owned by $u$, and $J_u\subseteq[k]$ for the active slots of its coherent inward basis $R_u=(r_j(u):j\in J_u)$.
At the root, put $J_u=\varnothing$.
For~$x\in S_u$, write~$\ell_u(x)\in\mathbb F_2^k$ for its row coordinates, padded by zero in the inactive slots.
Keep $\ell_u(x)$ as a finite unary label on every $x\in I_u$.

For an internal child $v$ of $u$, keep the $k\times k$ \emph{transition matrix}
\[
 M_v[i,j]=
 \begin{cases}
   \ell_u(r_i(v))_j,&i\in J_v,\\
   0,&i\notin J_v.
 \end{cases}
\]
For every~$x\in S_v$, we have
\begin{equation}\label{eq:generaltransition}
 \ell_u(x)=\ell_v(x)M_v.
\end{equation}
Indeed, the child-basis expression of the row of $x$ is valid on the whole exterior of $S_v$, and hence on the exterior of $S_u$.
Expressing the restricted child rows in the parent basis gives the displayed identity, by independence of the parent basis.

\input{figures/coordinate-transition}

In the local graph $H_u$, each internal child $v$ of $u$ gives a bounded port set $Z_v$ representing~$S_v$.
The leaf children of $u$ give the original vertices in $I_u$.
The type selected for~$Z_v$ gives each port~$p$ a profile~$c_v(p)\in\mathbb F_2^{J_v}$, its external row written in the basis~$R_v$.
These profiles span the whole space.
Select the first profile basis in the fixed finite index order of that gadget, and denote its ports by $p_1^v,\ldots,p_{a_v}^v$, where $a_v=|J_v|$.
Let $C_v$ be the invertible matrix with these profiles as rows.
For a row coordinate vector $a$ on the active slots define
\[
 b_v(a)=aC_v^{-1}.
\]
The index list of the selected ports and this linear map are finite data determined by the selected gadget type.
If $a_v=0$, the list and vector are empty, and the corresponding scalar products below are zero.

Suppose we have applied the local inverse and recovered $H_u$ on its distinguished vertex copies.
Let $x,y$ be original vertices, let $s,t$ be their owner nodes, and let $u$ be the least common ancestor of $s,t$.
Starting from the stored coordinate at its owner, \cref{eq:generaltransition} computes the coordinate of either vertex at every ancestor where it is needed.
We stop at the child of~$u$ whose territory contains the vertex, or use the vertex directly if~$u$ owns it.
This leaves three cases.

If $s=t=u$, then both vertices belong to $I_u$, and their adjacency is simply their adjacency in $H_u$.
No boundary needs to be crossed in this case.

If $s=u$ and $t$ lies in the child direction $v$, put $b=b_v(\ell_v(y)|_{J_v})$.
Then
\begin{equation}\label{eq:generalinteriorport}
 E_H(x,y)=\sum_{j=1}^{a_v} b_j E_{H_u}(x,p_j^v).
\end{equation}
The case $t=u$ is symmetric.
The identity holds because the chosen port profiles express the original boundary row of $y$ and $x$ lies outside $S_v$.

Finally suppose that $s,t$ lie in different child directions $v,w$ at $u$.
Put $b=b_v(\ell_v(x)|_{J_v})$ and $d=b_w(\ell_w(y)|_{J_w})$.
Then
\begin{equation}\label{eq:generalportbilinear}
 E_H(x,y)=
 \sum_{i=1}^{a_v}\sum_{j=1}^{a_w}
 b_i\, E_{H_u}(p_i^v,p_j^w)\,d_j.
\end{equation}
This is the simultaneous row-and-column expansion of the original cut matrix.
The identity also holds for zero profiles and for any partition realizing the chosen type in each direction.
Indeed, the types determine the port profiles, and these profiles together with the matrix~$A_H[R_v,R_w]$ determine the adjacencies between the two port sets.

\input{figures/decoder-three-cases}

We can express all these calculations in MSO.
An ancestor relation and a least common ancestor are definable on a rooted tree.
To evaluate \cref{eq:generaltransition} along a path, use one unary set for each of the finitely many states in $\mathbb F_2^k$, require exactly one state at each path node, and require the prescribed transition on every consecutive pair.
The initial state is the stored owner coordinate.
This determines a unique run.
The sums in \cref{eq:generalinteriorport} and \cref{eq:generalportbilinear} have bounded length and are fixed Boolean formulas.
The local graph relation~$E_{H_u}$ is given by the fixed local inverse, relativized to the domain recovered from its marked membership hub.

Thus, the assembly has one fixed MSO inverse.
It uses only the inward coordinates, the child-direction gadgets, and their local adjacencies.
Parent-direction gadgets are needed for the tangle argument, but their coordinates need not enter this final inverse.

\emph{Width and definability of the final output.} We keep the tree, the local domains, and the labels together while the local encoders run.
In the final graph, the marked membership paths tell us which vertices belong to each domain.
The combined graph has bounded cliquewidth, since it is obtained by a fixed MSO transduction from the class that gave us the enriched inputs.
Together with its bounded degeneracy, this gives bounded treewidth by \cref{sec:sparse-encodings}.
The combined graph itself, with its labels and distinguished copies, is the sparse output required by the lemma.
Its inverse recovers the graph reduct as above and reads the original support incidences on the separate layer.

Every construction mentioned above is a fixed MSO transduction, and there are at most~$k-r$ separating stages along each branch.
Their composition is therefore one finite-copy MSO transduction.
As in \cref{sec:sparse-encodings}, we can check that every accepted output recovers the input graph on its distinguished copies while the input is still present.
This completes the induction.
\end{proof}

\subsection{Completing the induction on rankwidth}
We now apply the sparse encoding and its decoder to prove the decomposition theorem.

\generaldecomposition*
\begin{proof}[Proof of \cref{thm:main,thm:generaldecomposition}]
We prove by induction on $k$ that one fixed MSO decomposer accepts every graph of rankwidth at most $k$.
For $k=1$, \cref{lem:rankonebase} gives a source-preserving encoding by finitely labeled forests.
We join the components under a new root and apply \cref{lem:treecliquecompiler} to the resulting tree and its decoder.

Let $k\geq 2$ and assume the result for smaller rankwidth.
If~$G$ has no order-$k$ tangle, then its rankwidth is less than~$k$ by tangle duality, so the induction hypothesis applies.
If~$G$ has exactly one order-$k$ tangle, \cref{thm:uniquetangleencoding} gives its decomposition.
The descriptions in \cref{cor:tangleparameters} let an MSO transduction distinguish these cases from the case of several tangles.

Suppose now that at least two order-$k$ tangles remain.
Graphs of rankwidth at most~$k$ have bounded cliquewidth by the inequalities of Oum and Seymour~\cite{rankwidth}.
We take the support family to be empty, so all order-$k$ tangles are uncaptured, and put~$r=0$ since every cut has nonnegative order.
Now \cref{lem:recursiveencoding} gives a fixed MSO transduction that encodes each such graph $G$ by a uniformly sparse graph $S$.
Its fixed MSO decoder recovers $G$ on the distinguished copies of its vertices in $S$.
The graphs $S$ have bounded cliquewidth by \cref{lem:transduction-preserves-cliquewidth} and hence bounded treewidth by their bounded degeneracy~\cite{gw}.

By \cref{prop:boundedtwencoder}, we can encode $S$ by a labeled forest from which a fixed MSO interpretation recovers $S$ and its labels on designated leaves.
We join the components under a new root and compose this interpretation with the decoder for $S$.
The resulting interpretation recovers $G$ on one leaf for each original vertex, with that vertex as its origin.
Now \cref{lem:treecliquecompiler} gives a clique expression for $G$ of uniformly bounded width, preserving these origins.
Each construction has a successful output on the inputs in question, and the source-preserving decoders ensure that every accepted output represents its input.
Their composition gives the decomposer in the case of several tangles.
Taking the union of the three guarded cases gives one decomposer for all graphs of rankwidth at most~$k$.

\input{figures/outer-induction}

Since rankwidth is at most cliquewidth~\cite{rankwidth}, the decomposer for rankwidth at most~$k$ accepts every graph of cliquewidth at most $k$.
Its output width gives the bound $\ell$ in \cref{thm:main}.
Taking $f(k)$ to be the size of this decomposer gives the bound on definable cliquewidth in \cref{thm:generaldecomposition}.
\end{proof}

\subsection{Recognizability and CMSO-definability}

We use \emph{recognizable} in the usual vertex-replacement sense.
A \emph{cliquewidth context} is an expression with one hole into which we may insert a labeled graph.
It can use the usual cliquewidth operations and fixed labeled graphs.
For a graph language $L$ and a fixed number of labels, two labeled graphs are equivalent if replacing one by the other in any cliquewidth context preserves membership in $L$.
The language is recognizable if there are only finitely many such equivalence classes for each fixed number of labels.
\emph{Counting~MSO}, or~\emph{CMSO}, extends MSO by predicates that test the cardinality of a set modulo a fixed positive integer.
Boja\'nczyk, Grohe, and Pilipczuk proved the following equivalence for graph languages of bounded linear cliquewidth~\cite[Theorem~3.5]{linear}.
Kalampakas proved a relative version for graphs of rankwidth at most two.
On that class, the restrictions of recognizable graph languages are exactly the languages definable there by CMSO~\cite[Theorem~1.1]{ranktwo}.
With the decomposer from \cref{thm:generaldecomposition}, the argument of Boja\'nczyk, Grohe, and Pilipczuk applies to every fixed cliquewidth bound.

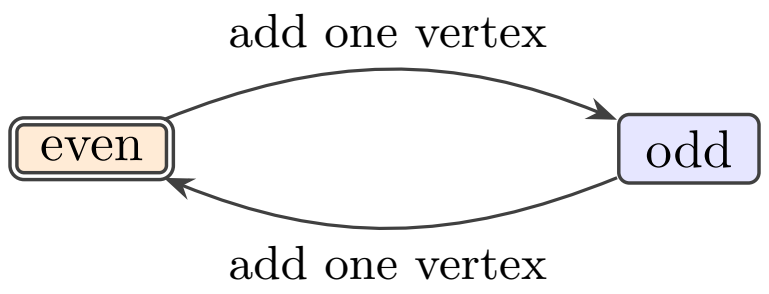
\begin{figure}[H]
\centering
\color{black}
\begin{tikzpicture}[manuscript illustration]
  \node[key,double,double distance=.65pt] (a) at (1.5,0.7) {even};
  \node[box] (b) at (5.45,0.7) {odd};
  \draw[arr,bend left=22] (a) to (b);
  \draw[arr,bend left=22] (b) to (a);
  \node[note] (n4) at (3.46,1.48) {add one vertex};
  \node[note] (n5) at (3.46,-0.05) {add one vertex};
\end{tikzpicture}
\caption{The double border marks the accepting state for the language of graphs with an even number of vertices.
A cliquewidth context adds a fixed number of vertices to the graph in its hole.
Graphs of the same parity therefore agree in every such context, and the CMSO test for even size reads the same information.}
\label{fig:counting-context}
\end{figure}

\begin{corollary}\label{cor:recognizability}
A graph language of bounded cliquewidth is recognizable if and only if it is definable in CMSO on the adjacency representation.
\end{corollary}
\begin{proof}
CMSO-definability implies recognizability by the usual composition theorem~\cite{courcelleengelfriet}.
Conversely, let $L$ be recognizable and have bounded cliquewidth.
By \cref{thm:generaldecomposition}, there is a decomposer of some fixed width $\ell$ that accepts every graph in $L$.
There is a CMSO sentence~$\psi_L$ on width-$\ell$ expression trees that holds if and only if the value of the expression belongs to $L$.
This follows from the tree-language claim of Boja\'nczyk, Grohe, and Pilipczuk \mbox{\cite[Claim~3.6]{linear}}.
By the backwards translation theorem for CMSO, we obtain a sentence $\varphi_L$ that holds on a graph $G$ if and only if some output of the decomposer on $G$ satisfies $\psi_L$~\cite[Theorem~7.10]{courcelleengelfriet}.
Indeed, to test the size of an output set modulo a fixed integer, we can test its size in each copy layer and add the residues.
The existence of an output for every graph in $L$ and correctness of every output show that the resulting sentence defines exactly $L$.
\end{proof}

\begin{figure}[H]
\centering
\color{black}
\begin{tikzpicture}[manuscript illustration]
\node[box] (g) at (0,1) {$G$};
  \node[key] (t) at (7.2,1) {expression $t$};
  \draw[arr] (g) -- node[above=2pt,note] {decomposer} (t);
  \node[box] (p) at (0,-.26) {graph sentence $\varphi_L$};
  \node[key] (q) at (7.2,-.26) {tree sentence $\psi_L$};
  \draw[arr] (q) -- node[above=2pt,note] {backwards translation} (p);
\end{tikzpicture}
\caption{Here $t$ is one possible output of the decomposer on $G$.
The tree sentence $\psi_L$ tests whether the value of $t$ belongs to $L$.
Its backwards translation $\varphi_L$ holds on $G$ if and only if some output on $G$ satisfies $\psi_L$.}
\label{fig:recognition-pullback}
\end{figure}
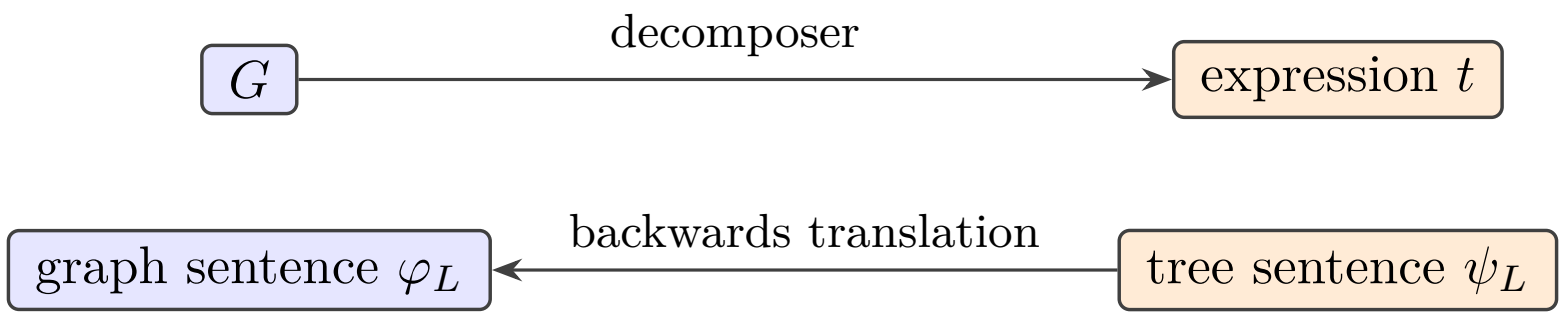

The decomposer itself uses ordinary MSO.
Counting enters when we test the acceptance condition on its unordered expression tree.

\section{Seese's conjecture}
\label{sec:consequences}

Seese asked whether every class of graphs with decidable MSO theory is interpretable in a class of trees~\cite{seese}.
For finite graphs, this is equivalent to asking whether every such class has bounded cliquewidth~\cite[Section~5.4]{courcelleoum}.
The MSO theory of a class consists of the sentences true in every member.
By negating the sentence, deciding this theory is equivalent to deciding whether a given MSO sentence holds in some graph of the class.
MSO quantifies over vertices and sets of vertices, with adjacency as the only graph relation.
Courcelle and Oum proved the conjectured bound under the stronger assumption that the theory remains decidable when the logic can also test whether a vertex set has even cardinality~\cite[Theorem~5.6]{courcelleoum}.
Seese proved that decidability for MSO with edge-set quantification implies bounded treewidth~\cite{seese}, as stated in \cite[Theorem~1.29(1)]{courcelleengelfriet}.
A sparse representation would allow us to apply this last result to the conjecture.

Suppose that a graph class $\mathscr C$ has a fixed source-preserving MSO encoder into finitely colored graphs, with a fixed MSO inverse.
Let $\mathscr D$ consist of all the encoder's accepted outputs on graphs in $\mathscr C$, and suppose that $\mathscr D$ is uniformly sparse.
If $\mathscr C$ has decidable MSO theory, backwards translation gives decidable MSO theory on $\mathscr D$.
Indeed, an output satisfying a given sentence exists precisely when some input satisfies its backwards translation.
On uniformly sparse graphs, Courcelle's incidence construction also gives a translation of sentences with edge-set quantification into MSO sentences on the graph~\cite[Theorem~4.1]{sparseincidence}.
The underlying graphs in $\mathscr D$ therefore have decidable MSO theory with edge-set quantification, so Seese's theorem gives them bounded treewidth.
They have bounded cliquewidth as well, and applying the fixed inverse to the colored graphs gives bounded cliquewidth for $\mathscr C$ by \cref{lem:transduction-preserves-cliquewidth}.

Our construction gives such a sparse representation under the assumption that $\mathscr C$ already has bounded cliquewidth.
To deduce the conjecture by this argument, one would need to obtain the representation from the decidability assumption alone.
It suffices to consider monadically dependent classes, that is, classes from which no FO transduction produces all finite graphs.
If an FO transduction produces all finite graphs from~$\mathscr C$, then deciding its MSO theory would also let us decide whether an MSO sentence has a finite graph model, by backwards translation.
The latter problem is undecidable.
It therefore remains to show that every monadically dependent class of unbounded cliquewidth has undecidable MSO theory.
For such a class, the uniform bound on cut-rank used in our construction is unavailable.


\appendix
\section{Deferred proofs}
\label[appendix]{app:transduction-lemmas}

We collect the proofs deferred from \cref{sec:preliminaries,sec:closure,sec:tangles}.

\subsection{Bounded treewidth to labeled forests}
\label[appendix]{app:boundedtwencoder}

\boundedtwencoder*

\begin{proof}
A \emph{star coloring} is a proper vertex coloring in which every two color classes induce a disjoint union of stars.
Fertin, Raspaud, and Reed proved that graphs of bounded treewidth admit a star coloring with a bounded number of colors~\cite{starcoloring}.
The same coloring argument used to encode orientations in~\cite[Corollary~7.4]{fotransductions} gives an FO transduction to the incidence representation.
Choose a coloring from a fixed palette and check that it is proper and has no path on four distinct vertices using only two colors.
Both conditions are first-order, and together they say exactly that the coloring is a star coloring.
For each pair of colors, consider the stars induced by those two color classes.
In each such star, represent an edge by a copy of its leaf endpoint, choosing the endpoint with the smaller color when the star has only one edge.
The pair of colors indexes the copy, whose two incidence neighbors are its original vertex and that vertex's unique neighbor of the other color.
We keep the original vertex layer and its labels.
We now apply the tree-decomposition transduction of Boja\'nczyk and Pilipczuk to this incidence representation, keeping the original vertices and their bag membership.
Their bounded-width decomposition suffices here~\cite[Theorem~2.4]{treewidthlogic}.
We mark the root of each component of the decomposition forest so that we can recognize the parent of each nonroot bag in the final forest.
Color the original vertices with a fixed number of colors so that the vertices of each bag have distinct colors.
To see that the maximum bag size gives enough colors, start at the root and color the bags in order of increasing depth.
In each bag the vertices already colored occur in its parent, so the remaining vertices can be given distinct unused colors.
The bags containing a vertex form a connected subtree, so exactly one of them is closest to the root.
At this bag we attach a designated copy of the vertex as a leaf.
Label each bag by the colors present in it and the adjacency relation on these colors, and each nonroot bag also by the colors of the vertices it shares with its parent.
On each designated leaf, keep its vertex color and its original unary labels.
Starting at the bag of a designated leaf, we can follow its color down the tree as long as each child marks the color as shared with its parent.
In this way we reach exactly the bags containing the vertex represented by the leaf.
Two distinct leaves represent adjacent vertices if and only if they reach a common bag whose label gives an edge between their colors.
After deleting the remaining relations, we obtain the required labeled forest.
All choices can be checked by MSO formulas while the input and the decomposition are still present, which ensures correctness of every accepted output.
\end{proof}

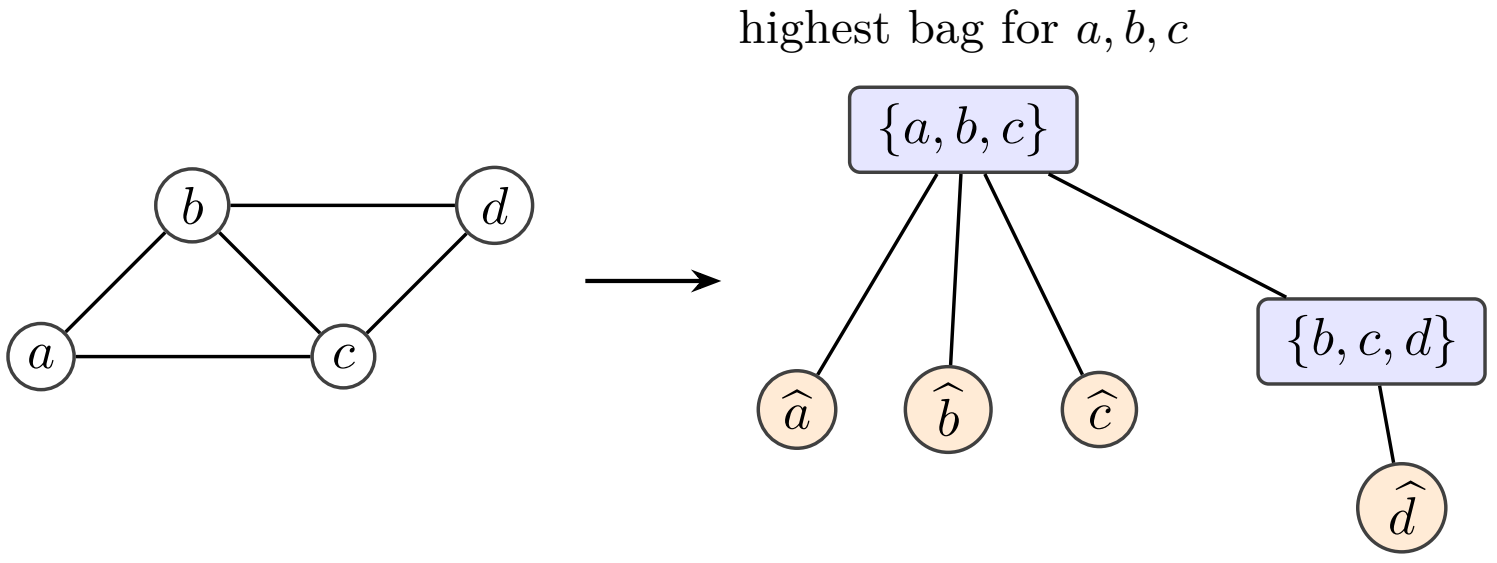
\begin{figure}[H]
\centering\color{black}
\begin{tikzpicture}[manuscript illustration]

\node[dot] (a) at (0,0) {$a$};\node[dot] (b) at (1,1) {$b$};\node[dot] (c) at (2,0) {$c$};\node[dot] (d) at (3,1) {$d$};
\draw (a)--(b)--(c)--(a) (b)--(d)--(c);
\draw[-{Stealth[length=2mm]},thick] (3.6,.5)--(4.5,.5);
\node[box] (u) at (6.1,1.5) {$\{a,b,c\}$};
\node[box] (v) at (8.8,.1) {$\{b,c,d\}$};
\draw (u)--(v);
\foreach \x/\lab in {5/a,6/b,7/c}{\node[dot,fill=orange!16] (\lab) at (\x,-.35) {$\widehat\lab$};\draw (u)--(\lab);}
\node[dot,fill=orange!16] (dd) at (9,-1) {$\widehat d$};\draw (v)--(dd);
\node[note] at (6.1,2.15) {highest bag for $a,b,c$};
\end{tikzpicture}
\caption{The graph on the left has a tree decomposition with bags $\{a,b,c\}$ and $\{b,c,d\}$.
The source copies are attached at their highest bags.
We write the vertex names in the drawing, but in the encoding each vertex in a bag is identified by its color.
Each bag describes the edges between its vertices, and the lower bag marks the colors of $b$ and $c$, the vertices it shares with its parent.
From these labels we can recover all five edges.}
\label{fig:treewidth-bags-example}
\end{figure}

\input{figures/highest-bag-leaf}

The same construction works for a fixed finite signature consisting of unary and binary relations when the Gaifman graph has bounded treewidth.
Store the finitely many relation matrices, including their directions, at each bag.

\subsection{Labeled trees to cliquewidth decompositions}
\label[appendix]{app:treecliquecompiler}

Courcelle and Engelfriet proved that, if a fixed MSO transduction constructs graphs from terms over a finite signature, then another fixed MSO transduction can construct cliquewidth expressions of bounded width for these graphs from the same terms~\cite[Theorem~7.34(1)]{courcelleengelfriet}.
We include the standard construction for unordered trees, using the Join operation of Boja\'nczyk, Grohe, and Pilipczuk~\cite{linear}, to make explicit that it preserves the origins of the graph vertices.

\treecliquecompiler*

\begin{proof}
We start with the input tree $T$ of the lemma, on which $\delta$ interprets the graph.
We first give each interpreted vertex a leaf of its own.
For each pair of leaves, we will decide whether to add the edge at the node where their branches meet.

Choose a root of $T$ by a unary predicate that marks exactly one vertex.
For each vertex~$x$ in the domain of $\delta$, create one new copy $x^\ast$, attach it to $x$ by an edge, and mark $x^\ast$ as a designated leaf.
We keep all original vertices and edges of $T$, so the resulting graph is again a tree, with the same root.
This modification is an MSO transduction using two copies of the input vertices, with the second copy kept only for vertices in the domain of $\delta$.
Each new leaf identifies its original vertex by its attachment edge.
We therefore adjust the interpretation to test adjacency of two designated leaves by applying the original formula to their neighbors, restricting its quantifiers to the original tree.
For the rest of the proof, we again write $T$ for the modified tree and $\delta$ for the adjusted interpretation.
There is at least one designated leaf, since these leaves represent the vertices of the interpreted graph.

Choose~$q$ at least as large as the quantifier rank of the adjusted adjacency formula, which we write as~$\delta(x,y)$.
The \emph{MSO \mbox{$q$-type}} of a structure with distinguished vertices is the collection of MSO formulas of quantifier rank at most~$q$ that hold at those vertices~\cite{courcelleengelfriet}.
For a fixed finite signature and a fixed number of distinguished vertices, there are only finitely many such types, and each is characterized by a single MSO formula of quantifier rank at most $q$.
Let $\Theta$ be the finite set of MSO $q$-types of rooted unary-labeled trees with one distinguished designated leaf.
For a node $u$ and a designated leaf $x$ below it, define $\operatorname{tp}_u(x)=\operatorname{tp}^{\mathrm{MSO}}_q(T_u,u,x)$, where~$T_u$ is the full descendant subtree.
To test~$\operatorname{tp}_u(x)=\theta$ in MSO, we take a formula characterizing~$\theta$ and restrict its quantifiers to~$T_u$.

We use two direct consequences of MSO composition for rooted trees~\cite[Section~5.3]{courcelleengelfriet}.
First, for a fixed node $u$, if $x$ and $y$ lie below different children $v,w$, the value of $\delta(x,y)$ depends only on $\operatorname{tp}_v(x)$ and $\operatorname{tp}_w(y)$.
The same test works for any two children of~$u$, so one Join can treat all the child subtrees together.
These types also determine the types of the rooted child subtrees without a distinguished leaf, since we may restrict attention to formulas that do not mention the leaf.
To compare two choices of the marked leaves, match the two marked child subtrees by their respective types.
The remaining children can then be matched by their unmarked types, since both choices removed the same two unmarked types from the same multiset.
MSO composition at $u$ and in the fixed tree above $u$ now gives the same truth value for the adjacency formula~\cite[Section~5.3]{courcelleengelfriet}.

Second, $\operatorname{tp}_u(x)$ depends only on $\operatorname{tp}_v(x)$, again uniformly over children $v$ of $u$.
Indeed, matching the marked child subtrees also matches their unmarked types, so the remaining children can again be paired by type.

At each internal node~$u$, let~$J_u$ consist of the unordered pairs of child types for which we add an edge between leaves in different child subtrees.
The types in a pair may agree, as in the definition of Join.
Let~$f_u:\Theta\to\Theta$ give the type at~$u$ of a leaf of each child type.
If no child realizes a type, we assign it a fixed value under $f_u$.
We define $J_u$ in MSO by asking whether two designated leaves of the given types lie below distinct children of $u$ and are adjacent in the interpreted graph.
For a type realized below a child of $u$, we define its image under $f_u$ by choosing one leaf of that type and testing its type at $u$.
Uniformity makes each realized entry unambiguous.
Since the tables have fixed finite size, their complete values can be stored as unary labels on $u$.

Now discard every subtree with no designated leaf.
At each remaining internal node~$u$, create a Join node labeled by~$J_u$ whose inputs are the expressions for its remaining children, and place a Recolor node labeled by~$f_u$ above it.
At a designated leaf~$x$, create a leaf of the cliquewidth expression representing a single vertex with label~$\operatorname{tp}_x(x)$.
Every edge is added exactly once, at the least common ancestor of its two designated leaves, and the colors after processing $u$ are the types $\operatorname{tp}_u(x)$.
Induction on the tree proves that the resulting expression evaluates to $\delta(T)$.

\input{figures/tree-type-join}

The construction uses two copies of each remaining internal node and one of each designated leaf, with all relations and operation labels defined by fixed MSO formulas.
The expression has width at most $|\Theta|$, and for each interpreted vertex $x$ it has one leaf with origin $x$.
\end{proof}

\input{figures/compiler-tree-example}

We can now compose a tree encoding with the transduction in \cref{lem:treecliquecompiler}, since both keep the origins of the graph vertices.
For a forest, one unary predicate chooses a vertex in each component.
A further unary predicate selects one vertex of the forest, and we use an additional copy of that vertex as a new marked root, joined to all chosen component vertices.
The inverse recovers the forest by deleting this root.

\subsection{Combining finitely many local encodings}
\label[appendix]{app:finitecover}

\finitecoverencoding*

\begin{proof}
A coloring chosen by the transduction may give a subgraph $G[c^{-1}(I)]$ outside $\mathscr D_I$, so we check the output of each local encoder before using it.
We fix a common bound $d$ on the degeneracy of the encodings in the hypothesis and reject any local output whose degeneracy exceeds $d$.
We also check that every vertex of the local input has exactly one designated copy in the output and that the MSO inverse gives the same adjacency on these copies as in the local input.
We can make these checks in MSO while we still have the local input, the output, and the relation that assigns to each copy its original vertex.
For each local graph in its intended class $\mathscr D_I$, at least one encoding passes these checks by the assumptions of the theorem.

Choose a coloring $c:V(G)\to[m]$ and construct disjoint copies of the local inputs $G[c^{-1}(I)]$ for which~$c^{-1}(I)$ is nonempty.
There are only finitely many indices $I$, so a fixed number of copies suffices.
Apply each local encoder to its own input with the checks above, and mark the vertices of its output by the index $I$.
Keep one further, edgeless copy of $V(G)$, with the colors chosen by $c$ and any original unary labels.
Join each designated vertex in a local output to the copy of its origin in this edgeless graph.
These links form a star forest, because every local source vertex has exactly one original vertex.
The local outputs are disjoint and have one common uniform sparsity bound, and adding a star forest preserves uniform sparsity.
All accepted outputs are sparse and source-preserving, even for unintended color choices.
A successful choice exists by the assumption on $\mathscr C$.

For two original vertices~$u,v$, take~$I=\{c(u),c(v)\}$, which is a singleton if the colors agree.
The corresponding local input contains both vertices.
The linking stars identify the two corresponding local copies, and we apply the inverse for $I$ on the vertices marked by $I$ to decide whether $uv$ is an edge.
A finite disjunction over the indices $I$ gives one MSO inverse interpretation.

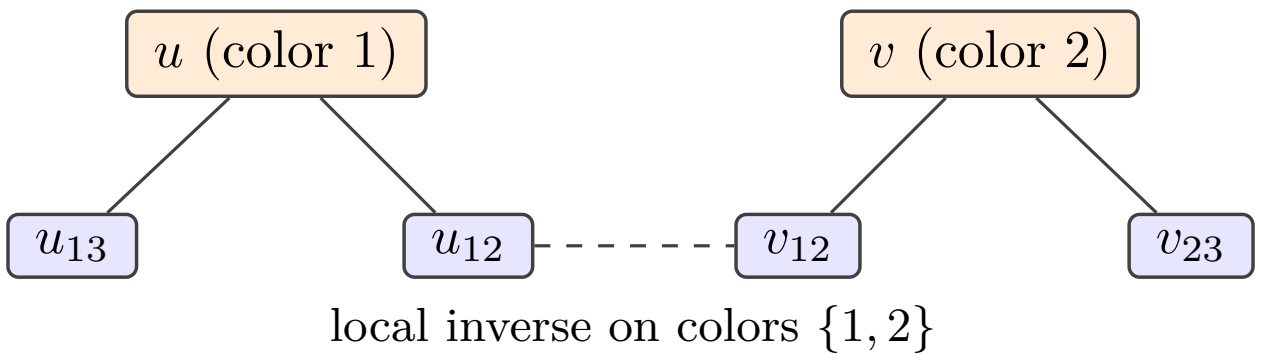
\begin{figure}[H]
\centering
\color{black}
\begin{tikzpicture}[manuscript illustration]
  \node[key] (x) at (2,1.27) {$u$ (color 1)};
  \node[key] (y) at (6.72,1.27) {$v$ (color 2)};
  \node[box] (a) at (0.65,0) {$u_{13}$};
  \draw[edge] (x) -- (a);
  \node[box] (b) at (3.27,0) {$u_{12}$};
  \draw[edge] (x) -- (b);
  \node[box] (c) at (5.45,0) {$v_{12}$};
  \draw[edge] (y) -- (c);
  \node[box] (d) at (8.05,0) {$v_{23}$};
  \draw[edge] (y) -- (d);
  \draw[edge,dashed] (b)--(c);
  \node[note] (n11) at (4.36,-0.55) {local inverse on colors $\{1,2\}$};
\end{tikzpicture}
\caption{Some source copies in a three-color cover.
Solid links identify which copies have the same origin in the construction of \cref{thm:finite-cover-encoding}.
To decide whether $u$ and $v$ are adjacent, we apply the local inverse to their copies in the encoding for colors $1$ and $2$, as indicated by the dashed line.}
\label{fig:two-cover-links}
\end{figure}

When the input class has bounded cliquewidth, the outputs have bounded cliquewidth by \cref{lem:transduction-preserves-cliquewidth}, so their uniform sparsity gives bounded treewidth.
\end{proof}

\subsection{Linear products of definably decomposable blocks}
\label[appendix]{app:linearclosure}

To prove \cref{lem:linearclosure}, we describe the addition of a block using the product of derivations introduced by Boja\'nczyk, Grohe, and Pilipczuk~\cite[Section~4]{linear}.
A \emph{$q$-derivation} is a triple $(H,\pi,\phi)$, where $H$ is a graph whose vertices have colors in $[q]$.
For each vertex~$v\in V(H)$, its \emph{profile} $\pi(v)\subseteq[q]$ specifies the colors of its neighbors in the part built so far.
Its \emph{recoloring} $\phi:[q]\to[q]$ specifies how to change the colors on that earlier part once $H$ has been added.

To multiply two derivations, we first take the disjoint union of their graphs.
We add an edge between a vertex $u$ of the first factor and a vertex $v$ of the second if and only if the color of $u$ belongs to $\pi(v)$, and then apply the recoloring of the second factor to all vertices of the first.
In our application, this recoloring lets us merge old color classes when their vertices have the same neighborhood in the part that still remains.

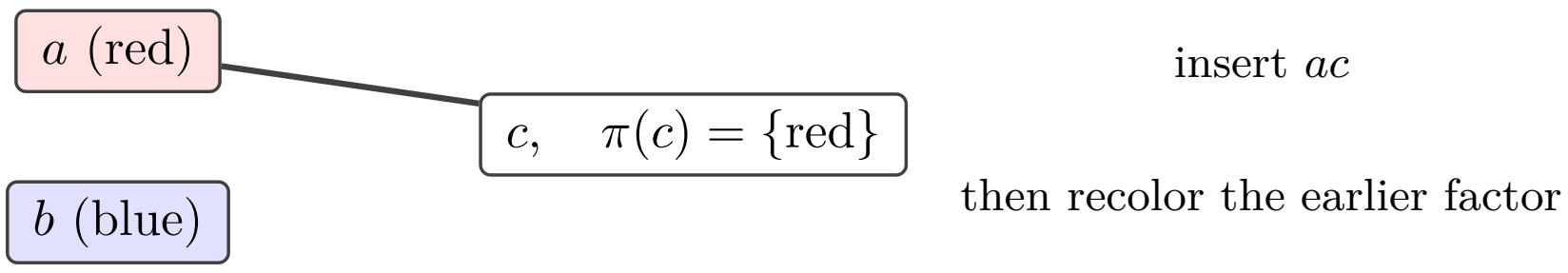
\begin{figure}[H]
\centering
\color{black}
\begin{tikzpicture}[manuscript illustration]
  \node[key,fill=red!12] (a) at (0,1) {$a$ (red)};
  \node[box,fill=blue!12] (b) at (0,-0.19) {$b$ (blue)};
  \node[box,fill=white] (c) at (4,0.42) {$c,\quad\pi(c)=\{\mathrm{red}\}$};
  \draw[edge,very thick] (a) -- (c);
  \node[note] (n4) at (7.95,0.92) {insert $ac$};
  \node[note] (n5) at (7.95,0.0) {then recolor the earlier factor};
\end{tikzpicture}
\caption{A product of two derivations.
The left factor has vertices $a,b$ with different colors.
The right factor has one vertex whose profile contains red.
Thus, $ac$ is the only edge we add between the factors, and we add it before recoloring the first factor.}
\label{fig:derivation-product}
\end{figure}

To describe a product of several factors, we also need the profiles and the recoloring of a product of two factors.
Write the two recolorings as $\alpha$ and $\beta$, in the order in which we take the factors.
A vertex of the first factor keeps its profile, and a vertex of the second factor with profile $Y$ gets the profile $\alpha^{-1}(Y)$, since a color arriving from an earlier factor passes through $\alpha$ before it reaches this vertex.
The recoloring of the product is $\beta\circ\alpha$.
These rules are associative, since at each later vertex they test the color obtained by applying all intervening recolorings in order.

A \emph{cell} is specified by a pair $c=(i,X)$ consisting of a color $i\in[q]$ and a profile $X\subseteq[q]$.
We write $\mathcal A=[q]\times 2^{[q]}$ for the set of cell names and $\sigma[c]$ for the vertices of cell $c$ in the derivation $\sigma$.
A \emph{flip} is specified by a set $Z$ of unordered pairs of cell names, where the two names in a pair may agree.
For distinct vertices in cells $c$ and $d$, we reverse their adjacency if and only if $\{c,d\}\in Z$.
We call the resulting graph the \emph{$Z$-flip}.

\input{figures/cell-flip}

The \emph{abstraction} of a derivation~$\sigma$ is a triple~$(L,R,\alpha)$, where~$L$ lists the cells that occur in~$\sigma$ and~$\alpha$ is its recoloring.
The set $R$ tells us, for each $Z$, each pair of cells $c,d$, and each set~$W\subseteq\mathcal A$, whether the $Z$-flip contains a path from cell $c$ to cell $d$ whose internal vertices all belong to cells in $W$.
We allow a path consisting of one vertex, so a path may stay in a single cell even when $W$ is empty.
There are only finitely many abstractions for each $q$.

The abstraction of a product is determined by the abstractions of its factors.
Indeed, a cell $(i,X)$ of the first factor becomes $(\beta(i),X)$ in the product, and a cell $(j,Y)$ of the second becomes $(j,\alpha^{-1}(Y))$.
These two maps tell us which flip to apply inside each factor, which cells may contain internal vertices, and which cells occur in the product.
Between the two old cells, adjacency is complete if $i\in Y$ and empty otherwise, and we reverse this choice if the pair of new cells belongs to the flip.
We want to determine whether a given flip of the product graph contains a path with endpoints in two specified cells and internal vertices in specified allowed cells.
Such a path splits into subpaths lying entirely in one factor, joined by edges between the two factors.
For each piece, we need only the path information in $R$ for its factor, with the additional requirement that every endpoint used in the interior of the whole path belongs to an allowed cell.
Conversely, such pieces can be joined whenever the indicated pair of cells is complete across the cut, independently of which vertices witness the pieces.
The resulting walk contains a path with the same endpoints and with its internal vertices in the allowed cells.
Thus, the abstractions of all $q$-derivations form a finite semigroup, which we denote by $S_q$.

\input{figures/abstraction-path}

A flip may split a block among several connected components.
We will then have to decompose the part of the block in just one component, given only the graph induced by that part.
Boja\'nczyk, Grohe, and Pilipczuk proved the following restriction lemma for definable cliquewidth on arbitrary graphs~\cite[Lemma~5.8]{linear}.
\begin{lemma}
\label{lem:blockrestriction}
Let $J$ be a nonempty induced subgraph of $B$.
On either side of the partition $(V(J),V(B)\setminus V(J))$, put two vertices in the same class if they have the same neighbors on the other side, and let $b$ be the total number of classes on the two sides.
Then $\operatorname{dcw}(J)$ is bounded by a function of $b$ and $\operatorname{dcw}(B)$.
\end{lemma}

\input{figures/block-neighborhoods}

Now let~$G$ be the underlying graph of~$\sigma_1\cdots\sigma_n$ and let~$B_i$ be the vertex set of~$\sigma_i$.
The \emph{block order} on $V(G)$ is the relation $x\preceq y$ that holds for $x\in B_i$ and $y\in B_j$ if and only if $i\leq j$.
An element $e$ of $S_q$ is \emph{idempotent} if $e^2=e$.
When every factor has the same idempotent abstraction~$e$, every nonempty consecutive product also has abstraction~$e$.
We can then recover the block order within each component of a suitable flip.
We use the definable order lemma of Boja\'nczyk, Grohe, and Pilipczuk~\cite[Lemma~5.2]{linear}.
Their application starts with factors on at most one vertex.
The proof of their definable order lemma uses only that the factors in this product have the same idempotent abstraction~\cite[Section~6]{linear}.
Indeed, idempotence of the recoloring ensures that the chosen flip acts in the same way whether we use the cells of a factor or those of the product~\cite[Lemmas~6.10 and~6.11]{linear}.
The common path information in~$R$ tells us whether a factor contains a path between the required cells with all its internal vertices in the allowed cells~\cite[Definition~4.3]{linear}.
Boja\'nczyk, Grohe, and Pilipczuk use this information to find such paths inside individual blocks~\cite[Lemma~6.15 and the proof of Claim~6.17]{linear}.
These arguments apply equally when each factor contains a whole block.
\begin{lemma}
\label{lem:blockordersincomponents}
Suppose that all the $q$-derivations $\sigma_1,\ldots,\sigma_n$ have the same idempotent abstraction.
There is a flip $Z$ on the cells of their product such that, with a bounded number of unary parameters, an MSO formula on $G$ defines
\[
  x\preceq y\quad\text{and}\quad
  x,y\text{ belong to the same connected component of the }Z\text{-flip}.
\]
The parameter bound and the choice from a finite list of formulas depend only on~$q$.
\end{lemma}

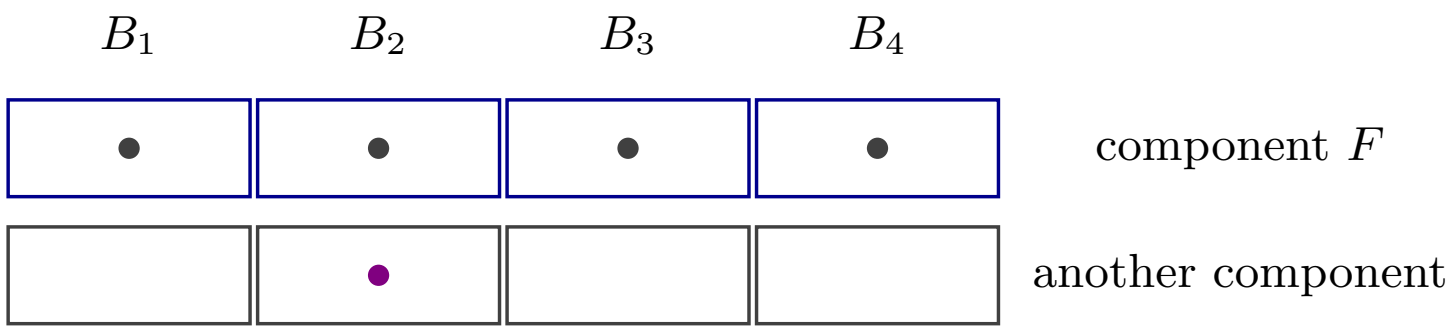
\begin{figure}[H]
\centering
\color{black}
\begin{tikzpicture}[manuscript illustration]
\foreach \i in {1,...,4}{
    \node[note] at (1.65*\i,1.78) {$B_{\i}$};
    \draw[blue!55!black] (1.65*\i-.8,.71) rectangle ++(1.6,.64);
    \draw[black!75] (1.65*\i-.8,-.13) rectangle ++(1.6,.64);
  }
  \foreach \i in {1,2,3,4} \fill[black!75] (1.65*\i,1.03) circle (2pt);
  \fill[violet] (3.30,.19) circle (2pt);
  \node[note] at (9,1.03) {component $F$};
  \node[note] at (9,.19) {another component};
\end{tikzpicture}
\caption{Columns show the original blocks.
A dot marks a nonempty intersection of a component of the flip with a block.
Here $F$ meets all four blocks, and the second component meets only $B_2$.
Within each row, the reconstructed relation orders the nonempty pieces by column and ties vertices in the same piece.
Vertices in different rows are incomparable in this relation.}
\label{fig:component-block-order}
\end{figure}

Once the block order is available, we use the combining lemma of Boja\'nczyk, Grohe, and Pilipczuk~\cite[Lemma~5.7]{linear}.

\begin{lemma}\label{lem:orderedcombiner}
For all positive integers~$q$ and~$K$, there is an MSO transduction with the following property.
Let $G$ be the underlying graph of a product of $q$-derivations whose underlying graphs have definable cliquewidth at most $K$, and let $\preceq$ be their block order.
Given $(G,\preceq)$, the transduction constructs a cliquewidth decomposition of $G$ whose width is bounded in terms of $q,K$.
Every accepted output is a decomposition of $G$ under the leaf-origin map, and there is at least one accepted output for every such input.
\end{lemma}

\input{figures/ordered-block-assembly}

Using \cref{lem:blockrestriction,lem:blockordersincomponents,lem:orderedcombiner}, we can now follow the factorization forest induction of Boja\'nczyk, Grohe, and Pilipczuk~\cite[Sections~4 and~5]{linear} to prove the following closure statement.
Their argument starts with factors containing at most one vertex.
Here the hypothesis gives a bound on the definable cliquewidth of each initial block, and the same induction combines these blocks using the preceding lemmas.

\begin{lemma}\label{lem:derivationclosure}
For each $q,m$, the underlying graphs of all finite products of $q$-derivations whose underlying graphs have definable cliquewidth at most $m$ have uniformly bounded definable cliquewidth.
\end{lemma}
\begin{proof}
Fix $q$ and apply Simon's factorization forest theorem~\cite{simon} to the word of abstractions of the given factors in the finite semigroup~$S_q$.
It gives a tree whose leaves are the factors in their given order and whose height is bounded by a function $h(q)$.
Each internal node either has two children or has children whose products all have the same idempotent abstraction.

At a leaf, the underlying graph has definable cliquewidth at most $m$ by hypothesis.
For the induction step, suppose that the graph represented by each child of an internal node has definable cliquewidth at most $K$.
We show that the graph represented by the parent has definable cliquewidth bounded in terms of $q$ and $K$.
We regard the products represented by the children as the factors in this step.

For two factors, one unary predicate can select the vertices of the first factor and thus give the block order.
We then apply \cref{lem:orderedcombiner} to combine the two factors.

When the children have a common idempotent abstraction, we use unary colors to choose the cells of their product and choose a flip $Z$ from the finite list of possible flips.
By \cref{lem:blockordersincomponents}, we can choose $Z$ and the parameters so as to define the block order within each component $F$ of the $Z$-flip.
If $B$ is a block meeting $F$, vertices on the same side of the cut between $B\cap F$ and $B\setminus F$ with the same product cell have the same neighbors across the cut, since the flip removes all edges across it.
There are thus at most $2|\mathcal A|$ neighborhood classes, and \cref{lem:blockrestriction} bounds the definable cliquewidth of the piece induced by $B\cap F$ in terms of $q$ and $K$.
Restricting each factor to its vertices in $F$ gives a product representation of $G[F]$.
An empty factor only changes colors, so its recoloring can be incorporated into the next factor by the product rule, or omitted if there is no next factor and only the underlying graph is needed.
We now apply \cref{lem:orderedcombiner} in parallel to the original graphs $G[F]$, using the block order just defined within each component.
We use the transfer and color-enforcement lemmas of Boja\'nczyk, Grohe, and Pilipczuk~\cite[Lemmas~5.4 and~5.5]{linear} to give each vertex its product cell as a color at the root of the component expression.
Indeed, the leaf still has the original vertex as its origin, so we can refine its construction color by its cell, keep the cell at every recoloring, and at the root recolor by the cell alone.
One final Join now adds all edges between components prescribed by $Z$, since there are no edges between them in the flip.

In both cases, we evaluate the proposed expression on its leaves and check that the leaf-origin map is an isomorphism to the input graph~\cite[Lemma~5.3]{linear}.
The choices described above give at least one expression that passes this check, and both the transduction and the check have size bounded in terms of $q$ and $K$.
Let $f_q(K)\geq K$ be a common nondecreasing bound for the definable cliquewidth obtained in the two cases.
Starting at~$m$, we apply~$f_q$ at most~$h(q)$ times, so the bound at the root depends only on~$q$ and~$m$.
The factorization tree establishes this bound, while at each step the transduction chooses the colors needed to apply the relevant lemmas to the graph itself.
Up to logical equivalence and renaming of the fixed signatures, there are only finitely many transductions of this bounded size.
We can view all of their output expressions over a common palette by adding unused colors.
Taking the finite union of the decomposers then gives one decomposer for the whole class~\cite[Lemma~2.3]{linear}.
\end{proof}

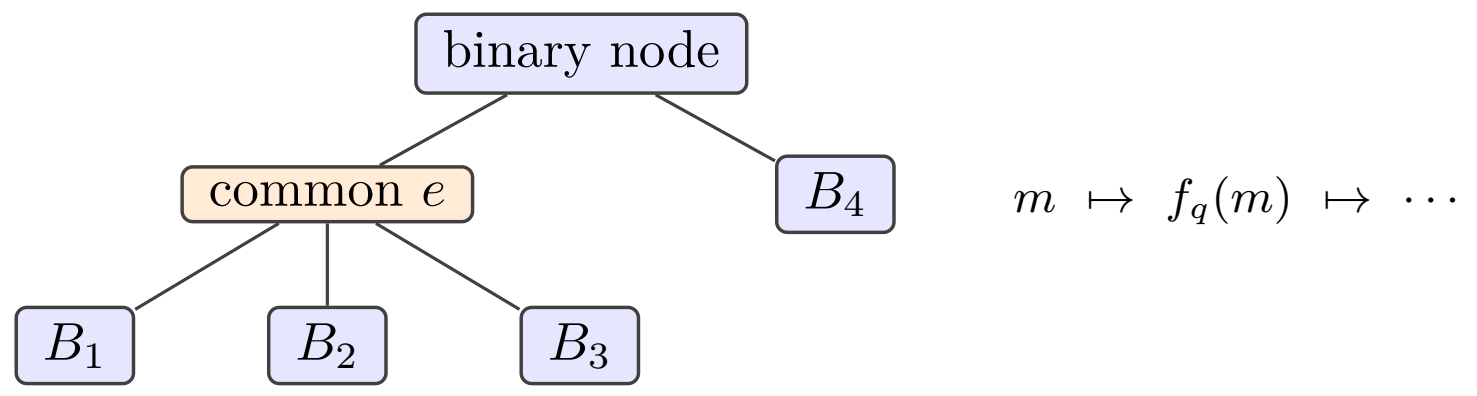
\begin{figure}[H]
\centering
\color{black}
\begin{tikzpicture}[manuscript illustration]
  \node[box] (r) at (4.12,1.66) {binary node};
  \node[key] (i) at (2.44,0.73) {common $e$};
  \node[box] (t) at (5.8,0.73) {$B_4$};
  \draw[edge] (i) -- (r);
  \draw[edge] (r) -- (t);
  \node[box] (s0) at (0.77,-0.27) {$B_1$};
  \draw[edge] (i) -- (s0);
  \node[box] (s1) at (2.44,-0.27) {$B_2$};
  \draw[edge] (i) -- (s1);
  \node[box] (s2) at (4.11,-0.27) {$B_3$};
  \draw[edge] (i) -- (s2);
  \node[note] (n11) at (8.48,0.69) {$m\ \mapsto\ f_q(m)\ \mapsto\ \cdots$};
\end{tikzpicture}
\caption{A node of the factorization tree may have many children with the same idempotent abstraction.
At the lower node, we recover the block order inside each component of the flip by \cref{lem:blockordersincomponents}, and at the binary node we need only choose its first factor.
The height bounds the number of times we apply $f_q$ to the initial bound $m$.}
\label{fig:factorization-height}
\end{figure}

\linearclosure*

\begin{proof}
We fix a partition and an order witnessing the hypothesis and use them to show that~$G$ has a product representation to which \cref{lem:derivationclosure} applies.

Put $q=2^r$ and $P_i=W_1\cup\cdots\cup W_i$.
Partition $P_i$ by equality of neighborhoods in~$V(G)\setminus P_i$.
There are at most $q$ classes, since the corresponding cut matrix has rank at most~$r$.
Assign these classes distinct colors from $[q]$, independently for every $i$.

\input{figures/prefix-row-colors}

When passing from $P_{i-1}$ to $P_i$, each old color class lies in one new class, because the outside vertex set only shrinks.
Let $\phi_i$ be this recoloring, extended arbitrarily to unused colors.
Color each vertex of $W_i$ by its class in $P_i$, and give it the profile consisting of the old classes to which it is adjacent.
This profile is well-defined because the vertices in an old color class have the same neighbors outside~$P_{i-1}$, including in~$W_i$.

Let $\sigma_i$ be the resulting $q$-derivation with underlying graph $G[W_i]$.
In the product, the profile of each new vertex is tested against the colors on $P_{i-1}$ before $\phi_i$ changes them to their colors on $P_i$.
Induction on $i$ therefore shows that $\sigma_1\cdots\sigma_i$ has underlying graph $G[P_i]$ and exactly the colors chosen for $P_i$.
Thus, $G$ is the underlying graph of a product of $q$-derivations whose underlying graphs have definable cliquewidth at most $m$.
By \cref{lem:derivationclosure}, one fixed MSO decomposer works for all such products, with its size depending only on~$q$ and~$m$.
The induction on $i$ verifies the product representation, while this single decomposer acts on~$G$ alone.
Its number of copying and coloring operations is bounded independently of the number $t$ of blocks.
Since $q=2^r$, we obtain the required bound $F(r,m)$.
\end{proof}

\subsection{The rank-one base case}
\label[appendix]{app:rankonebase}

\rankonebase*

\begin{proof}
Consider a connected component with at least three vertices.
We apply the transduction of Campbell and K\"ohler to obtain its canonical split tree, whose leaves are the vertices of the component~\cite[Corollary~7(3) and Appendix~A]{laminar}.
Each internal node carries a graph, called its bag, with one vertex for each incident tree edge.
Since the component is distance-hereditary, each bag is a clique or a star~\cite[Section~3]{gioanpaul}.
We keep the original adjacency on the leaves until we have determined the labels of the bags.

We call the bag vertex corresponding to an incident tree edge its \emph{port}.
A path between two original leaves is \emph{accessible} if, at every bag it visits, the two ports used by the path are adjacent in that bag.
The split tree represents the graph by declaring two distinct leaves adjacent exactly when their path is accessible~\cite[Definition~2.2]{gioanpaul}.
Replace each bag node by a hub with one adjacent vertex for each of its ports.
Join the two ports belonging to an internal tree edge, and join a leaf-side port to its original vertex.
This graph is a subdivision of the decomposition tree, and hence is a tree.
Mark each hub as a clique or star, and at a star hub mark its unique center port.
Keep the original vertices as designated source leaves.

These labels are definable even if the split transduction returns only the tree and the copy of the input graph.
Two ports of one bag are adjacent in its bag graph if and only if there is an original graph edge with endpoints in their two tree branches.
To see the forward direction, continue from either port through adjacent bag ports until reaching an original leaf.
Such a continuation exists because every clique or star bag is connected, and it cannot return along the edge just used.
Finiteness of the tree therefore gives an accessible original leaf on each side.
Conversely, an edge between the two branches has an accessible path, so its two ports in this bag must be adjacent.
We can therefore define the bag type and the center port in MSO.

Leaves in different components are nonadjacent.
For two distinct original leaves $x,y$ in the same component, consider their unique tree path.
The vertices $x$ and $y$ were adjacent in the input graph if and only if, at every star hub on their path, one of the two ports on the path is the center port.
Clique hubs impose no further condition.
An MSO formula can test this condition on the unique path, giving the required inverse.

The construction uses only finite copying.
After choosing a root, each nonroot node names its parent edge, and two fixed copies of that node suffice for the ports at the two ends.
The total number of tree nodes and ports is linear, since every internal decomposition node has degree at least three.
In a component of size two, subdivide its edge once to give a tree with both original vertices as leaves.
An isolated vertex forms a one-vertex tree.
Apply the construction in parallel to connected components, obtaining a forest.
Any given finite unary labels are kept on the source vertices.
\end{proof}

\subsection{Pair witnesses inside a prescribed big set}
\label[appendix]{app:localpairwitness}

\localpairwitness*
\begin{proof}
The set~$A$ itself is a pair witness, because any three big sets have nonempty intersection.
Choose an inclusion-minimal pair witness $W\subseteq A$.

To bound the size of $W$, we first show that a chain that repeatedly removes a new minimal big set has length at most $k$.
We then use minimality of $W$ to arrange its vertices in a binary recursion of this bounded depth.

We use the intersection-chain argument of Seibt~\cite[Proposition~5]{witnesses}, and include the argument to make the localization explicit.
Let $B_1,\ldots,B_s$ be inclusion-minimal big sets such that
\[
 V\supsetneq V\setminus B_1
 \supsetneq V\setminus(B_1\cup B_2)
 \supsetneq\cdots\supsetneq V\setminus(B_1\cup\cdots\cup B_s).
\]
Put $U_i=B_1\cup\cdots\cup B_i$.
We claim $\lambda(U_i)\leq k-i$.
The claim holds for $i=1$.
Strictness implies that $B_{i+1}\cap U_i$ is a proper subset of $B_{i+1}$.
If it had order less than $k$, the two big sets $B_{i+1},U_i$ would force their intersection to be big, contrary to minimality of $B_{i+1}$.
Its order is therefore at least $k$, and submodularity gives
\[
 \lambda(U_{i+1})
 \leq\lambda(U_i)+\lambda(B_{i+1})
       -\lambda(U_i\cap B_{i+1})
 \leq(k-i)+(k-1)-k=k-i-1.
\]
Here $U_i$ is big whenever its order is less than $k$, because it contains the big set $B_1$.
Nonnegativity proves $s\leq k$.

For each $w\in W$, minimality gives big sets $B_w,C_w$ such that $W\cap B_w\cap C_w=\{w\}$.
We may shrink both to inclusion-minimal big sets: the pair-witness property ensures that their intersection with $W$ remains nonempty, and hence remains exactly $\{w\}$.

Now start with $I=V$.
If $I\cap W$ is nonempty, select $w\in I\cap W$.
Every other point of $I\cap W$ lies in $(I\setminus B_w)\cup(I\setminus C_w)$.
Both of these are proper subsets of $I$, since they omit~$w$.
Repeating this operation gives a binary recursion whose successive regions are strict intersections of complements of minimal big sets.
By the preceding bound, a region reached after~$k$ strict intersections contains no point of~$W$, since such a point would allow one more strict step.
The recursion therefore has at most $1+2+\cdots+2^{k-1}=2^k-1$ selected points, and covers $W$.
This proves the claim.
\end{proof}

\end{document}